\pdfoutput=1
\documentclass[11pt,a4paper]{article}
\usepackage[T1]{fontenc}
\usepackage{lmodern}
\usepackage[margin=2.5cm]{geometry}
\usepackage{amsmath}
\usepackage{amssymb}
\usepackage{amsthm}
\usepackage{microtype}
\usepackage{graphicx}
\usepackage{booktabs}
\usepackage{array}
\usepackage{float}
\usepackage{adjustbox}
\usepackage{setspace}
\usepackage[font=small,labelfont=bf,labelsep=period,skip=5pt]{caption}
\usepackage{natbib}
\bibpunct[, ]{(}{)}{,}{a}{}{,}

\usepackage{xcolor}
\definecolor{linkblue}{rgb}{0.10,0.20,0.50}
\usepackage{xurl}
\usepackage[colorlinks=true,linkcolor=linkblue,citecolor=linkblue,urlcolor=linkblue]{hyperref}
\usepackage{orcidlink}
\usepackage{algorithm}
\usepackage{algpseudocode}
\hypersetup{pdftitle={Certified benefit intervals for network changes under user equilibrium},pdfauthor={Yue Li, Shujuan Chen, Ying Jin},pdfkeywords={traffic assignment, user equilibrium, a posteriori error bounds, transport appraisal, verified computation}}
\graphicspath{{figures/}}
\theoremstyle{plain}
\newtheorem{theorem}{Theorem}
\newtheorem{lemma}{Lemma}
\newtheorem{proposition}{Proposition}
\newtheorem{corollary}{Corollary}
\newtheorem{applemma}{Lemma}[section]
\theoremstyle{definition}
\newtheorem{definition}{Definition}
\newtheorem{assumption}{Assumption}
\newtheorem{example}{Example}
\newenvironment{pf}[1][Proof]{\begin{proof}[#1]}{\end{proof}}
\newcommand{\Links}{\mathcal{A}}
\newcommand{\ODs}{\mathcal{W}}
\newcommand{\Routes}{\mathcal{K}}
\newcommand{\Feas}{X}
\newcommand{\R}{\mathbb{R}}
\newcommand{\T}{^{\top}}
\newcommand{\xs}{{x^{*}}}
\newcommand{\cs}{{c^{*}}}
\newcommand{\Js}{{J^{*}}}
\newcommand{\xh}{\hat{x}}
\newcommand{\ch}{\hat{c}}
\newcommand{\cb}{\bar{c}}
\newcommand{\Bh}{\hat{B}}
\newcommand{\Bm}{\tilde{B}}
\newcommand{\TSTT}{\mathrm{TSTT}}
\newcommand{\SPTTm}{\mathrm{SPTT}}
\newcommand{\TC}{\mathrm{TC}}
\newcommand{\Gc}{G_{\mathrm{c}}}
\newcommand{\epsabs}{\varepsilon_{\mathrm{abs}}}
\newcommand{\epsrel}{\varepsilon_{\mathrm{rel}}}
\newcommand{\appref}[1]{Appendix~\ref{#1}}
\newcommand{\floatnote}[1]{\par\vspace{3pt}{\leftskip=0pt\rightskip=0pt\parfillskip=0pt plus 1fil\footnotesize\noindent
  \textit{Note.}\enskip#1\par}}
\newcommand{\Mref}[1]{\ref{#1}}
\newcommand{\Meqref}[1]{\eqref{#1}}
\newcommand{\tocnote}[1]{{\par\noindent\leftskip=2.2em\rightskip=2.55em\small\itshape #1\par\vspace{2pt}}}
\makeatletter
\renewcommand{\@seccntformat}[1]{\ifcsname #1@cntformat\endcsname\csname #1@cntformat\endcsname
  \else\csname the#1\endcsname\quad\fi}
\newcommand{\section@cntformat}{\thesection\quad}
\g@addto@macro\appendix{\renewcommand{\section@cntformat}{Appendix~\thesection\quad}}
\renewcommand*{\l@section}{\@dottedtocline{1}{0em}{2.2em}}
\renewcommand*{\l@subsection}{\@dottedtocline{2}{2.2em}{2.6em}}
\newcommand{\suppcontents}{\begingroup\setstretch{1.05}\@starttoc{stoc}\endgroup}
\newcommand{\suppstart}{\renewcommand{\section@cntformat}{\thesection\quad}
  \setcounter{section}{0}\def\thesection{S\arabic{section}}\def\theHsection{S\arabic{section}}
  \setcounter{table}{0}\def\thetable{S\arabic{table}}\def\theHtable{S\arabic{table}}
  \setcounter{figure}{0}\def\thefigure{S\arabic{figure}}\def\theHfigure{S\arabic{figure}}
  \setcounter{equation}{0}\def\theequation{S\arabic{equation}}\def\theHequation{S\arabic{equation}}
  \setcounter{proposition}{0}\def\theproposition{S\arabic{proposition}}\def\theHproposition{S\arabic{proposition}}
  \let\arx@addcontentsline\addcontentsline
  \def\addcontentsline##1##2##3{\def\arx@a{##1}\def\arx@b{toc}\ifx\arx@a\arx@b
    \arx@addcontentsline{stoc}{##2}{##3}\else\arx@addcontentsline{##1}{##2}{##3}\fi}}
\makeatother

\begin{document}

\title{\parbox{0.7\textwidth}{\centering Certified benefit intervals for network changes under user equilibrium}}
\author{Yue Li\,\orcidlink{0000-0002-0947-5374}, Shujuan Chen\,\orcidlink{0000-0001-6829-6547}\thanks{Corresponding author: Shujuan Chen, \href{mailto:sc2343@cam.ac.uk}{sc2343@cam.ac.uk}}\kern0.3em{} and Ying Jin\,\orcidlink{0000-0003-2683-6829}\\[1.5ex]
\small Martin Centre for Architectural and Urban Studies, University of Cambridge\\
\small 1\textendash 5 Scroope Terrace, Cambridge, CB2 1PX, United Kingdom\\
\small Centre for Smart Infrastructure and Construction\\
\small Department of Engineering, University of Cambridge\\
\small Civil Engineering Building, 7a JJ Thomson Avenue, Cambridge, CB3 0FA, United Kingdom}
\date{}
\maketitle

\begin{abstract}
Transport appraisal relies on differences in equilibrium travel costs between alternative network configurations. These differences can be small relative to the total cost of a network, making benefit estimates sensitive to numerical errors in traffic assignment. A small convergence gap does not guarantee that a benefit estimate is accurate.
We derive computable two-sided bounds on equilibrium total travel cost at any exactly feasible assignment, without requiring strong monotonicity, nondegeneracy or uniqueness of equilibrium flows. These bounds provide guaranteed intervals for scheme benefits and enable certification of their signs and rankings. We establish the tightest possible bounds among rules based only on the convergence gap and an externality weight, and prove that, for Bureau of Public Roads cost functions, no such rule can certify an error smaller than a multiple of the square root of the gap.
Building on these results, we develop an incremental appraisal procedure that certifies a shared base equilibrium and controls the numerical accuracy of each scheme through verified computation. Experiments on regional networks with up to 2.3 million origin\textendash destination pairs show that conventional gap-based stopping produces inaccurate benefit estimates, with 25 of 55 held-out schemes failing to meet the required accuracy at a relative gap of 0.0001. In a prespecified blind test of 72 schemes, the proposed method achieved the required accuracy for every scheme and certified which scheme had the largest benefit on each network. For schemes with benefits above 0.1\% of the total cost, certification required a median of 0.75 times the scheme-level cost of a cold assignment stopped at a relative gap of 0.000001.
The results establish a mathematical foundation for controlling numerical error in network appraisal and provide a computational procedure that links assignment accuracy directly to the decisions it supports.

\end{abstract}

\section{Introduction}\label{sec:introduction}

\subsection{Background}\label{sec:background}

Traffic assignment models play an important role in the appraisal of road network changes. These models predict how travelers respond to changes in the network and how their responses affect congestion and travel times. The difference in travel costs with and without a proposed scheme determines its travel-cost benefit and its ranking among alternative schemes \citep{DfT2026TAGM31}. At the core of these models is the concept of static user equilibrium introduced by \citet{Wardrop1952}, under which no traveler can reduce their travel time by changing routes \citep{BeckmannEtAl1956,Smith1979,Dafermos1980,Patriksson1994}. 

Under fixed demand, the user benefit of a scheme is given by the reduction in equilibrium total travel cost, which equals the demand-weighted sum of origin\textendash destination (OD) travel costs. In practice, the equilibria with and without the scheme are approximated using iterative assignment algorithms that typically terminate when the relative convergence gap falls below a prescribed tolerance \citep{BoyceEtAl2004,PerederieievaEtAl2015}. The relative gap compares the total system travel time (TSTT) with the total travel time that would result if all travelers used their shortest routes at the current link costs. It therefore measures the extent to which a computed assignment satisfies the equilibrium conditions.

The benefit of a scheme can be small relative to the total travel cost of the network, particularly when the scheme affects only a few links in a large regional network. Numerical errors in the two equilibrium solutions can consequently be comparable to or larger than the estimated benefit. Such errors can lead to incorrect conclusions about whether a scheme reduces travel costs or which of several alternative schemes provides the greatest benefit.

Current appraisal practice addresses this problem through convergence requirements. UK guidance recommends that assignment models achieve sufficient convergence to distinguish scheme benefits from model noise \citep[paragraph 3.4.13]{DfT2026TAGM31}. Ideally, the benefit as a proportion of network cost should be at least ten times the relative gap achieved in the assignments with and without the scheme, although smaller ratios may also be acceptable \citep[paragraph 3.4.15]{DfT2026TAGM31}. The guidance also recognizes that numerical errors arising from insufficient convergence can obscure the benefits of schemes with small impacts on large networks \citep[paragraph D.5.2]{DfT2026TAGM31}. These recommendations provide practical criteria for controlling convergence error, but the relationship between the convergence gap and the accuracy of the resulting benefit estimates remains unresolved.

\subsection{Related Literature}\label{sec:literature}

Previous studies of convergence error in traffic assignment have focused primarily on the accuracy of flow estimates. \citet{RoseEtAl1988} associated flow estimation errors with overlapping alternative routes, while \citet{BoyceEtAl2004} showed that differences in link flows between build and no-build scenarios stabilize only at tight convergence tolerances. \citet{InoueMaruyama2012} found that link-flow errors were approximately proportional to the duality gap, and \citet{PatilEtAl2021} examined the behavior of different convergence metrics. More recently, \citet{BagchiBoyles2025} derived local error bounds for path flows and TSTT under logit stochastic user equilibrium to guide early termination. Meanwhile, tighter convergence has become computationally feasible through origin-based and bush-based algorithms \citep{BarGera2002,Dial2006,Nie2010,Gentile2014}, paired alternative segments \citep{BarGera2010,XieXie2016} and decomposition methods \citep{JafariEtAl2017}.

The mathematical relationship between convergence and solution accuracy has been studied extensively for variational inequalities (VIs), which provide a general formulation of user equilibrium \citep{Dafermos1980}. Under strong monotonicity, the gap provides a bound on the squared distance to the equilibrium solution \citep{Pang1987,LarssonPatriksson1994,FacchineiPang2003}. For monotone linear complementarity problems, error bounds generally depend on the square root of the residual \citep{MangasarianShiau1986,LuoEtAl1994}, while nondegeneracy permits linear bounds \citep{Mangasarian1990}. Local error bounds establish relationships between residuals and solution distances in a neighborhood of the solution set \citep{LuoTseng1993,Pang1997,DrusvyatskiyLewis2018}, whereas weak sharpness provides such relationships over the feasible set \citep{BurkeFerris1993,MarcotteZhu1998}. Gap and merit functions provide equivalent optimization formulations of VIs \citep{Auchmuty1989,Fukushima1992,LarssonPatriksson1994}, and conjugate duality provides a theoretical basis for their interpretation and construction \citep{ChenEtAl1997,AltangerelEtAl2007}.

Sensitivity analysis provides a related approach to studying equilibrium responses to network changes. Existing methods characterize the derivatives of equilibrium flows and costs with respect to network parameters \citep{TobinFriesz1988,Patriksson2004,JosefssonPatriksson2007}. \citet{DafermosNagurney1984} established bounds on changes in strongly monotone equilibria resulting from perturbations of their cost maps, while \citet{BarGeraEtAl2013} investigated the numerical accuracy of equilibrium sensitivities in network design and road pricing. The computation of perturbed equilibria is also related to Newton-type methods, which solve successive affine VIs obtained by linearizing the original problem \citep{Josephy1979,Robinson1980,IzmailovSolodov2010}. Truncated Newton methods with forcing terms have been applied to traffic equilibrium problems \citep{DemboEtAl1982,DemboTulowitzki1988}.

Direct estimation of numerical errors in quantities of interest has been developed in other fields. In computational mechanics, two-sided bounds and error estimates have been derived for scalar quantities evaluated at approximate solutions \citep{PragerSynge1947,ParaschivoiuEtAl1997,BeckerRannacher2001}. Verified computation provides rigorous numerical bounds in the presence of floating-point errors, with established applications in linear programming \citep{NeumaierShcherbina2004,KeilJansson2006,Rump2010}.

\subsection{Research Gap and Contributions}\label{sec:gap}

Despite advances in traffic assignment algorithms and mathematical error bounds, the numerical accuracy of scheme benefit estimates remains unresolved. Existing error bounds for VIs primarily characterize the distance between an approximate solution and the solution set. Explicit bounds are available under strong monotonicity, while for monotone affine and analytic problems the existence of error bounds has been established without explicit constants \citep[Sections 6.3 to 6.6]{FacchineiPang2003}. Standard traffic assignment models do not generally satisfy strong monotonicity because Bureau of Public Roads (BPR) cost functions can have zero slopes at zero flow, and centroid connectors often have constant costs. Equilibrium solutions can also be degenerate when unused routes have the same travel costs as used routes. Moreover, large errors in flows can translate into small errors in travel costs when link cost functions are flat. These properties call for error bounds directly on equilibrium total travel cost.

A related question is how the convergence gap determines the accuracy that can be guaranteed for equilibrium total travel cost. Without an explicit relationship between these quantities, the convergence required for a prescribed appraisal accuracy cannot be determined. Current practice therefore relies on heuristic convergence criteria \citep{DfT2026TAGM31}. A further challenge is to obtain such guarantees at a computational cost suitable for evaluating multiple schemes. Sensitivity analysis characterizes equilibrium responses to network changes, and warm starts reduce the computational effort of subsequent assignments. However, neither provides guaranteed bounds on finite changes in equilibrium total travel cost. A comparison of the properties of existing methods is provided in Section~S1 of the Supplementary Material.

To address these gaps, this study makes three contributions.

First, we derive computable two-sided bounds on equilibrium total travel cost at any exactly feasible link flow, without requiring strong monotonicity, nondegeneracy, uniqueness of equilibrium flows or proximity to equilibrium. The interval width is known exactly, and the bounds for two network configurations give a guaranteed interval for a scheme benefit. These intervals provide computable criteria for certifying benefit signs and scheme rankings from approximate equilibria.

Second, we establish the fundamental limits of certification based on the convergence gap. We prove that the refined bounds are the tightest possible among rules based solely on the convergence gap and an externality weight. For BPR cost functions, we establish that a relative certified accuracy of $\varepsilon$ requires a relative gap of order $\varepsilon^2$ within this class of rules. A decomposition of the convergence gap further identifies the component that limits the accuracy of the certificates and the additional equilibrium information required for tighter bounds.

Third, we develop a certified incremental appraisal procedure for evaluating multiple schemes against a shared base network. The procedure certifies the base equilibrium once and warm-starts each scheme, using certified benefit intervals to determine when the prescribed accuracy has been achieved. Verified arithmetic ensures that the mathematical guarantees remain valid in numerical computation. We establish the validity of the returned intervals, derive conditions for finite termination of the idealized stopping rule and develop an accuracy allocation rule for the shared base and individual schemes. The method is evaluated on instances with known equilibria, development schemes across five networks and two prespecified blind tests on regional networks with up to 2.3 million OD pairs. These experiments assess the accuracy of conventional gap-based appraisal, the certification of benefit estimates and scheme rankings, and its computational cost.

The study focuses on the numerical accuracy of travel-cost benefits under static user equilibrium with fixed demand. Section~\ref{sec:problem} formulates the appraisal problem. Sections~\ref{sec:enclosure} and~\ref{sec:price} establish the bounds and their theoretical limits. Section~\ref{sec:algorithm} develops the certified appraisal procedure, and Section~\ref{sec:experiments} presents the numerical experiments. Sections~\ref{sec:discussion} and~\ref{sec:conclusions} discuss the implications and conclude the study, respectively. The appendices and Supplementary Material provide proofs, implementation details, extensions and additional experimental results.
\section{The Appraisal Problem}\label{sec:problem}

\subsection{Network and User Equilibrium}\label{sec:network}

Consider a road network with a finite set of links $\Links$ and a set of OD pairs $\ODs$. Each OD pair $w\in\ODs$ has a fixed demand $d_w>0$ and a finite nonempty set of routes $\Routes_w$. A route flow vector $h\ge 0$ is feasible if $\sum_{k\in\Routes_w}h_k=d_w$ for every $w$. The corresponding link flow vector is $x=\Lambda h$, where $\Lambda$ is the link\textendash route incidence matrix. The set of feasible link flows $\Feas$ is a nonempty polytope. Given a vector of link costs $c\in\R^{\Links}$, define
\begin{equation}\label{eq:S}
S(c) := \min_{v\in\Feas} c\T v
\end{equation}
as the total cost of assigning all demand to shortest routes at costs $c$. When $\Routes_w$ contains all routes between the corresponding origin and destination and link costs are nonnegative, $S(c)$ can be computed using one shortest-path tree per origin. As the minimum of linear functions, $S$ is concave and superadditive, satisfying $S(c+c')\ge S(c)+S(c')$ for all cost vectors $c$ and $c'$.

Each link $a$ has a cost function $t_a$ of its own flow $x_a$, and $t(x)$ denotes the vector of link costs.

\begin{assumption}\label{ass:costs}
Each link cost $t_a:[0,\infty)\to\R$ is continuously differentiable, convex and nondecreasing.
\end{assumption}

The BPR function $t_a(x_a)=t_a^0\bigl(1+\alpha_a(x_a/C_a)^{\beta_a}\bigr)$, with free-flow time $t_a^0\ge 0$, $\alpha_a\ge 0$, capacity $C_a>0$ and power $\beta_a\ge 1$, satisfies Assumption~\ref{ass:costs}. Its derivative is zero at zero flow when $\beta_a>1$, and $t_a^0=0$ represents a constant zero-cost link, such as a centroid connector. Assumption~\ref{ass:costs} thus accommodates standard link cost functions without requiring strong monotonicity.

A flow $\xs\in\Feas$ is a user equilibrium \citep{Wardrop1952} if it solves the VI \citep{Dafermos1980}
\begin{equation}\label{eq:vi}
t(\xs)\T(v-\xs)\ge 0 \qquad\text{for all } v\in\Feas.
\end{equation}
Under Assumption~\ref{ass:costs}, an equilibrium exists. The equilibrium link costs $\cs:=t(\xs)$ and the equilibrium total cost
\begin{equation}\label{eq:Jstar}
\Js := \cs\T\xs = S(\cs)
\end{equation}
are unique even when equilibrium link flows are not (Lemma~\ref{lem:unique} in \appref{app:proofs}). The second equality in \eqref{eq:Jstar} follows directly from the equilibrium condition \eqref{eq:vi}.

At any feasible flow $x$, the TSTT and the shortest-path travel time (SPTT) are defined as
\begin{equation}\label{eq:tstt}
\TSTT(x):=t(x)\T x, \qquad \SPTTm(x):=S\bigl(t(x)\bigr).
\end{equation}
The convergence gap $\Gc(x):=\TSTT(x)-\SPTTm(x)$ is nonnegative and vanishes if and only if $x$ is an equilibrium. We define the relative gap as $\Gc(x)/\SPTTm(x)$, consistent with the assignment solver used in Section~\ref{sec:experiments}. Since $\SPTTm(x)$ is the demand-weighted sum of shortest OD travel costs at $t(x)$, it corresponds to the total travel cost estimated from the OD cost matrices in network appraisal.
\subsection{Appraisal Programs and Benefits}\label{sec:program}

An appraisal program consists of a base network, indexed by $\theta=0$, and $m$ schemes, indexed by $s=1,\dots,m$. Each scheme modifies the cost functions of a subset of links, for example through changes in capacity, while keeping demand and route sets unchanged. We denote the link cost functions, equilibrium total cost and shortest-path travel time of network $\theta$ by $t^{(\theta)}$, $\Js_\theta$ and $\SPTTm_\theta$, respectively. The benefit of scheme $s$ is
\begin{equation}\label{eq:benefit}
B_s := \Js_0 - \Js_s.
\end{equation}
Under fixed demand, $B_s$ represents the change in user benefit because $\Js_\theta$ equals the demand-weighted sum of equilibrium OD costs.

In practice, each equilibrium is approximated by a computed feasible flow $\xh_\theta$, and the scheme benefit is estimated as
\begin{equation}\label{eq:estimate}
\Bh_s := \SPTTm_0(\xh_0) - \SPTTm_s(\xh_s).
\end{equation}
We call this estimate accurate when its error is within a tolerance that combines absolute and relative requirements:
\begin{equation}\label{eq:tau}
\tau_s := \epsabs\,\TC_0 + \epsrel\,|B_s|,
\end{equation}
where $\TC_0>0$ is a fixed normalizing constant, such as an estimate of the base network's total travel cost. A fixed constant keeps the tolerance computable, and the lower bound $a_0$ of the base network's certified interval in Section~\ref{sec:algorithm} is a conservative choice. The absolute term specifies the accuracy requirement at the network level, while the relative term accounts for the magnitude of the scheme benefit. Section~\ref{sec:experiments} specifies the values of $\epsabs$ and $\epsrel$.

The sign of $B_s$ determines whether scheme $s$ reduces equilibrium total travel cost, while the sign of $B_i-B_j$ determines which of schemes $i$ and $j$ provides the greater benefit. An interval that is guaranteed to contain $B_s$ and excludes zero certifies the sign of the benefit. Similarly, an interval that is guaranteed to contain $B_i-B_j$ and excludes zero certifies the ranking of the two schemes.

\subsection{Error Propagation from a Shared Base}\label{sec:identity}

Let $e_\theta := \SPTTm_\theta(\xh_\theta) - \Js_\theta$ denote the error in the estimated total cost of network $\theta$. The following proposition shows how the errors of the base and scheme assignments enter the benefit estimate.

\begin{proposition}\label{prop:identity}
For every scheme $s$, $\Bh_s - B_s = e_0 - e_s$.
\end{proposition}

The result follows from $\Bh_s - B_s = \bigl(\SPTTm_0(\xh_0)-\Js_0\bigr) - \bigl(\SPTTm_s(\xh_s)-\Js_s\bigr)$.
When the base solution $\xh_0$ is shared across all $m$ schemes, the same error $e_0$ enters every benefit estimate. As each scheme assignment converges, $e_s$ approaches zero and the benefit error approaches $e_0$. Increasing the accuracy of individual scheme assignments therefore cannot eliminate the error inherited from the base. To guarantee benefit accuracy across the appraisal program, the base error must be controlled below the smallest scheme tolerance. Section~\ref{sec:allocation} develops an accuracy allocation rule based on this property, and Section~\ref{sec:practice} examines its effect on held-out schemes.

\subsection{Gap and Total-Cost Error in a Two-Route Example}\label{sec:example}

\begin{figure}[t]
\centering
\includegraphics[width=\textwidth]{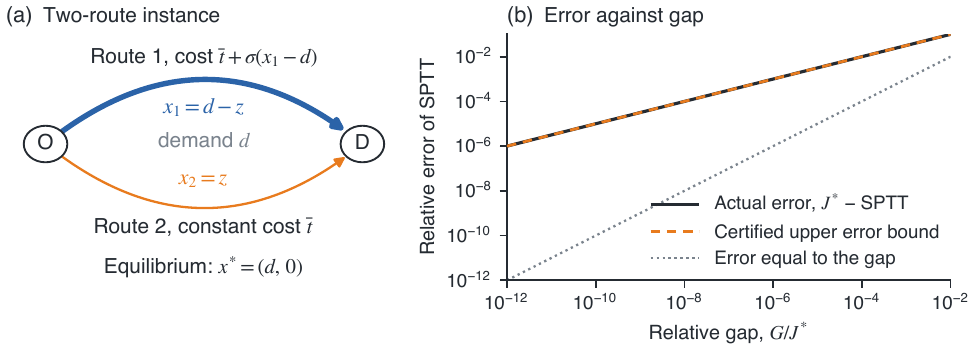}
\caption{Square-root relationship between gap and total-cost error in the two-route example.}\label{fig:example}
\floatnote{(a) Network configuration and feasible flows $x=(d-z,z)$ in Example~\ref{ex:tworoute}. (b) Total-cost estimation error against the convergence gap for $d=\sigma=\bar t=1$, both normalized by $\Js=1$ and shown on logarithmic axes. The certified upper error bound $\psi G^{1/2}+G$, obtained from Theorem~\ref{thm:plain}, coincides with the actual error.}
\end{figure}

The two-route example in Figure~\ref{fig:example}(a) demonstrates that a small convergence gap does not guarantee a small total-cost error.

\begin{example}\label{ex:tworoute}
One OD pair with demand $d$ is served by two routes, each consisting of a single link. Route~1 has the affine cost $t_1(x_1)=\bar t+\sigma(x_1-d)$ with $\sigma>0$, and route~2 has the constant cost $\bar t$. At equilibrium, all demand uses route~1, both routes have cost $\bar t$, and $\Js=d\bar t$. Thus, the unused route has the same cost as the used route. At the feasible point $x=(d-z,z)$ with $0<z<d$, route~1 has cost $\bar t-\sigma z$, giving
\begin{align}
\SPTTm(x)&=d(\bar t-\sigma z), \nonumber\\
\TSTT(x)&=d\bar t-\sigma z d+\sigma z^2, \label{eq:example}\\
\Gc(x)&=\sigma z^2. \nonumber
\end{align}
The error in the total-cost estimate is therefore $\Js-\SPTTm(x)=\sigma z d=d\sqrt{\sigma}\,\sqrt{\Gc(x)}$.
\end{example}

In Example~\ref{ex:tworoute}, the total-cost error is proportional to the square root of the convergence gap. Relative to the equilibrium total cost $d\bar t$, the error equals $(\sigma d/\bar t)^{1/2}$ times the square root of $\Gc(x)/(d\bar t)$. When $\sigma d$ is comparable to $\bar t$, a gap of $10^{-10}$ of the total cost therefore gives a relative total-cost error of order $10^{-5}$. This behavior arises because the unused route has the same cost as the used route at equilibrium. Assigning a small flow to the unused route makes the gap quadratic in the flow error, while the total-cost error remains linear in it. Figure~\ref{fig:example}(b) illustrates this relationship. Section~\ref{sec:price} establishes the resulting limits of gap-based certification.

\section{Certified Intervals for the Equilibrium Total Cost}\label{sec:enclosure}

We derive computable bounds on the equilibrium total cost $\Js$ at any exactly feasible link flow. We first establish the bounds at the computed link costs and then extend them to general backward costs. Combining the bounds for two network configurations provides guaranteed intervals for scheme benefits and criteria for certifying appraisal decisions.
\subsection{A Certificate at the Computed Link Costs}\label{sec:plain}

Let $\xh\in\Feas$ be an exactly feasible link flow, such as the repaired output of an assignment algorithm, with computed link costs $\ch:=t(\xh)$ and convergence gap $G:=\Gc(\xh)$. To bound the equilibrium total cost, we introduce a quantity that characterizes the sensitivity of link costs to changes in flow around $\xh$. For each link, let $\bar z_a\ge0$ be the smallest step satisfying $\bigl(t_a(\xh_a+\bar z_a)-t_a(\xh_a)\bigr)\bar z_a\ge G$, and define $\bar\sigma_a:=\bigl(t_a(\xh_a+\bar z_a)-t_a(\xh_a)\bigr)/\bar z_a$ as the secant slope over this step. We set $\bar\sigma_a:=t_a'(\xh_a)$ when $\bar z_a=0$, and $\bar\sigma_a:=0$ when $t_a$ is constant. We define the externality weight at $\xh$ as
\begin{equation}\label{eq:psi}
\psi:=\Bigl(\sum_{a\in\Links}\bar\sigma_a\,\xh_a^2\Bigr)^{1/2}.
\end{equation}
The term externality weight reflects its relationship to the marginal congestion externality $t_a'(x_a)\,x_a$, the additional travel cost that a unit increase in link flow imposes on other travelers. For a BPR link with a small step $\bar z_a$,
$\bar\sigma_a\xh_a^2\approx t_a'(\xh_a)\,\xh_a^2=\beta_a\,\xh_a\bigl(t_a(\xh_a)-t_a^0\bigr)$, which equals $\beta_a$ times the total delay on the link. When all BPR powers equal $\beta$, $\psi^2$ is therefore close to $\beta$ times the total delay.

\begin{theorem}\label{thm:plain}
Let Assumption~\ref{ass:costs} hold and let $\xh\in\Feas$. Then
\begin{equation}\label{eq:plain}
\begin{aligned}
\Js&\ge\SPTTm(\xh)-\psi G^{1/2},\\
\Js&\le\TSTT(\xh)+\psi G^{1/2},
\end{aligned}
\end{equation}
that is, $\Js\in I_0:=\bigl[\SPTTm(\xh)-\psi G^{1/2},\,\TSTT(\xh)+\psi G^{1/2}\bigr]$. Moreover, $\Js\ge\SPTTm(\xh)-p^*(\psi,G)$, where $p^*(\psi,G):=\psi G^{1/2}-G$ if $\psi\ge2G^{1/2}$ and
$p^*(\psi,G):=\psi^2/4$ otherwise.
\end{theorem}

\begin{pf}
Let $\xs$ be an equilibrium and define $\delta:=\xh-\xs$, $\Delta:=\ch-\cs$, the energy $E:=\Delta\T\delta$ and the margin $M:=\cs\T\delta$. Since each $t_a$ is nondecreasing, $E\ge0$. The equilibrium condition \eqref{eq:vi} at $v=\xh$ gives $M\ge0$. Since $\xs\in\Feas$, $S(\ch)\le\ch\T\xs$, and hence
\begin{equation}\label{eq:EM}
E+M=\ch\T\delta\le\ch\T\xh-S(\ch)=G.
\end{equation}
Write $\Delta_a=\sigma_a\delta_a$, where $\sigma_a\ge0$ is the secant slope of $t_a$ between $\xs_a$ and $\xh_a$, or $t_a'(\xh_a)$ if the two flows coincide. Since $E\le G$, Lemma~\ref{lem:secant} in \appref{app:proofs} gives $\sigma_a\le \bar\sigma_a$ on every link. By the Cauchy\textendash Schwarz inequality,
$|\xh\T\Delta|\le\bigl(\sum_a\sigma_a\xh_a^2\bigr)^{1/2}E^{1/2}\le\psi E^{1/2}$.
Using $\TSTT(\xh)-\Js=\ch\T\xh-\cs\T\xs=\xh\T\Delta+M$, together with $M\ge0$ and $E\le G$, gives the upper bound $\Js\le\TSTT(\xh)+\psi G^{1/2}$.

For the lower bound, substituting $\SPTTm(\xh)=\TSTT(\xh)-G$ and applying \eqref{eq:EM} gives
\begin{equation*}
\begin{aligned}
\SPTTm(\xh)-\Js&=\xh\T\Delta+M-G\\
&\le\psi E^{1/2}-E\le p^*(\psi,G),
\end{aligned}
\end{equation*}
where $p^*(\psi,G)$ is the maximum of $\psi e^{1/2}-e$ over $0\le e\le G$. The first lower bound follows from $p^*(\psi,G)\le\psi G^{1/2}$.
\end{pf}

The interval $I_0$ has width $2\psi G^{1/2}+G$ and contains the point estimate $\SPTTm(\xh)$. Using the tighter lower bound reduces the width to $2\psi G^{1/2}$ when $\psi\ge2G^{1/2}$.

Theorem~\ref{thm:plain} does not require strong monotonicity, nondegeneracy, uniqueness of equilibrium link flows or proximity to equilibrium. It therefore accommodates BPR links at zero flow, constant-cost connectors and routes tied at zero flow. Exact feasibility of $\xh$ is required to apply the equilibrium inequality~\eqref{eq:vi} at $v=\xh$. The route sets used to evaluate $S$ must be those defining the equilibrium. The interval width depends on the square root of the convergence gap, and Section~\ref{sec:price} shows that this dependence is unavoidable for certificates based solely on the gap and the externality weight.

\subsection{General Backward Costs and the Exact Width}\label{sec:quantities}

We generalize the certificate using a second cost vector $\cb\in\R^{\Links}$, referred to as a backward cost. Examples include the costs of a linearized assignment problem and the link costs at a rounded point, both of which are used in Section~\ref{sec:algorithm}. Following backward error analysis, we define the gap of the computed flow under the backward cost as
\begin{equation}\label{eq:G}
G := \cb\T\xh - S(\cb) \ge 0.
\end{equation}
This gap measures the extent to which $\xh$ satisfies the equilibrium condition of the assignment problem with fixed link costs $\cb$. The residual $r:=\ch-\cb$ measures the discrepancy between the actual link costs and the backward costs at $\xh$. When $\cb=\ch$, the residual vanishes and $G$ reduces to the gap $\Gc(\xh)$ of Theorem~\ref{thm:plain}, which we refer to as the raw gap.

Let $\mathcal{P}:=\{a\in\Links:\xh_a>0\}$ denote the links carrying positive flow at $\xh$, and $\mathcal{Z}:=\Links\setminus\mathcal{P}$ the remaining links. For each $a\in\mathcal{P}$, define $\mu_a:=\bigl(t_a(\xh_a)-t_a(0)\bigr)/\xh_a$ as the average slope of the link cost function between zero flow and $\xh_a$. These slopes are used to define a weighted measure of the cost residual $R$, an upper bound on the energy $E_{\max}$ and the backward residual $\varrho$:
\begin{equation}\label{eq:R}
\begin{aligned}
R^2&:=\sum_{a\in\mathcal{P},\,r_a\neq 0} \frac{r_a^2}{\mu_a},\\
E_{\max}&:=\tfrac14\bigl(R+(R^2+4G)^{1/2}\bigr)^2,\\
\varrho&:=\bigl(R^2/4+G\bigr)^{1/2},
\end{aligned}
\end{equation}
with $R:=\infty$ if $\mu_a=0$ on some link of $\mathcal{P}$ with $r_a\neq0$. The steps $\bar z_a$, secant slopes $\bar\sigma_a$ and externality weight $\psi$ are those of Section~\ref{sec:plain} with $E_{\max}$ in place of $G$. At $r=0$, $R=0$, $E_{\max}=G$ and $\varrho=G^{1/2}$.

\begin{theorem}\label{thm:enclosure}
Let Assumption~\ref{ass:costs} hold, let $\xh\in\Feas$, and let $\cb$ be a backward cost with $r_a\ge0$ for every
$a\in\mathcal{Z}$ and $R<\infty$. Then, for every equilibrium $\xs$, with $\delta:=\xh-\xs$ and $\Delta:=\ch-\cs$, the
following hold.
\begin{enumerate}
\item The energy $E:=\Delta\T\delta$ satisfies $0\le E\le E_{\max}$.
\item $\Js\in I_{\mathrm{S}}:=\bigl[\SPTTm(\xh)-\tfrac12 r\T\xh-\psi\varrho,\quad
\SPTTm(\xh)+\tfrac12 r\T\xh+\psi\varrho+G-S(r)\bigr]$.
\item $\Js\in I_{\mathrm{T}}:=\bigl[\TSTT(\xh)-\tfrac12 r\T\xh-\psi\varrho-R^2/4-G,\quad
\TSTT(\xh)-\tfrac12 r\T\xh+\psi\varrho\bigr]$.
\item Hence $\Js\in I:=I_{\mathrm{S}}\cap I_{\mathrm{T}}$.
\end{enumerate}
\end{theorem}

The proof in \appref{app:proofs} extends the argument of Theorem~\ref{thm:plain}. Combining the gap inequality $\cb\T(\xs-\xh)\ge-G$ with the equilibrium inequality $\cs\T(\xh-\xs)\ge0$ gives $E\le r\T\delta+G$. This is the perturbation inequality of \citet{DafermosNagurney1984} applied to an inexact solution of the perturbed problem. Since $r_a\delta_a\le0$ on unused links and convexity ensures $\sigma_a\ge\mu_a$ on used links, we obtain $r\T\delta\le R\,E^{1/2}$ and hence the energy bound in item~1. The remaining steps bound each secant slope $\sigma_a$ by $\bar\sigma_a$, complete the square in $E\le r\T\delta+G$ and bound $\xh\T\Delta$ by the Cauchy\textendash Schwarz inequality, which gives the two intervals and their intersection $I$.

The inequality $r\T\delta\le R\,E^{1/2}$ avoids the need for strong monotonicity by using the link-specific lower bound $\mu_a\le\sigma_a$. Under Assumption~\ref{ass:costs}, $\mu_a$ can be arbitrarily small, and it is zero when the link cost is constant on $[0,\xh_a]$. The condition $R<\infty$ requires the residual to vanish on such links. Beyond the assumptions of Theorem~\ref{thm:plain}, Theorem~\ref{thm:enclosure} requires only $r_a\ge0$ on $\mathcal{Z}$ and $R<\infty$.

The following lemma characterizes the width of the intersection $I$.

\begin{lemma}\label{lem:width}
Under the assumptions of Theorem~\ref{thm:enclosure}, the upper end of $I$ is the upper end of $I_{\mathrm{T}}$, and
\begin{equation}\label{eq:width}
|I| = 2\psi\varrho + \min\bigl(\Gc(\xh),\, G+R^2/4\bigr).
\end{equation}
In particular, $|I|\ge 2\psi\varrho\ge 2\psi\,G^{1/2}$.
\end{lemma}

The width is thus $2\psi\varrho$, twice the product of the externality weight and the backward residual, plus a term bounded above by the raw gap. The next two results show when the interval contains the point estimate and how its lower bound can be tightened as in Theorem~\ref{thm:plain}.

\begin{lemma}\label{lem:inside}
Under the assumptions of Theorem~\ref{thm:enclosure}, if $S(r)\ge0$, which holds when $r\ge0$, then
$\SPTTm(\xh)\in I$.
\end{lemma}

\begin{proposition}\label{prop:sharp}
Under the assumptions of Theorem~\ref{thm:enclosure}, $\Js\ge\ell^*:=\TSTT(\xh)-\tfrac12 r\T\xh-p^*_{\mathrm{T}}$, where
$p^*_{\mathrm{T}}:=\psi\varrho$ if $\psi\ge2\varrho$ and $p^*_{\mathrm{T}}:=\varrho^2+\psi^2/4$ otherwise. Hence $\Js$ lies in the refined
interval $I^*:=\bigl[\max(\ell^*,\ell_{\mathrm{S}}),\,u_{\mathrm{T}}\bigr]\subseteq I$, with $\ell_{\mathrm{S}}$ the lower end of $I_{\mathrm{S}}$ and $u_{\mathrm{T}}$ the
upper end of $I_{\mathrm{T}}$. Its width is $|I^*|=\min\bigl(\psi\varrho+p^*_{\mathrm{T}},\,\Gc(\xh)+2\psi\varrho\bigr)\ge2\psi\varrho$, with
equality when $\psi\ge2\varrho$.
\end{proposition}

When $r=0$, $I$ reduces to $I_0$ and $I^*$ to the interval of Theorem~\ref{thm:plain} with its tighter lower bound.
\subsection{Benefit Intervals and Certified Decisions}\label{sec:benefitint}

Two certified intervals give a guaranteed interval for a scheme benefit by interval subtraction.

\begin{corollary}\label{cor:benefit}
Let $\Js_0\in[a_0,b_0]$ and $\Js_s\in[a_s,b_s]$. Then $B_s\in[L_s,U_s]:=[a_0-b_s,\,b_0-a_s]$, and the estimate $\Bh_s$
of \eqref{eq:estimate} satisfies $|\Bh_s-B_s|\le\mathcal{E}_s:=\max\bigl(\Bh_s-L_s,\,U_s-\Bh_s\bigr)$.
\end{corollary}

We refer to $\mathcal{E}_s$ as the certified error bound of the benefit estimate. When both intervals are obtained from Theorem~\ref{thm:enclosure} with $r\ge0$, Lemma~\ref{lem:inside} ensures that $\Bh_s$ lies within $[L_s,U_s]$, giving $\tfrac12(U_s-L_s)\le\mathcal{E}_s\le U_s-L_s$. The midpoint $\Bm_s:=\tfrac12(L_s+U_s)$ minimizes the maximum possible estimation error over the benefit interval. Its certified error bound is $\tfrac12(U_s-L_s)=w_0+w_s$, where $w_0$ and $w_s$ are the half-widths of the base and scheme intervals, and is therefore no greater than $\mathcal{E}_s$. We retain $\Bh_s$ as the conventional benefit estimate and also report the midpoint in Section~\ref{sec:experiments}.

The base interval $[a_0,b_0]$ is computed once and shared across all schemes in the appraisal program. A benefit interval that excludes zero certifies the sign of the benefit, as described in Section~\ref{sec:program}. When comparing two schemes against the same base, the equilibrium total cost of the base cancels, allowing their ranking to be certified directly from their respective cost intervals.

\begin{corollary}\label{cor:compare}
Let $\Js_i\in[a_i,b_i]$ and $\Js_j\in[a_j,b_j]$ for two schemes $i$ and $j$ with the same base. Then
$B_i-B_j=\Js_j-\Js_i\in[a_j-b_i,\,b_j-a_i]$. In particular, scheme $i$ has the larger benefit whenever $a_j>b_i$.
\end{corollary}

Since the comparison is independent of the width of the base interval, the ranking of two schemes can be certified even when their individual benefits cannot be certified to the required accuracy.

Every certified interval for a network contains its equilibrium total cost, as does the intersection of several such intervals. Intervals computed at successive checkpoints of an assignment run can therefore be intersected, and Corollaries~\ref{cor:benefit} and~\ref{cor:compare} apply to the intersections. The conventional estimate $\Bh_s$ may then lie outside $[L_s,U_s]$, but $\mathcal{E}_s$ remains bounded above by the width of any benefit interval containing both $\Bh_s$ and $[L_s,U_s]$.

Corollary~\ref{cor:benefit} requires only a certified interval for each network. It therefore also applies to schemes that add or remove links, provided that every OD pair remains connected (Section~\ref{sec:limitations}).
\section{Sharpness and Limits of Certification}\label{sec:price}

The interval widths derived in Section~\ref{sec:enclosure} depend on the square root of the convergence gap. We now establish that this dependence is intrinsic to certification based on the gap, determine the convergence required to guarantee a prescribed accuracy for BPR costs, and examine the extent to which alternative backward costs can improve the bounds.

\subsection{Square-Root Error Scaling}\label{sec:barrier}

Example~\ref{ex:tworoute} attains the upper bound of Theorem~\ref{thm:plain}. At the computed flow, $G=\sigma z^2$. Since route~2 has constant cost and route~1 has affine cost, $\bar\sigma_2=0$ and $\bar\sigma_1=\sigma$. Hence $\psi=\sqrt{\sigma}\,(d-z)$, and the upper end of $I_0$ is $\TSTT(\xh)+\psi G^{1/2}=d\bar t=\Js$. The error of the point estimate is proportional to the square root of the gap, and the bound in Theorem~\ref{thm:plain} is exact for this instance.

Similar square-root relationships are established for monotone linear complementarity problems and monotone affine VIs, with a linear bound available under nondegeneracy
\citep[][Theorems 6.4.1 and 6.4.6]{MangasarianShiau1986,Mangasarian1990,FacchineiPang2003}. User equilibrium with affine separable costs belongs to this class of problems when formulated in route flows, with $\Gc$ corresponding to the complementarity product.

\subsection{Optimal Rules from the Gap and the Externality Weight}\label{sec:optimal}

The interval in Theorem~\ref{thm:plain} depends on the point estimate $\SPTTm(\xh)$, the convergence gap $G$ and the externality weight $\psi$. To establish its optimality, we consider all valid interval rules based on them.

\begin{definition}\label{def:rule}
A $(\psi,G)$-rule assigns to every instance satisfying Assumption~\ref{ass:costs} and every point $\xh\in\Feas$,
with gap $G=\Gc(\xh)$ and externality weight $\psi$, the interval
$\bigl[\SPTTm(\xh)-p_-(\psi,G),\,\SPTTm(\xh)+p_+(\psi,G)\bigr]$, where $p_-$ and $p_+$ are functions of $\psi$ and $G$
only. The rule is valid if the interval contains $\Js$ on every such instance and point.
\end{definition}

\begin{corollary}\label{cor:optimal}
With $p^*$ as in Theorem~\ref{thm:plain}, every valid $(\psi,G)$-rule satisfies, for all $\psi,G>0$,
\begin{equation}\label{eq:optimal}
\begin{aligned}
p_+(\psi,G)&\ge\psi G^{1/2}+G,\\
p_-(\psi,G)&\ge p^*(\psi,G).
\end{aligned}
\end{equation}
The interval of Theorem~\ref{thm:plain} with its tighter lower bound,
$\bigl[\SPTTm(\xh)-p^*,\,\SPTTm(\xh)+\psi G^{1/2}+G\bigr]$, is therefore the optimal valid $(\psi,G)$-rule, and its
width is $2\psi G^{1/2}$ when $\psi\ge2G^{1/2}$. The upper bound in \eqref{eq:optimal} is attained for every
$(\psi,G)$ on two-route instances, also when all costs are BPR functions of one fixed power $\beta\ge1$. The lower
bound is attained for every $(\psi,G)$ on instances with affine and constant costs.
\end{corollary}

The upper bound is attained on the instances of Example~\ref{ex:tworoute}. For the lower bound, two OD pairs share a link with affine cost, and one pair has a constant-cost bypass that, at zero flow, costs the same as its route through the shared link (\appref{app:proofs}). Both constructions involve routes that carry no flow at equilibrium but cost the same as used routes, which is common in traffic networks.

At the computed link costs, the tighter interval of Theorem~\ref{thm:plain} is therefore the narrowest possible among valid $(\psi,G)$-rules. The interval $I_0$ has the same upper bound, and its lower bound is lower by at most $G$. When every nonconstant link cost function is strictly convex, as for BPR links with $\beta_a>1$, the optimal lower bound is not attained but can be approached arbitrarily closely. Consequently, it cannot be improved even within this more restrictive class of networks (\appref{app:proofs}).
\subsection{Gap Requirement for BPR Costs}\label{sec:law}

For BPR cost functions, the externality weight is bounded below in terms of total network delay, which gives the convergence gap required to certify a prescribed accuracy.

\begin{proposition}\label{prop:law}
Let all link costs be BPR functions, let $\beta_{\min}$ be the smallest power over the links with $\xh_a>0$ and
$t_a^0\alpha_a>0$, with $\beta_{\min}:=1$ if there is no such link, and let
$D(\xh):=\sum_a\bigl(t_a(\xh_a)-t_a^0\bigr)\xh_a$ be the total delay. Then $\psi^2\ge\beta_{\min}D(\xh)$. Every interval of Theorem~\ref{thm:enclosure}, and the refined interval $I^*$, has
half-width at least $\bigl(\beta_{\min}D(\xh)\,G\bigr)^{1/2}$, and so does every valid $(\psi,G)$-rule. Hence a certified half-width of at most $\varepsilon\,\TSTT(\xh)$ requires
$\beta_{\min}D(\xh)\,G\le\varepsilon^2\,\TSTT(\xh)^2$, that is, when $D(\xh)>0$,
\begin{equation}\label{eq:law}
\frac{G}{\TSTT(\xh)} \le \frac{\varepsilon^2}{\beta_{\min}\,f}, \qquad f:=\frac{D(\xh)}{\TSTT(\xh)}.
\end{equation}
\end{proposition}

The proof follows from the identity $x\,t_a'(x)=\beta_a\bigl(t_a(x)-t_a^0\bigr)$, which relates the marginal congestion externality of a BPR link to its delay. Condition~\eqref{eq:law} is necessary for the prescribed accuracy under every admissible backward cost in Theorem~\ref{thm:enclosure} and for every valid $(\psi,G)$-rule. When the residual $R$ is sufficiently small relative to $G^{1/2}$ and all BPR powers equal $\beta$, the condition is also close to sufficient: the interval width approaches $2\psi\varrho$ (Lemma~\ref{lem:width}) and $\psi^2$ approaches $\beta D(\xh)$ as the secant steps become small.

For example, with $\beta_{\min}=4$ and a delay share near $0.3$, certifying an accuracy of $10^{-5}$ of the total cost requires $G/\TSTT(\xh)$ below about $10^{-10}$. As $\varepsilon\to0$, certification therefore requires about twice as many orders of magnitude of gap reduction as an uncertified estimate whose error is proportional to the gap. The time to reach such a gap depends on the assignment algorithm.

\subsection{Gap Decomposition and the Residual-Minimizing Backward Cost}\label{sec:source}

Different backward costs at the same computed flow produce different values of the backward residual $\varrho$ and hence different interval widths. To determine the smallest attainable residual and the components of the gap that limit certification, we first decompose the raw gap.

\begin{lemma}\label{lem:parts}
For every $\xh\in\Feas$ and every equilibrium $\xs$,
\begin{equation}\label{eq:parts}
\Gc(\xh) = E + M + N,
\end{equation}
where $M:=\cs\T\xh-\Js$, $N:=\ch\T\xs-S(\ch)$ and $E$ is the energy of Theorem~\ref{thm:enclosure}, and $E,M,N\ge0$. The term $M$ does not depend on the choice of
$\xs$.
\end{lemma}

For any route flow inducing $\xh$, the margin $M$ is the sum of route flows weighted by their excess costs over the corresponding shortest OD costs at equilibrium. It vanishes exactly when $\xh$ assigns flow only to routes that are shortest at equilibrium. The energy $E$ is of second order in the flow error, whereas $N$, the gap of the equilibrium flow evaluated at the current costs, is generally of first order. The inequality $\Gc\ge E+M$ appears in \eqref{eq:EM} and is a standard starting point for gap-based error bounds. The margin $M$ is also central to weak sharpness \citep{MarcotteZhu1998}.

The backward cost that minimizes the residual $\varrho$ can be characterized through a regularized gap function. Let $\mu_a:=0$ on $\mathcal{Z}$, and define
\begin{equation}\label{eq:gamma}
\begin{aligned}
\Phi(v)&:=\ch\T(\xh-v)-\sum_{a\in\Links}\mu_a(\xh_a-v_a)^2,\\
\gamma(\xh)&:=\max_{v\in\Feas}\Phi(v).
\end{aligned}
\end{equation}
Since $\Phi$ is concave and continuous and $\Feas$ is compact, the maximum is attained, with $0\le\gamma(\xh)\le\Gc(\xh)$. The function $\gamma$ is the merit function of \citet{Auchmuty1989} in the form of \citet{LarssonPatriksson1994}, weighted by the secant slopes from zero flow. \citet{Fukushima1992} studied the case of a fixed positive definite metric. Let $\mathbb{F}(\xh)$ denote the set of backward costs admissible under Theorem~\ref{thm:enclosure}, namely those with $r_a\ge0$ on $\mathcal{Z}$ and $r_a=0$ on every link of $\mathcal{P}$ with $\mu_a=0$.

\begin{theorem}\label{thm:best}
Let Assumption~\ref{ass:costs} hold and let $\xh\in\Feas$. Then $\min_{\cb\in\mathbb{F}(\xh)}\varrho^2(\cb)=\gamma(\xh)$.
The minimum is attained at $\cb^*:=\ch-2\mu\circ(\xh-v^*)$ for any maximizer $v^*$ of $\Phi$, where $\circ$ is the
entrywise product, and $\Phi(v)\le\gamma(\xh)\le\Phi(v)+G_v$ for every $v\in\Feas$, where $G_v$ is the gap of $v$ in
the affine assignment with link costs $x_a\mapsto\ch_a+2\mu_a(x_a-\xh_a)$.
\end{theorem}

The value identity follows from the Fenchel-dual gap function of \citet{AltangerelEtAl2007} applied to the merit function~\eqref{eq:gamma}. Theorem~\ref{thm:best} identifies the dual variable as the backward cost of Theorem~\ref{thm:enclosure}, establishes that the dual feasible set coincides with the set of admissible backward costs, and provides computable bounds on the minimum residual from any approximate maximizer $v$. The minimizing backward cost $\cb^*$ is obtained from an affine assignment initialized at $\xh$, the cost-approximation subproblem of \citet{Patriksson1994} with a diagonal metric (\appref{app:proofs}).

Since $G+R^2/4=\varrho^2$, the widths of $I$ and $I^*$ depend on the backward cost through $\psi$ and $\varrho$ and, for fixed $\psi$, are nondecreasing in $\varrho$ (Lemma~\ref{lem:width} and Proposition~\ref{prop:sharp}). Minimizing $\varrho$ therefore minimizes the width for fixed $\psi$, although $\psi$ depends on the backward cost through the steps $\bar z_a$.

Every admissible backward cost leaves a minimum width. Let $\psi_0:=\bigl(\sum_a t_a'(\xh_a)\,\xh_a^2\bigr)^{1/2}$ denote the smallest possible value of $\psi$.

\begin{corollary}\label{cor:margin}
For every $\cb\in\mathbb{F}(\xh)$, the interval $I$ of Theorem~\ref{thm:enclosure} and the interval $I^*$ of
Proposition~\ref{prop:sharp} each have width at least $2\psi_0\gamma(\xh)^{1/2}$. Moreover
$\gamma(\xh)\ge\max\bigl(M,\,M/2+E/4\bigr)$, so each of these intervals has half-width at least
$\psi_0\max\bigl(M,\,M/2+E/4\bigr)^{1/2}$. Finally, $E\le4\gamma(\xh)$ for every equilibrium, and the constant~4 is
attained.
\end{corollary}

Thus, the margin $M$ limits the accuracy attainable by any single interval, regardless of the backward cost. Intersecting intervals obtained from different backward costs can produce narrower bounds. Section~S2 of the Supplementary Material develops a further certificate using the supergradient inequality $S(\cs)\le\cs\T y$ at a second feasible flow $y$.
\section{Certified Incremental Appraisal}\label{sec:algorithm}

\begin{figure}[t]
\centering
\includegraphics[width=\textwidth]{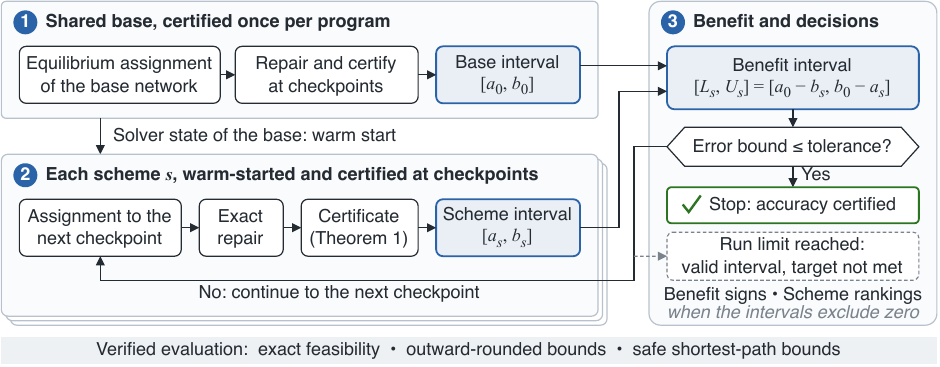}
\caption{Certified incremental appraisal in three stages.}\label{fig:workflow}
\floatnote{Stage~1 is performed once per appraisal program. Stages~2 and~3 are repeated at successive checkpoints for each scheme until the certified error bound $\mathcal{E}_s$ meets the computable tolerance $\hat\tau_s$ (Algorithm~\ref{alg:cia}). If a run reaches its computational limit first, the last interval remains valid but does not certify the required accuracy.}
\end{figure}

Building on Theorems~\ref{thm:plain} and~\ref{thm:enclosure}, we develop a certified incremental appraisal method with three stages (Figure~\ref{fig:workflow}). The base network is solved and certified once, with its interval and solver state shared across all schemes. Each scheme is warm-started from the stored state and certified at successive checkpoints. The resulting benefit intervals determine when the required accuracy is achieved and which decisions can be certified. We refer to this approach as the plain route. An alternative using Newton steps from a certified point, the Newton route, is described in Section~S3 of the Supplementary Material and compared with the plain route in Section~\ref{sec:cost}.

\subsection{Certification of the Shared Base}\label{sec:base}

The base network is solved once for the entire appraisal program using a bush-based assignment algorithm \citep{Dial2006,Nie2010}. The solver is paused at a decreasing sequence of relative gaps. At each checkpoint, the computed flow is repaired to an exactly feasible point and certified at its own link costs using Theorem~\ref{thm:plain}. The base solution is selected at the first checkpoint whose interval half-width meets the accuracy requirement established in Section~\ref{sec:allocation}, or at the final checkpoint if the requirement is not met. The resulting feasible flow $\xh_0$, certified interval $[a_0,b_0]$ with half-width $w_0$, and solver state are stored for subsequent scheme assignments.

\subsection{Warm-Started Certification of Each Scheme}\label{sec:schemes}

Each scheme $s$ is solved using its link cost functions $t^{(s)}$, with the assignment algorithm initialized from the stored base solver state. As for the base, the solver is paused at checkpoints defined by decreasing relative gaps $g_1>g_2>\dots>g_n$. At each checkpoint, the computed flow is repaired to an exactly feasible point $\xh$ and certified at its own link costs $\ch=t^{(s)}(\xh)$ using Theorem~\ref{thm:plain}, giving the interval $I_0$ of width $2\psi G^{1/2}+G$. We refer to this as the plain certificate, which requires neither linearization of the assignment problem nor an auxiliary assignment.

To account for floating-point rounding, the implementation uses the Josephy\textendash Newton cost
\begin{equation}\label{eq:jn}
\cb_a = t_a(y_a)+t_a'(y_a)\,(\xh_a-y_a), \qquad a\in\Links,
\end{equation}
where $y$ is the double-precision rounding of $\xh$. Theorem~\ref{thm:enclosure} applies at this backward cost without requiring $r=0$. The residual is of second order in the rounding error, and its effect on the interval width is negligible relative to the outward rounding of the interval endpoints (\appref{app:verified}).

The resulting interval $[a_s,b_s]$ for $\Js_s$ is combined with the base interval using Corollary~\ref{cor:benefit}. At each checkpoint, the method intersects the new benefit interval with those obtained at previous checkpoints. The resulting interval $[L_s,U_s]$ contains $B_s$ (Section~\ref{sec:benefitint}) and provides the certified error bound $\mathcal{E}_s$ for the current estimate $\Bh_s$.

Since the tolerance $\tau_s$ in \eqref{eq:tau} depends on the unknown benefit, we introduce the computable tolerance
\begin{equation}\label{eq:tauhat}
\hat\tau_s := \epsabs\,\TC_0 + \epsrel\,\hat m_s,
\end{equation}
where $\hat m_s:=\min\bigl(|L_s|,|U_s|\bigr)$ if $0\notin[L_s,U_s]$ and $\hat m_s:=0$ otherwise. Because $\hat m_s$ is a lower bound on $|B_s|$, we have $\hat\tau_s\le\tau_s$. Thus, $\mathcal{E}_s\le\hat\tau_s$ guarantees that the estimate meets the prescribed tolerance $\tau_s$.

\subsection{Validity and Termination of the Certified Stop}\label{sec:stop}

The scheme assignment stops at the first checkpoint at which the certified error bound meets the computable tolerance, $\mathcal{E}_s\le\hat\tau_s$. Validity holds for any run, whereas termination is established for an idealized procedure in which the checkpoint gaps tend to zero, the solver reaches each checkpoint in finitely many iterations without computational limits, every checkpoint yields a certificate, the gap $G$ of the repaired point tends to zero, and all certificates are evaluated in exact arithmetic.

\begin{proposition}\label{prop:stop}
\begin{enumerate}
\item Every interval returned contains $B_s$, and every returned estimate satisfies $|\Bh_s-B_s|\le\mathcal{E}_s$,
whatever the checkpoints and the limits of the run.
\item Along the certificates of the idealized procedure run without the test, the half-width of the scheme's
interval tends to zero. Hence, if $\epsabs\TC_0>2w_0$, the idealized procedure with the test stops after finitely
many certificates.
\item For the idealized procedure with $\epsabs=0$ and $\epsrel>0$, the rule stops after finitely many certificates
when $B_s\neq0$ and $\epsrel|B_s|/(1+\epsrel)>2w_0$. If $B_s=0$, it stops only if $[L_s,U_s]=\{0\}$.
\item Every interval that Algorithm~\ref{alg:cia} returns, with its finitely many checkpoints and its limits on the
solve, is valid and is flagged as certified to the tolerance or not. The algorithm may end without meeting
$\hat\tau_s$.
\end{enumerate}
\end{proposition}

Item~1 follows from Theorem~\ref{thm:enclosure}, which does not require proximity to equilibrium, and item~2 from the boundedness of the externality weight on the feasible set and the vanishing gap. Items~2 and~3 motivate the inclusion of an absolute tolerance, since a purely relative tolerance requires exact certification when the benefit is zero. In floating-point computation, rounding errors may prevent further reduction in the convergence gap. Item~4 ensures that the returned intervals remain valid even when the required accuracy cannot be achieved within the computational limits.

\subsection{Accuracy Allocation Between the Base and the Schemes}\label{sec:allocation}

As established in Proposition~\ref{prop:identity}, the error in the base assignment affects every benefit estimate. Similarly, the certified error bound for each scheme is at least the half-width $w_0+w_s$ of its benefit interval, where $w_s$ is the half-width of the intersection of the scheme's intervals (Section~\ref{sec:benefitint}). To determine how computational effort should be allocated between the base and individual schemes, suppose that achieving a half-width $w$ on network $\theta$ requires work $W_\theta(w)=c_\theta\log(\bar w_\theta/w)$, with $c_\theta>0$. This logarithmic cost model is consistent with linear convergence of the gap and the square-root dependence of the plain certificate's width on the gap (Lemma~\ref{lem:width}).

\begin{proposition}\label{prop:allocation}
Let $c_\theta>0$ and $W_\theta(w)=c_\theta\log(\bar w_\theta/w)$ for every $w\in(0,\max_s\tau_s]$. The problem of minimizing $W_0(w_0)+\sum_{s=1}^m W_s(w_s)$ subject to $w_0+w_s\le\tau_s$ for every $s$ has a unique
solution. Every constraint is active, and $w_0$ is the unique root in $(0,\min_s\tau_s)$ of
\begin{equation}\label{eq:allocation}
\frac{c_0}{w_0} = \sum_{s=1}^m \frac{c_s}{\tau_s-w_0}.
\end{equation}
If all $c_\theta$ are equal and every $\tau_s$ equals $\tau$, then $w_0=\tau/(m+1)$, and the work on the base grows like
$\log m$.
\end{proposition}

Using the minimum possible tolerance $\tau:=\epsabs\TC_0$ and assuming equal costs per order of magnitude of width reduction, Proposition~\ref{prop:allocation} gives a target base half-width of $\tau/(m+1)$ for a program of $m$ schemes. Compared with a target of $\tau$, this requires reducing the base interval width by an additional $\log_{10}(m+1)$ orders of magnitude, with the computational cost incurred only once. We use this target to select the base checkpoint in Section~\ref{sec:base}. The allocation rule is optimal under the assumed cost model and does not affect the validity of the certificates. Since $w_0+w_s\le\tau_s$ is necessary for $\mathcal{E}_s\le\tau_s$ but sufficient only for the midpoint estimate $\Bm_s$, the rule provides guidance on the required accuracy of the base (\appref{app:proofs}).

\begin{algorithm}[t]
\caption{Certified incremental appraisal of scheme $s$}\label{alg:cia}
\begin{algorithmic}[1]
\Require base interval $[a_0,b_0]$, estimate $\SPTTm_0(\xh_0)$ and stored solver state, scheme costs
$t^{(s)}$, checkpoint gaps $g_1>\dots>g_n$, limits on iterations, time and gap stalling, $\epsabs$, $\epsrel$ and $\TC_0$
\State start the solver with costs $t^{(s)}$ from the stored base solver state, and set $[L_s,U_s]\gets\R$
\For{$j=1,\dots,n$}
  \State run the solver until its relative gap is at most $g_j$, and exit the loop if a limit is reached first
  \State repair the iterate to $\xh\in\Feas$ and evaluate Theorem~\ref{thm:plain} at $\xh$, and proceed to the next checkpoint if no certificate is obtained
  \State intersect $[L_s,U_s]$ with $[a_0-b_s,\,b_0-a_s]$ and compute $\Bh_s$, $\mathcal{E}_s$ and $\hat\tau_s$
  \If{$\mathcal{E}_s\le\hat\tau_s$}
    \State \Return $[L_s,U_s]$, $\Bh_s$ and $\mathcal{E}_s$, certified to the tolerance
  \EndIf
\EndFor
\If{the solver terminated by its own convergence criterion}
  \State certify the final iterate in the same way and \Return the result as certified if $\mathcal{E}_s\le\hat\tau_s$
\EndIf
\State \Return the last certificate's $[L_s,U_s]$, $\Bh_s$ and $\mathcal{E}_s$, if any, valid but not certified to the tolerance
\end{algorithmic}
\end{algorithm}

\subsection{Verified Evaluation}\label{sec:verified}

Theorems~\ref{thm:plain} and~\ref{thm:enclosure} are established in exact arithmetic, whereas assignment algorithms operate in floating-point arithmetic. All quantities entering a certificate are therefore evaluated using verified computation, so that rounding errors can only widen the interval \citep{NeumaierShcherbina2004,KeilJansson2006,Rump2010}. Flows are repaired in exact arithmetic to ensure $\xh\in\Feas$. Link quantities are evaluated using interval arithmetic with outward rounding, and a rigorous lower bound on $S(\cb)$ is obtained using Dijkstra's algorithm with every addition rounded down. If any of these steps cannot be completed, no certificate is returned (\appref{app:verified}).

\subsection{The Algorithm for One Scheme}\label{sec:algobox}

Algorithm~\ref{alg:cia} summarizes the procedure for a single scheme. Since the base solution is computed once and shared across all schemes, individual schemes can be evaluated independently, in any order or in parallel. For the midpoint estimate $\Bm_s$, the stopping criterion becomes $U_s-L_s\le2\hat\tau_s$. This criterion is satisfied no later than that for the conventional estimate $\Bh_s$, since $\tfrac12(U_s-L_s)\le\mathcal{E}_s$.
\section{Numerical Experiments}\label{sec:experiments}

We assess the accuracy of gap-based appraisal, the validity and sharpness of the bounds, and the method's performance, cost and certified decisions in two prespecified blind tests.

\subsection{Experimental Design}\label{sec:setup}

\subsubsection{Networks and Schemes.}\label{sec:networks}

The five networks in Table~\ref{tab:networks} are obtained from the Transportation Networks for Research repository \citep{TNTP2026}. Sioux Falls \citep{LeBlancEtAl1975}, Anaheim and Chicago-Sketch are standard benchmark networks, while Chicago regional and Philadelphia are derived from regional planning models. All links with positive delay have BPR cost functions with $\beta_a=4$. The three larger networks also contain constant zero-cost links, which shows the practical relevance of constant link costs. Where the source data prohibit routes through zone centroids, the same restriction is applied in the equilibrium assignment and shortest-path computations, as required by Theorem~\ref{thm:enclosure}.

\begin{table}[t]
\centering
\caption{Networks and scheme sets.}\label{tab:networks}
\footnotesize
{\itshape Panel A. Network characteristics}\par\smallskip
\begin{adjustbox}{max width=\linewidth}
\begin{tabular}{@{}lrrrrr@{}}
\toprule
Network & Nodes & Links & Zero-cost links & Zones & OD pairs \\
\midrule
Sioux Falls & 24 & 76 & 0 & 24 & 528 \\
Anaheim & 416 & 914 & 0 & 38 & 1,406 \\
Chicago-Sketch & 933 & 2,950 & 774 & 387 & 93,135 \\
Chicago regional & 12,982 & 39,018 & 3,650 & 1,790 & 2,296,227 \\
Philadelphia & 13,389 & 40,003 & 9,802 & 1,525 & 1,149,795 \\
\bottomrule
\end{tabular}
\end{adjustbox}
\par\medskip{\itshape Panel B. Number of schemes in each set}\par\smallskip
\begin{adjustbox}{max width=\linewidth}
\begin{tabular}{@{}lrrrr@{}}
\toprule
Network & Development sets & Held-out set & First blind set & Second blind set \\
\midrule
Sioux Falls & 40 & \textemdash & \textemdash & \textemdash \\
Anaheim & 50 & \textemdash & \textemdash & \textemdash \\
Chicago-Sketch & 85 & \textemdash & \textemdash & \textemdash \\
Chicago regional & 20 & 35 & 48 & 36 \\
Philadelphia & 20 & 20 & 48 & 36 \\
\bottomrule
\end{tabular}
\end{adjustbox}
\floatnote{Zero-cost links have constant cost zero. A dash indicates that the corresponding scheme set is not applicable.}
\end{table}

A single-link scheme changes the capacity $C_a$ of one link by a specified multiplier, while an area scheme applies a common multiplier to the capacities of links nearest a seed link. The development sets were used to develop the method. On the three benchmark networks, they consist of single-link schemes covering different levels of congestion. On the two regional networks, they comprise 24 single-link schemes with capacity multipliers of 1.1, 1.5, 0.5 and 0.01, and 16 area schemes, eight per network. The area schemes scale the capacities of the 10 to 1,000 development-pool links nearest a congested seed link by 1.5 or 0.5.

The regional networks have three separate test sets. The held-out set comprises 55 schemes on 40 links, selected before any appraisal protocol was run and evaluated once to assess conventional gap-based appraisal. The first blind set comprises 96 schemes drawn once from a pool of links disjoint from the development pool and kept blind under the protocol in Section~\ref{sec:impl}. Most schemes increase the capacity of a link in the top decile of the volume-to-capacity ratio by a factor of 1.1. Five links in Chicago regional are also evaluated at multipliers of 1.5, 0.5 and 0.01, the last representing near closure. On each network, eight links from the bottom decile are evaluated at a multiplier of 1.1 as near-null schemes with benefits close to zero. The second blind set comprises 72 schemes drawn once from the same pool, excluding the links in the first blind set, and kept blind under the same protocol. On each network, 24 are area schemes involving 30, 100 or 300 links with multipliers of 1.5 and 0.5, and 12 are single-link schemes in the top decile with a multiplier of 1.1.

\subsubsection{References, Tolerance and Comparison Protocols.}\label{sec:refs}

The reference benefit $B_{\mathrm{ref}}$ of each blind scheme comes from Algorithm~B \citep{Dial2006}, with cold starts on the base and scheme networks and a relative gap target of $10^{-14}$, subject to a time limit. A second run with a different solver setting gives a second estimate. The difference between the two high-accuracy estimates, the reference uncertainty $u_{\mathrm{ref}}$, measures their agreement and is not an error bound. For all 168 blind schemes, $u_{\mathrm{ref}}$ is below $\tau_s/10$. The held-out and development references are similarly accurate (Table~\ref{tab:settings}).

Throughout the experiments, the tolerance in \eqref{eq:tau} uses $\epsabs=10^{-5}$ and $\epsrel=0.10$, with $\TC_0$ fixed before the runs at the reference TSTT of the base network. Replacing $\TC_0$ with the lower bound $a_0$ of the certified base interval removes the need for a reference solution. This changes $\TC_0$ by at most $3.4\times10^{-7}$ in relative terms and does not alter any stopping checkpoint in either blind test.

We compare the proposed method with two conventional gap-based protocols. A$(g)$ solves the base and scheme networks from cold starts, terminating each assignment when its relative gap first reaches or falls below $g$. W$(g)$ solves the base to gap $g$ once and warm-starts each scheme from the stored base solver state, using the same stopping criterion. W$(g)$ corresponds to the shared-base procedure described in Section~\ref{sec:identity} and was the least expensive protocol meeting the accuracy requirement in the held-out test. Both protocols report the benefit estimate in \eqref{eq:estimate}.

\subsubsection{Implementation and Cost Accounting.}\label{sec:implementation}

Table~\ref{tab:settings} in \appref{app:settings} documents the settings, hardware and software. We use Algorithm~B as implemented in the open-source solver tap-b, extended to accept affine link costs and to store solver states at checkpoints. Certificates are evaluated by a compiled program that performs the flow repair described in Section~\ref{sec:verified} using exact integer arithmetic. Repeated runs on the same machine reproduce every certificate bit for bit.

The computational cost of a stopping criterion includes the assignment time and the evaluation of all certificates up to the stopping checkpoint. Per-scheme costs exclude the cost of computing the shared base for both the proposed method and the comparison protocols. Whole-program costs are reported separately in Section~\ref{sec:cost}. To record subsequent checkpoints, the experimental runs continued beyond their stopping criteria as specified in Table~\ref{tab:settings}, but only work completed up to the relevant stopping checkpoint is included in the reported cost.

\subsubsection{Blind-Test Protocols.}\label{sec:impl}

For each blind test, the protocol, code, inputs and analysis scripts were frozen with checksums before execution. Each test was run once on a single workstation, and the reference values were accessed by the frozen analysis scripts only after completion.

The first blind test evaluated the Newton route on the first blind set, taking 25.9 hours (Section~S3 of the Supplementary Material, with settings in Table~S3). Three hypotheses were prespecified: the validity of every certified benefit interval (H1), satisfaction of the computable tolerance $\hat\tau_s$ (H2), and the computational cost relative to the paired A$(10^{-6})$ protocol (H3). As specified in the protocol, the five schemes with a capacity multiplier of 0.01 were included only in H1, leaving 91 schemes for H2 and H3. The base equilibrium was certified after two Newton steps. We refer to this solution as the Newton base and to the base obtained using the procedure in Section~\ref{sec:base} as the plain base.

The second blind test evaluated the plain route on the second blind set, taking 13.6 hours. Five hypotheses were prespecified, with acceptance criteria given in Table~\ref{tab:blind}. H1 concerns the validity of every computed certificate, and H2 concerns satisfaction of $\hat\tau_s$ for all 72 schemes. H3 compares the cost of meeting this tolerance on the 24 single-link schemes with that of the paired A$(10^{-6})$ protocol. H4 evaluates accuracy and computational cost under a purely relative tolerance with $\epsabs=0$ and $\epsrel=0.1$, denoted by T10, for schemes satisfying $|B_{\mathrm{ref}}|\ge10^{-3}\TC_0$. H5 tests whether a synchronous elimination procedure certifies the scheme with the largest $B_{\mathrm{ref}}$ as having the largest benefit on each network. In each round, every remaining scheme advances to its next certificate, and a scheme is eliminated when Corollary~\ref{cor:compare} certifies that another scheme has a larger benefit.

The original protocol for the second blind test used the Newton base to construct benefit intervals, as the plain base was adopted during the run. Before any results were accessed, the same acceptance criteria were specified for an analysis using the plain base interval. Both analyses satisfy all five criteria, and Section~\ref{sec:blind} reports the results using the plain base. In both analyses, the scheme assignments were initialized from the solver state of the Newton base. To assess the effect of this initialization, all 40 regional development schemes were rerun from the plain base's own solver state. Every interval agreed with its reference benefit, and all schemes met $\hat\tau_s$, at a median cost of 0.96 times that of the runs from the Newton base (Section~S4.4 of the Supplementary Material).

The evaluation distinguishes mathematical validity, numerical implementation and empirical agreement. Interval validity is established under the stated assumptions by the proofs in \appref{app:proofs}. Correct numerical evaluation is supported by the exact flow repair and outward rounding described in Section~\ref{sec:verified}, bit-for-bit agreement between the compiled program and a separate implementation (\appref{app:verified}), and tests against exact solutions (Section~\ref{sec:designed}). Agreement with the high-accuracy references provides additional empirical evidence but does not independently establish validity, since the references are themselves approximate solutions.
\subsection{Accuracy of Gap-Stopped Appraisal}\label{sec:practice}

Table~\ref{tab:practice} compares the accuracy of cold and shared-base warm starts at three relative gap tolerances on the 55 held-out schemes. At a relative gap of $10^{-4}$, 25 benefit estimates from cold assignments and 50 from warm-started assignments fell outside their prescribed tolerances. At tighter gaps, both protocols became substantially more accurate, while the median cost of a warm-started scheme assignment was only 11.7\% of that of a cold assignment. The sensitivity of warm-started appraisal to the convergence of the shared base is consistent with Proposition~\ref{prop:identity}. Although tighter convergence improves benefit accuracy, neither protocol provides a guarantee that the prescribed tolerance has been met. Table~S4 of the Supplementary Material reports four additional protocols.

\begin{table}[t]
\centering
\caption{Accuracy of gap-stopped appraisal on the 55 held-out schemes.}\label{tab:practice}
\footnotesize
\begin{adjustbox}{max width=\linewidth}
\begin{tabular}{@{}lrrrr@{}}
\toprule
 & Cold start, A$(g)$ & \multicolumn{3}{c@{}}{Shared-base warm start, W$(g)$} \\
\cmidrule(lr){2-2}\cmidrule(l){3-5}
Target relative gap $g$ & Inaccurate benefits & Inaccurate benefits & Unfinished runs & Cost relative to A$(g)$ \\
\midrule
$10^{-4}$ & 25 & 50 & 1 & 0.094 \\
$10^{-5}$ & 1 & 0 & 1 & 0.117 \\
$10^{-6}$ & 0 & 0 & 1 & 0.117 \\
\bottomrule
\end{tabular}
\end{adjustbox}
\floatnote{A benefit is inaccurate if its estimate differs from the reference benefit by more than $\tau_s$. A run is unfinished if it fails to reach gap $g$ within its computational limits. All A$(g)$ runs finished. The cost ratio is the median cost of a warm-started scheme assignment under W$(g)$ relative to its paired cold assignment under A$(g)$.}
\end{table}

Conventional appraisal was also accurate at tight convergence tolerances on the blind sets. On the second blind set, A$(10^{-6})$, W$(10^{-6})$ and W$(10^{-7})$ met $\tau_s$ on all 72 schemes. Their maximum errors were 0.18, 0.31 and 0.06 times $\tau_s$, with median errors of 0.008, 0.018 and 0.004 times $\tau_s$. Relative to the benefits themselves, however, the maximum errors were 68\%, 30\% and 11\%, each occurring on a single-link scheme with a benefit below $2\times10^{-6}\TC_0$. On the first blind set, the maximum errors were 0.222, 0.295 and 0.074 times $\tau_s$, respectively. Tightly converged cold and warm-started assignments are therefore accurate comparators for the computational cost of certification.

\subsection{Validity and Sharpness of the Bounds}\label{sec:designed}

We evaluated the results of Sections~\ref{sec:enclosure} and~\ref{sec:price} on instances with known exact equilibria, in exact rational arithmetic except for one check in 60-digit floating-point arithmetic. None of the bounds was violated in 133,297 evaluations (Table~S5 of the Supplementary Material). The interval $I^*$ attained a width of exactly $2\psi\varrho$ in all 24,796 cases with $\psi\ge2\varrho$. The constructions in Corollary~\ref{cor:optimal} attained the upper offset $\psi G^{1/2}+G$ exactly, including BPR instances with $\beta=4$, $\psi=6{,}250$ and $G=10^{-4}$, the regime of the Chicago regional network. The lower offset $p^*$ was also attained exactly on instances with affine and constant costs.

\subsection{Performance of Certified Appraisal in Blind Tests}\label{sec:blind}

The second blind test satisfied all five prespecified acceptance criteria (Table~\ref{tab:blind}). All 577 certificates agreed with the reference benefits within the reference uncertainty (H1), and all 72 schemes met the computable tolerance $\hat\tau_s$ (H2). Certification of the 24 single-link schemes required a median of 1.90 times the cost of the paired cold assignments A$(10^{-6})$ (H3). For the 29 schemes with larger benefits, all met the relative tolerance T10 at a median cost of 0.75 times that of A$(10^{-6})$ (H4).

The elimination procedure also certified which scheme had the largest benefit on each network (H5). On Chicago regional, this required only the first certificate of each scheme, while Philadelphia required four additional rounds. The corresponding computational costs were 0.25 and 0.33 times the aggregate cost of A$(10^{-6})$ across the 36 schemes on each network. The analysis using the Newton base yields the same counts and cost statistics.

\begin{table}[t]
\centering
\caption{Prespecified hypotheses and results of the second blind test.}\label{tab:blind}
\footnotesize
\begin{adjustbox}{max width=\linewidth}
\begin{tabular}{@{}>{\raggedright\arraybackslash}p{0.43\textwidth}>{\raggedright\arraybackslash}p{0.14\textwidth}rrr@{}}
\toprule
Hypothesis & Acceptance criterion & Chicago regional & Philadelphia & Both networks \\
\midrule
H1. Certificates that contain $B_{\mathrm{ref}}$ within $u_{\mathrm{ref}}$ & All & 290 of 290 & 287 of 287 & 577 of 577 \\
H2. Schemes that meet $\hat\tau_s$ & At least 95\% & 36 of 36 & 36 of 36 & 72 of 72 \\
H3. Single-link schemes that meet $\hat\tau_s$ within 4.0 times the cost of A$(10^{-6})$ & At least 95\% & 12 of 12 & 12 of 12 & 24 of 24 \\
\quad Median cost ratio of the single-link schemes & At most 2.5 & 2.43 & 1.84 & 1.90 \\
H4. Schemes with $|B_{\mathrm{ref}}|\ge10^{-3}\TC_0$ that meet T10 & At least 90\% & 14 of 14 & 15 of 15 & 29 of 29 \\
\quad Median cost ratio of these schemes & At most 1.0 & 0.73 & 0.82 & 0.75 \\
H5. Elimination certifies the scheme with the largest $B_{\mathrm{ref}}$ & On each network & Yes & Yes & Yes \\
\bottomrule
\end{tabular}
\end{adjustbox}
\floatnote{Cost ratios are evaluated at the certified stop relative to the paired cold assignment A$(10^{-6})$. The maximum ratios on Chicago regional and Philadelphia were 3.21 and 2.45 for H3, and 1.10 and 1.39 for H4, respectively. T10 denotes the tolerance with $\epsabs=0$ and $\epsrel=0.1$. Benefit intervals use the plain base (Section~\ref{sec:impl}).}
\end{table}

The first blind test of the Newton route also met its three prespecified acceptance criteria (Table~S6). All 2,133 certificates across the 96 schemes agreed with the references within $u_{\mathrm{ref}}$, and all 91 schemes in scope met $\hat\tau_s$ within ten times the cost of A$(10^{-6})$, at a median cost ratio of 2.64.

After this test, a descriptive comparison evaluated the plain route on the same 96 schemes with the same bases, references and tolerance. All 940 certificates agreed with the references within $u_{\mathrm{ref}}$, and all 91 schemes in scope met $\hat\tau_s$. The median cost of the plain route was 1.73 times that of A$(10^{-6})$ and 0.70 times that of the Newton route. At the certified stops of the Newton route, the certified error bound exceeded the observed error relative to the reference by a median factor of $1.4\times10^{4}$ and a minimum factor of 117 (Section~\ref{sec:conservative}).
\subsection{Computational Cost of Certification}\label{sec:cost}

\begin{figure}[t]
\centering
\includegraphics[width=\textwidth]{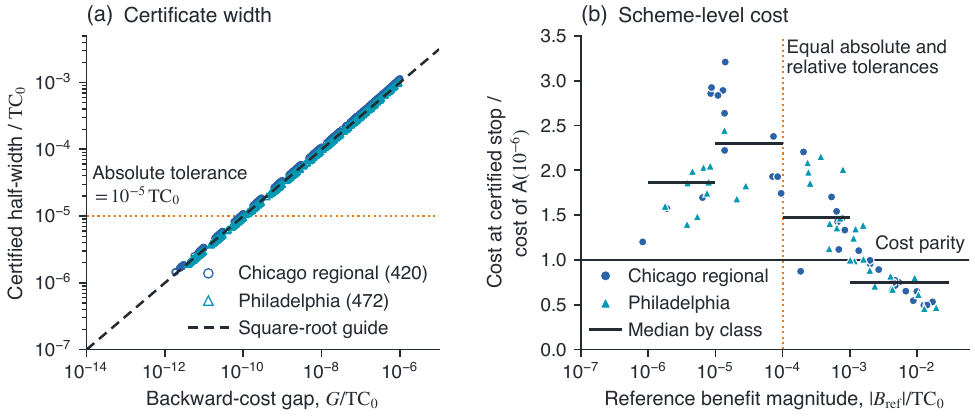}
\caption{Certificate width and scheme-level cost of certified appraisal.}\label{fig:price}
\floatnote{(a) Certified half-width against the gap at the backward cost, both normalized by $\TC_0$, for 892 certificates from the descriptive evaluation of the plain route on the first blind set. The dashed line $(G/\TC_0)^{1/2}$ is a guide and is not fitted. (b) Scheme-level cost at the certified stop relative to the paired cold assignment A$(10^{-6})$ for the 72 schemes of the second blind set. Horizontal segments indicate median cost ratios over each order of magnitude of $|B_{\mathrm{ref}}|/\TC_0$ from $10^{-6}$ to $10^{-3}$ and over all values above $10^{-3}$. Benefit intervals use the plain base (Section~\ref{sec:impl}).}
\end{figure}

Figure~\ref{fig:price}(a) shows the relationship between the certified half-width and the gap at the backward cost for 892 certificates obtained using the plain route on the 91 schemes in scope of the first blind set. On both regional networks, the half-width decreases with the square root of the gap over five orders of magnitude, consistent with Theorem~\ref{thm:plain} and Lemma~\ref{lem:width}. The reference line $(G/\TC_0)^{1/2}$ equals $\psi G^{1/2}/\TC_0$ when $\psi^2=\TC_0$. On Chicago regional, the median value of $\psi^2/\TC_0$ is 1.17, close to four times the delay share of the base network, 1.16, as expected from Proposition~\ref{prop:law} for BPR costs with power~4. A certified half-width of $10^{-5}\TC_0$ is reached only at a gap of approximately $10^{-10}\TC_0$, consistent with the necessary condition in Proposition~\ref{prop:law}. Evaluating a certificate takes a median of 10.7~s and a maximum of 18.4~s on the two regional networks.

Figure~\ref{fig:price}(b) relates the computational cost of certification to the magnitude of the benefit for the 72 schemes of the second blind set. For benefits below $10^{-4}\TC_0$ in magnitude, the absolute term dominates the tolerance, and every scheme stopped at a relative gap between $3\times10^{-11}$ and $3\times10^{-10}$. The median cost ratios relative to A$(10^{-6})$ are 1.86 for benefits between $10^{-6}\TC_0$ and $10^{-5}\TC_0$ and 2.30 between $10^{-5}\TC_0$ and $10^{-4}\TC_0$. For larger benefits, the relative term increasingly determines the tolerance, allowing certification at larger gaps. The median cost ratio decreases to 1.47 for benefits between $10^{-4}\TC_0$ and $10^{-3}\TC_0$, and to 0.75 for benefits above $10^{-3}\TC_0$. Of the 29 schemes in the latter group, 23 are certified at a lower cost than A$(10^{-6})$.

Program-level costs include the shared base. At a relative gap of $10^{-13}$, the base half-width remains above the allocation target $\tau/(m+1)$ for $m=48$ schemes on four of the five networks, including both regional networks (Table~S7 of the Supplementary Material), so the base is selected at the final checkpoint for programs of this size. Its certification requires 4,781~s on Chicago regional and 2,307~s on Philadelphia, against 498~s and 276~s for the cold base assignment of A$(10^{-6})$.

\begin{figure}[t]
\centering
\includegraphics[width=\textwidth]{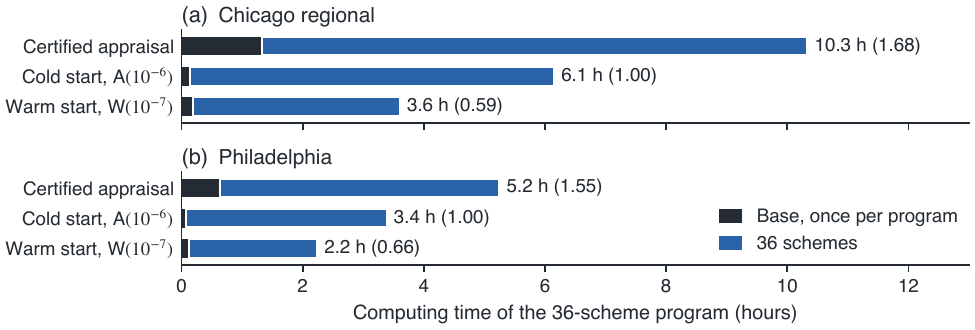}
\caption{Computing time of an appraisal program of 36 schemes under three protocols.}\label{fig:program}
\floatnote{Each bar combines the cost of the shared base and the 36 scheme assignments of the second blind set on each network. Labels indicate total computing time and its ratio to A$(10^{-6})$. Times are sums of single-thread run times. The certified program cost is reconstructed using the plain base and the scheme costs at their certified stops, with scheme assignments initialized from the Newton base's solver state (Section~\ref{sec:impl}).}
\end{figure}

For the programs of 36 schemes in the second blind set, the total cost of certified appraisal is 1.68 and 1.55 times that of A$(10^{-6})$ on Chicago regional and Philadelphia, respectively, and 2.87 and 2.35 times that of W$(10^{-7})$. Base certification accounts for 12.9\% and 12.2\% of the total cost (Figure~\ref{fig:program}). The relative cost is higher for smaller programs, over which the cost of the base is distributed among fewer schemes. On the first blind set, the programs of 43 and 48 schemes in scope cost 1.95 and 1.56 times as much as A$(10^{-6})$, compared with 4.06 and 3.54 times for single-scheme programs (Table~S8). The allocation rule of Proposition~\ref{prop:allocation} selects a base relative gap of $10^{-11}$ for a single scheme and $10^{-13}$ for the full programs. Across the program sizes considered, its selected checkpoint achieves a computational cost within 5.1\% of the best fixed checkpoint.

We also compared the plain and Newton routes on 28 regional development schemes from a common solver state and base interval. Both routes met the tolerance on every scheme, and the plain route cost a median of 0.71 times as much as the Newton route, with quartiles of 0.50 and 0.95. The prespecified selection rule therefore chose the plain route (Table~S9). On ten of these schemes, cold-started certification by the plain route cost a median of 2.23 times as much as warm-started certification, which shows the benefit of reusing the base solver state.

\begin{table}[t]
\centering
\caption{Decisions certified on the second blind set at the certified stop and after further convergence.}\label{tab:decisions}
\footnotesize
\begin{adjustbox}{max width=\linewidth}
\begin{tabular}{@{}>{\raggedright\arraybackslash}p{0.37\textwidth}>{\centering\arraybackslash}p{0.11\textwidth}>{\centering\arraybackslash}p{0.11\textwidth}c>{\centering\arraybackslash}p{0.11\textwidth}>{\centering\arraybackslash}p{0.11\textwidth}@{}}
\toprule
 & \multicolumn{2}{c}{At the stop} & & \multicolumn{2}{c@{}}{After continuing the run} \\
\cmidrule(lr){2-3}\cmidrule(l){5-6}
 & Chicago regional & Philadelphia & & Chicago regional & Philadelphia \\
\midrule
Signs certified, of 36 schemes & 34 & 28 & & 34 & 35 \\
Rankings certified, of 630 pairs & 573 & 567 & & 607 & 592 \\
Scheme with the largest benefit certified & Yes & No & & Yes & Yes \\
Contradictions with the reference & 0 & 0 & & 0 & 0 \\
Median scheme cost relative to A$(10^{-6})$ & 1.38 & 1.40 & & 2.28 & 2.48 \\
\bottomrule
\end{tabular}
\end{adjustbox}
\floatnote{Benefit signs are certified when the benefit intervals exclude zero, and pairwise rankings are certified using Corollary~\ref{cor:compare}. The right-hand columns report decisions certified at subsequent checkpoints beyond the certified stops (Table~\ref{tab:settings}). All counts were made after the blind test and are descriptive.}
\end{table}

\begin{figure}[t]
\centering
\includegraphics[width=\textwidth]{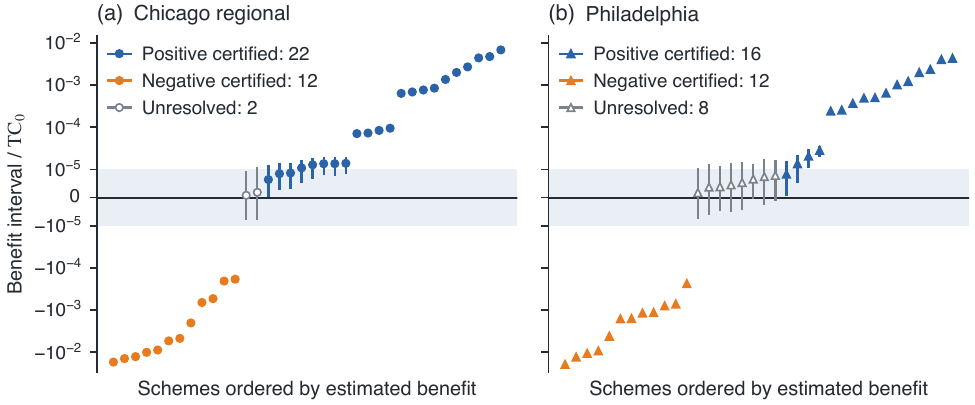}
\caption{Certified benefit intervals for the 72 schemes of the second blind set.}\label{fig:decisions}
\floatnote{Benefit intervals at the certified stop are normalized by $\TC_0$, with markers indicating the estimates $\Bh_s$. The horizontal scale is linear within the shaded band of $\pm10^{-5}\TC_0$ and logarithmic outside it. Negative benefits correspond to the 12 area schemes with a capacity multiplier of 0.5 on each network. Benefit intervals use the plain base (Section~\ref{sec:impl}).}
\end{figure}

\subsection{Certified Appraisal Decisions}\label{sec:decisions}

Table~\ref{tab:decisions} summarizes the decisions certified on the second blind set at the certified stop and after further convergence. At the stop, the method certified 34 and 28 of the 36 benefit signs and 573 and 567 of the 630 pairwise rankings on Chicago regional and Philadelphia, respectively. Continuing the assignments increased the number of certified decisions and established which scheme had the largest benefit on both networks. No certified sign or ranking contradicted the reference.

All benefits whose signs remained unresolved at the stop were associated with single-link schemes with $|B_{\mathrm{ref}}|$ below $10^{-5}\TC_0$ (Figure~\ref{fig:decisions}). The method certified the signs of all 58 benefits with magnitudes above $10^{-5}\TC_0$, four of the 13 between $10^{-6}\TC_0$ and $10^{-5}\TC_0$, and none below $10^{-6}\TC_0$. This pattern reflects the absolute component of the tolerance, which allows the accuracy requirement to be satisfied without resolving the signs of sufficiently small benefits.
\section{Discussion}\label{sec:discussion}

The results establish a different approach to numerical error in network appraisal. Current practice prescribes a common convergence tolerance for all assignments, whereas effort can be allocated according to the accuracy required of each benefit estimate and of the decisions it supports.

\subsection{Implications for Appraisal Practice}\label{sec:practiceimpl}

The held-out and blind tests demonstrate that tightly converged assignments can provide accurate benefit estimates, but the convergence required for a new scheme is not generally known in advance. Certified appraisal allows the analyst to specify the required accuracy directly in terms of the benefit and terminate each assignment once that accuracy is guaranteed. The benefit interval provides a rigorous alternative to heuristic checks of numerical accuracy, such as the recommendation that the benefit as a proportion of network cost should ideally be ten times the relative gap.

The accuracy required of a benefit estimate also depends on how it is used. An appraisal concerned with whether a scheme reduces travel costs may require only certification of the benefit sign, while a comparison of alternatives may require certification of their ranking. These decisions can be certified while the equilibria remain approximate, as demonstrated in Sections~\ref{sec:blind} and~\ref{sec:decisions}. Since the certificate requires only an exactly feasible link flow, it can be combined with different assignment algorithms, provided their outputs can be repaired to exact feasibility.

The theoretical results also guide the allocation of computational effort. Proposition~\ref{prop:allocation} sets the half-width of the shared base at about $1/(m+1)$ of the absolute tolerance for a program of $m$ schemes, at a cost incurred once. Proposition~\ref{prop:law} gives a necessary gap of approximately $\varepsilon^2/(\beta f)$ of the total cost for a half-width of $\varepsilon$ of the total cost when BPR powers equal $\beta$ and the delay share is $f$. This relationship can guide the selection of checkpoints in Algorithm~\ref{alg:cia} from the prescribed tolerance and an estimate of the network's delay share.

The computational results support complementary roles for certified and conventional appraisal. Tightly converged shared-base warm starts provided accurate benefit estimates at low cost on both blind sets (Sections~\ref{sec:practice} and~\ref{sec:cost}) and are suited to screening many alternatives. Certification then applies to schemes for which numerical uncertainty could affect a consequential decision, such as the selection of a preferred investment, the preparation of a business case or the audit of an appraisal. The additional computational cost provides an explicit guarantee on the numerical accuracy of the estimates used in these decisions.

\subsection{Sharpness and Conservatism of the Bounds}\label{sec:conservative}

The tighter interval of Theorem~\ref{thm:plain} is optimal among valid rules based on the gap and the externality weight (Corollary~\ref{cor:optimal}). The attaining instances involve routes that have zero flow at equilibrium but remain shortest routes. In such cases, the gap is quadratic in the flow error, while the total-cost error is linear in it (Example~\ref{ex:tworoute}). Other configurations of flow error can produce a first-order relationship between the gap and the flow error (Lemma~\ref{lem:firstorder} in \appref{app:proofs}), allowing total-cost errors to be substantially smaller than the square-root bound.

The conservatism observed in the experiments reflects this difference. At the certified stops of the Newton route in the first blind test, the certified error bound exceeded the observed error by a median factor of $1.4\times10^4$ (Section~\ref{sec:blind}). Although Corollary~\ref{cor:optimal} does not apply to Newton backward costs, their interval width depends similarly on the backward residual (Lemma~\ref{lem:width}).

\subsection{Implications for Equilibrium Computation}\label{sec:research}

Advances in assignment algorithms can reduce the computational cost of reaching a prescribed convergence gap. Improving the accuracy that can be certified at a given gap, however, requires a different approach. Corollary~\ref{cor:optimal} and Proposition~\ref{prop:law} establish the limits of certificates based on the gap and the externality weight, while the decomposition of Lemma~\ref{lem:parts} and Corollary~\ref{cor:margin} identify the margin term as the quantity that limits individual backward-cost intervals. Existing error bounds under nondegeneracy and weak sharpness \citep{Mangasarian1990,BurkeFerris1993,MarcotteZhu1998} suggest that information about the equilibrium face and the cost margins of unused routes could support tighter certification. An open question is whether such information can be verified from approximate solutions using interval arithmetic.

The alternative certificates examined in this study exploit additional equilibrium information. The supergradient certificate in Section~S2 of the Supplementary Material removes the margin term under conditions involving the unknown equilibrium, with a square-root dependence on the gap of an auxiliary problem (Proposition~S1). Similarly, Newton steps yield a second-order remainder under the conditions of Proposition~S2, and the affine gap was the larger part of the backward residual for 84 of the 91 schemes at their certified stops (Section~S4.3). These results show that such certificates require a small convergence error of the auxiliary problem as well as a small remainder.

A further direction concerns the allocation of computational effort according to the decision being made. For scheme ranking, an adaptive procedure could refine only those assignments whose relative order remains unresolved under Corollary~\ref{cor:compare}. The elimination procedure of Section~\ref{sec:blind} certified the scheme with the largest benefit without solving every scheme to the accuracy required for its own benefit estimate, and continuing the runs certified additional rankings (Table~\ref{tab:decisions}).

Extending certification beyond static user equilibrium with fixed demand is another direction for research. The construction in Theorem~\ref{thm:enclosure} relies on separable, convex link costs, exact feasibility and the shortest-route value $S$. Whether comparable computable bounds can be established for stochastic user equilibrium, elastic demand and multiclass assignment remains an open question. Developing certified equilibrium sensitivities would also be relevant to network design and road pricing \citep{BarGeraEtAl2013,JosefssonPatriksson2007}.

\subsection{Limitations}\label{sec:limitations}

The guarantee is limited in three ways. First, it holds for static user equilibrium with fixed demand, one
user class and separable convex costs, at an exactly feasible point. Junction models and other non-separable costs are
outside Theorem~\ref{thm:enclosure}, because its energy bound uses the costs link by link.

Second, the guarantee has a computational cost. On the regional networks it is a multiple of the cost of the tight
shared-base warm start (Section~\ref{sec:cost}), and for a single scheme the certified base costs several times the scheme's own
appraisal. Proposition~\ref{prop:law} shows that every certificate built from the gap and
the externality weight needs a relative gap of order $\varepsilon^2$, and the extra cost pays for reaching that gap.
For single-link schemes with benefits of a few millionths of the total cost, the absolute term sets the tolerance, so the certified interval does not always establish the sign of the benefit. By Proposition~\ref{prop:law},
certifying a benefit of one millionth of the total cost to 10\% on these networks would need a gap of about $10^{-14}$ of the total cost, near the limit of double-precision arithmetic.

Third, the guarantee covers the numerical part of an appraisal, and its evidence is bounded. The benefit certified is
the change in equilibrium travel cost under fixed demand, the numerical travel-cost component of an appraisal. A
certified interval bounds the numerical error of this model output, not the error of the model, so demand response,
scheme costs, environmental effects and the uncertainty of model inputs lie outside the guarantee. The benefit
interval also covers schemes that add or remove links, while the incremental method and the experiments change link
costs on a fixed network. The blind tests changed the capacities of single links and, in the second, of areas of up to
300 links, on two regional networks and with one assignment algorithm. Larger
responses need more iterations, and the near-closures of a link, which produce the largest
responses, counted in the first test's validity hypothesis only, where all their intervals agreed with the references.

\section{Conclusions}\label{sec:conclusions}

The benefit of a network change under user equilibrium is the difference of two equilibrium total costs, each known
only to the accuracy at which its run was stopped, so a solver's gap does not bound the error of the benefit. This
paper has established what a computed assignment does guarantee. A two-sided bound on the equilibrium
total cost can be computed at any exactly feasible flow, without strong monotonicity, nondegeneracy or uniqueness of
link flows, and two such bounds give a guaranteed benefit interval. The limits of gap-based certification are established as well. The refined bound is the best that any rule based only on the gap and the externality weight can achieve, and for BPR costs every such rule needs a relative gap
of order $\varepsilon^2$ to certify a relative accuracy $\varepsilon$.

The certified appraisal method makes these results operational. It certifies the base once and each scheme along a warm-started run, and its intervals are valid whatever the run's limits. On two regional planning
networks, conventional stopping at a relative gap of $10^{-4}$ left 25 of 55 held-out benefits outside their tolerance. In a prespecified blind test on 72 new schemes, all 577 certified intervals agreed with high-accuracy reference benefits within the reference uncertainty, every scheme met a tolerance of $10^{-5}$ of the total cost plus 10\% of the benefit, and elimination certified the scheme with the largest benefit on each network. For benefits of at least a thousandth of the total cost, a certified accuracy of 10\% cost a median 0.75 times as much as a cold assignment stopped at $10^{-6}$.

These results connect the convergence of a traffic assignment to the reliability of the decisions drawn from
it. With a certified interval, the accuracy that an appraisal needs is stated for the benefit itself, a run stops when it is reached, and computable criteria certify the sign of a benefit and the ranking of two schemes. Such an interval can accompany every reported benefit and replace rules of thumb for convergence. For research, these limits point to the information that a sharper certificate would need, such as the equilibrium face and the margins of unused routes.

\section*{Acknowledgments}
The authors thank the Transportation Networks for Research Core Team for maintaining the benchmark networks used in
this study. Computational resources were provided by the Cambridge Service for Data Driven Discovery (CSD3), operated
by the University of Cambridge Research Computing Service, with support from Dell EMC and Intel under EPSRC Tier-2
funding (EP/T022159/1) and DiRAC funding from the Science and Technology Facilities Council.

\section*{Funding}
This work was supported by the Cambridge Martin Centre for Architectural and Urban Studies and Cambridge Public
Health. Yue Li and Shujuan Chen were supported by the Cambridge Commonwealth, European and International Trust.

\section*{Disclosure statement}
The authors report there are no competing interests to declare.

\section*{Declaration of generative AI use}
During the preparation of this work, the authors used ChatGPT (OpenAI) and Claude Code (Anthropic) for language
enhancement and code development. The authors reviewed and edited the output as needed and take full responsibility
for the content of the publication.

\section*{Data and code availability}
The benchmark networks are available from the Transportation Networks for Research repository
(\url{https://github.com/bstabler/TransportationNetworks}). An open-source software package that implements the method, CertiFlow, is available at \url{https://github.com/yueli901/CertiFlow}. The scheme definitions, the frozen protocols of the two blind tests
with their checksums, the experiment configurations, the records and the analysis code that support the findings are
available from the corresponding author upon reasonable request.

\bibliographystyle{plainnat}
\bibliography{references}

\clearpage
\appendix
\section{Notation}\label{app:notation}

Table~\ref{tab:notation} summarizes the principal notation used throughout the paper.

\begin{table}[!ht]
\centering
\caption{Summary of notation.}\label{tab:notation}
\footnotesize
\begin{adjustbox}{max width=\linewidth}
\begin{tabular}{@{}>{\raggedright\arraybackslash}p{0.21\textwidth}>{\raggedright\arraybackslash}p{0.75\textwidth}@{}}
\toprule
Symbol & Definition \\
\midrule
\multicolumn{2}{@{}l}{\textit{Network and equilibrium}} \\
$\Links$, $\ODs$, $\Routes_w$ & Set of links, set of OD pairs, set of routes for OD pair $w$ \\
$d_w$, $h$, $\Lambda$ & Demand of OD pair $w$, route flow vector, link\textendash route incidence matrix \\
$\Feas$, $x$, $x_a$ & Set of feasible link flows, link flow vector, flow on link $a$ \\
$t_a$, $t(x)$ & Link cost function, vector of link costs \\
$t_a^0$, $\alpha_a$, $C_a$, $\beta_a$ & Free-flow time, BPR coefficient, capacity and power of link $a$ \\
$S(c)$ & Total cost of assigning all demand to shortest routes at costs $c$, $\min_{v\in\Feas}c\T v$ \\
$\xs$, $\cs$, $\Js$ & Equilibrium link flow, equilibrium link costs, equilibrium total cost \\
$\TSTT(x)$, $\SPTTm(x)$ & Total system travel time $t(x)\T x$, shortest-path travel time $S(t(x))$ \\
$\Gc(x)$ & Convergence gap $\TSTT(x)-\SPTTm(x)$, called the raw gap at $\xh$, and relative gap $\Gc(x)/\SPTTm(x)$ \\
$D(x)$, $f$, $\beta_{\min}$ & Total delay, delay share $D/\TSTT$, minimum BPR power among links with positive flow and delay \\
\midrule
\multicolumn{2}{@{}l}{\textit{Appraisal and accuracy}} \\
$\theta$, $t^{(\theta)}$, $s$, $m$ & Network index (0 for the base), network-specific link cost functions, scheme index, number of schemes \\
$B_s$, $\Bh_s$, $\Bm_s$ & Benefit of scheme $s$, conventional benefit estimate, midpoint of the benefit interval \\
$e_\theta$ & Error in the total-cost estimate of network $\theta$ \\
$\TC_0$, $\epsabs$, $\epsrel$ & Fixed normalizing constant, absolute and relative tolerance parameters \\
$\tau_s$, $\hat\tau_s$, $\hat m_s$ & Prescribed tolerance, computable tolerance, computable lower bound on $|B_s|$ \\
$[a_0,b_0]$, $[a_s,b_s]$ & Certified equilibrium total-cost intervals for the base and scheme $s$ \\
$[L_s,U_s]$, $\mathcal{E}_s$ & Certified benefit interval (intersection over checkpoints in Algorithm~\ref{alg:cia}), certified error bound \\
\midrule
\multicolumn{2}{@{}l}{\textit{Certificates}} \\
$\xh$, $\ch$ & Exactly feasible link flow and its link costs \\
$\bar z_a$, $\bar\sigma_a$ & Step on link $a$ and corresponding secant slope bounding $\sigma_a$ (Section~\ref{sec:plain}) \\
$\psi$, $\psi_0$ & Externality weight $\bigl(\sum_a\bar\sigma_a\xh_a^2\bigr)^{1/2}$ and its derivative-based lower bound \\
$I_0$, $p^*$ & Interval of Theorem~\ref{thm:plain}, optimal lower offset from $\SPTTm(\xh)$ \\
$\delta$, $\Delta$, $\sigma_a$, $E$ & Flow error, cost error, secant slope between $\xs_a$ and $\xh_a$, energy $\Delta\T\delta$ \\
$\cb$, $r$, $\mathbb{F}(\xh)$ & Backward cost, residual $\ch-\cb$, set of admissible backward costs \\
$G$ & Gap $\cb\T\xh-S(\cb)$ at the backward cost, equal to $\Gc(\xh)$ when $\cb=\ch$ \\
$\mathcal{P}$, $\mathcal{Z}$ & Sets of links with positive and zero flow at $\xh$, respectively \\
$\mu_a$, $R$, $E_{\max}$ & Secant slope from zero flow ($\mu_a:=0$ on $\mathcal{Z}$), weighted cost residual, upper bound on energy \\
$\varrho$ & Backward residual $(R^2/4+G)^{1/2}$ \\
$I_{\mathrm{S}}$, $I_{\mathrm{T}}$, $I$, $I^*$ & Intervals based on $\SPTTm$ and $\TSTT$, their intersection, refined interval \\
$\ell^*$, $p^*_{\mathrm{T}}$ & Refined lower bound and its offset (Proposition~\ref{prop:sharp}) \\
$p_-$, $p_+$ & Lower and upper offsets of a $(\psi,G)$-rule (Definition~\ref{def:rule}) \\
$M$, $N$ & Margin and equilibrium-gap components of the raw gap (Lemma~\ref{lem:parts}) \\
$\Phi$, $\gamma$, $\cb^*$, $v^*$ & Regularized objective, regularized gap, residual-minimizing backward cost, maximizer of $\Phi$ \\
$G_v$ & Gap of $v$ in the affine assignment of Theorem~\ref{thm:best} \\
\midrule
\multicolumn{2}{@{}l}{\textit{Algorithm and experiments}} \\
$g_j$, $n$ & Checkpoint relative gaps $g_1>\dots>g_n$ and number of checkpoints \\
$w_0$, $w_s$ & Half-widths of the base interval and the intersection of scheme intervals \\
$W_\theta$, $c_\theta$ & Computational work to achieve a given half-width on network $\theta$, logarithmic cost coefficient (Proposition~\ref{prop:allocation}) \\
$g$, A$(g)$, W$(g)$ & Target relative gap, cold-start and shared-base warm-start protocols (Section~\ref{sec:refs}) \\
$B_{\mathrm{ref}}$, $u_{\mathrm{ref}}$ & Reference benefit and reference uncertainty (Section~\ref{sec:refs}) \\
\bottomrule
\end{tabular}
\end{adjustbox}
\floatnote{A hat generally denotes a computed quantity, and a star denotes an equilibrium or optimal quantity. Uppercase $G$ denotes a gap in units of total cost, and lowercase $g$ denotes a relative gap.}
\end{table}
\section{Proofs}\label{app:proofs}

Throughout this appendix, let $\xs$ be any equilibrium, $\delta:=\xh-\xs$ and $\Delta:=\ch-\cs$. Let $\sigma_a$ denote the secant slope of $t_a$ between $\xs_a$ and $\xh_a$, with $\sigma_a:=t_a'(\xh_a)$ when the two flows coincide. Then $\Delta_a=\sigma_a\delta_a$ and $E=\Delta\T\delta=\sum_a\sigma_a\delta_a^2\ge0$.

\subsection{The Equilibrium Total Cost}

\begin{applemma}\label{lem:unique}
Under Assumption~\ref{ass:costs}, an equilibrium exists, and the equilibrium link costs $\cs$ and the total cost $\Js=\cs\T\xs=S(\cs)$ are the same for every equilibrium.
\end{applemma}

\begin{pf}
The function $F(x):=\sum_a\int_0^{x_a}t_a(u)\,du$ is convex and continuously differentiable, with gradient $t(x)$. The equilibria are therefore the minimizers of $F$ over the compact polytope $\Feas$ \citep{BeckmannEtAl1956,Dafermos1980,Patriksson1994}, and a minimizer exists.

Let $x$ and $x'$ be two equilibria. Adding \eqref{eq:vi} at $x$ with $v=x'$ and at $x'$ with $v=x$ gives
$\sum_a\bigl(t_a(x_a)-t_a(x'_a)\bigr)(x_a-x'_a)\le0$.
Since each $t_a$ is nondecreasing, every term in the sum is nonnegative and must therefore vanish. It follows that $t_a(x_a)=t_a(x'_a)$ for every link, establishing the uniqueness of $\cs$. The equilibrium condition \eqref{eq:vi} also gives $\cs\T\xs\le\cs\T v$ for every $v\in\Feas$, so that $\cs\T\xs=S(\cs)$, which depends only on $\cs$.
\end{pf}

\subsection{The Secant Bound}

The following lemma provides a computable upper bound on the unknown secant slopes $\sigma_a$ used in both certificates.

\begin{applemma}\label{lem:secant}
Let Assumption~\ref{ass:costs} hold, let $\xh\in\Feas$, and let the steps $\bar z_a$ and the slopes $\bar\sigma_a$ be defined as in Section~\ref{sec:plain} with some $\bar E\ge E$ in place of $G$. Then $\sigma_a\le \bar\sigma_a$ on every link.
\end{applemma}

\begin{pf}
Each term $\sigma_a\delta_a^2$ is bounded above by $E\le\bar E$. If $\xs_a\le\xh_a$, convexity gives $\sigma_a\le t_a'(\xh_a)\le\bar\sigma_a$.

Now suppose that $\xs_a>\xh_a$. Define $E_a(z):=\bigl(t_a(\xh_a+z)-t_a(\xh_a)\bigr)z=\bar\sigma_a(z)z^2$, where $\bar\sigma_a(z)$ is the secant slope of $t_a$ from $\xh_a$ over a step $z$. By convexity, $\bar\sigma_a(z)$ is nondecreasing in $z$, and $E_a(\xs_a-\xh_a)=\sigma_a\delta_a^2\le\bar E$.

If $t_a$ is constant, then $\sigma_a=0=\bar\sigma_a$. If $\bar E=0$, then $\sigma_a\delta_a^2=0$, so $\sigma_a=0\le\bar\sigma_a$. Otherwise, $E_a(z)\to\infty$ as $z\to\infty$, ensuring the existence of $\bar z_a$. Moreover, $\bar\sigma_a(\bar z_a)>0$ because $E_a(\bar z_a)\ge\bar E>0$. Hence $E_a(z)>E_a(\bar z_a)\ge\bar E$ for every $z>\bar z_a$. It follows that $\xs_a-\xh_a\le\bar z_a$ and, by the monotonicity of the secant slope, $\sigma_a=\bar\sigma_a(\xs_a-\xh_a)\le\bar\sigma_a(\bar z_a)=\bar\sigma_a$.
\end{pf}

\subsection{Proof of Theorem~\ref{thm:enclosure}}

\begin{pf}[Proof of Theorem~\ref{thm:enclosure}]
The proof proceeds in three parts: bounding the energy, controlling the cost error and constructing the interval.

\emph{Part I. Bounding the energy (item~1).} Step 1. By the definition $G=\cb\T\xh-S(\cb)$, we have $\cb\T(v-\xh)\ge-G$ for every $v\in\Feas$. Substituting $v=\xs$ and $\cb=\ch-r$ gives $(\ch-r)\T(\xs-\xh)\ge-G$. The equilibrium inequality~\eqref{eq:vi} at $v=\xh$ gives $\cs\T(\xh-\xs)\ge0$. Adding these inequalities yields
\begin{equation}\label{eq:step1}
E\le r\T\delta+G.
\end{equation}

Step 2. We use three properties of the link cost functions and residuals.

(F1) On $\mathcal{Z}$, $\delta_a=-\xs_a\le0$ and $r_a\ge0$, so $r_a\delta_a\le0$.

(F2) On $\mathcal{P}$, $\sigma_a\ge\mu_a$. If $\xs_a<\xh_a$, this follows from the monotonicity of the secant slopes of a convex function in both endpoints, since $\xs_a\ge0$. Otherwise, the secant slope from $\xh_a$ to $\xs_a\ge\xh_a$ is at least $t_a'(\xh_a)$, which is at least the secant slope from $0$ to $\xh_a$.

(F3) If $a\in\mathcal{P}$ and $\sigma_a=0$, then $t_a$ is constant on $[0,\xh_a]$. Consequently, $\Delta_a=0$ and $\mu_a=0$, and the condition $R<\infty$ implies $r_a=0$.

Step 3. By (F1), $r\T\delta\le\sum_{a\in\mathcal{P}}r_a\delta_a$. Applying the Cauchy\textendash Schwarz inequality and (F2) gives
$\sum_{\mathcal{P}}r_a\delta_a\le\bigl(\sum_{\mathcal{P},r_a\ne0}r_a^2/\mu_a\bigr)^{1/2}\bigl(\sum_{\mathcal{P}}\mu_a\delta_a^2\bigr)^{1/2}\le R\,E^{1/2}$.
Substituting into \eqref{eq:step1} yields $E\le R\,E^{1/2}+G$. Solving this quadratic inequality in $E^{1/2}$ gives $E\le E_{\max}$, establishing item~1.

\emph{Part II. Controlling the cost error.} Step 4. Since $E\le E_{\max}$, Lemma~\ref{lem:secant} with $\bar E=E_{\max}$ gives $\sigma_a\le\bar\sigma_a$ on every link.

Step 5. Let $\mathcal{P}^+:=\{a\in\mathcal{P}:\sigma_a>0\}$. By (F1) and (F3), inequality \eqref{eq:step1} gives
$\sum_a\sigma_a\delta_a^2\le\sum_{\mathcal{P}^+}r_a\delta_a+G$.
Completing the square on $\mathcal{P}^+$ and using $\sigma_a\ge\mu_a$ yields
\begin{align}
&\sum_{a\in\mathcal{P}^+}\sigma_a\Bigl(\delta_a-\frac{r_a}{2\sigma_a}\Bigr)^2+\sum_{a\in\mathcal{Z}}\sigma_a\delta_a^2
\nonumber\\
&\qquad\le G+\sum_{a\in\mathcal{P}^+}\frac{r_a^2}{4\sigma_a}\le G+\frac{R^2}{4}=\varrho^2. \label{eq:square}
\end{align}
Define $\omega_a:=\Delta_a-r_a/2$ on $\mathcal{P}$ and $\omega_a:=\Delta_a$ on $\mathcal{Z}$. Then \eqref{eq:square} gives $q\le\varrho$, where $q:=\bigl(\sum_{\sigma_a>0}\omega_a^2/\sigma_a\bigr)^{1/2}$. Moreover, $\omega_a=0$ wherever $\sigma_a=0$, by (F3) on $\mathcal{P}$ and by $\Delta_a=\sigma_a\delta_a$ on $\mathcal{Z}$. For any vector $v$, we have $v\T\Delta=\tfrac12\sum_{\mathcal{P}}r_av_a+\sum_a v_a\omega_a$. Applying the Cauchy\textendash Schwarz inequality and using $\sigma_a\le\bar\sigma_a$ gives
\begin{equation}\label{eq:cs}
\Bigl|v\T\Delta-\tfrac12\textstyle\sum_{a\in\mathcal{P}}r_av_a\Bigr|\le\Bigl(\sum_a\bar\sigma_av_a^2\Bigr)^{1/2}\varrho.
\end{equation}

\emph{Part III. Constructing the interval (items~2 to~4).} Step 6. To obtain the upper bound on $\SPTTm(\xh)-\Js$, note that $S(\ch)\le\ch\T\xs$ and $S(\cs)=\cs\T\xs$. Hence $\SPTTm(\xh)-\Js\le\Delta\T\xs=\xh\T\Delta-E$.

For the lower bound, superadditivity gives $S(\ch)=S(\cb+r)\ge S(\cb)+S(r)=\cb\T\xh-G+S(r)$. Since $S(\cs)\le\cs\T\xh$, it follows that $\SPTTm(\xh)-\Js\ge\xh\T\Delta-r\T\xh-G+S(r)$.

For $\TSTT$, we have $\TSTT(\xh)-\Js=\xh\T\Delta+\cs\T\delta$, where $\cs\T\delta\ge0$ by \eqref{eq:vi}. The gap inequality in Step~1 also gives $\cs\T\delta=\ch\T\delta-E\le r\T\delta-E+G\le R^2/4+G$, with the final inequality obtained by completing the square as in Step~5.

Applying \eqref{eq:cs} with $v=\xh$, for which $\sum_a\bar\sigma_a\xh_a^2=\psi^2$ and $\sum_{\mathcal{P}}r_a\xh_a=r\T\xh$, establishes items~2 and~3, using $E\ge0$ in the upper bound on $\SPTTm(\xh)-\Js$.

Step 7. By Lemma~\ref{lem:unique}, both intervals contain the same equilibrium total cost $\Js$. Their intersection therefore also contains $\Js$, establishing item~4.
\end{pf}
\subsection{Proofs of Lemma~\ref{lem:width}, Lemma~\ref{lem:inside} and Proposition~\ref{prop:sharp}}

\begin{pf}[Proof of Lemma~\ref{lem:width}]
Let $\ell_{\mathrm{S}},u_{\mathrm{S}}$ and $\ell_{\mathrm{T}},u_{\mathrm{T}}$ denote the lower and upper endpoints of $I_{\mathrm{S}}$ and $I_{\mathrm{T}}$, respectively. By superadditivity, $S(\ch)\ge S(\cb)+S(r)$, so $\Gc\le G+r\T\xh-S(r)$. It follows that $u_{\mathrm{S}}-u_{\mathrm{T}}=-\Gc+r\T\xh-S(r)+G\ge0$. Thus, the upper endpoint of $I$ is $u_{\mathrm{T}}$, and its width is
$|I|=\min(u_{\mathrm{T}}-\ell_{\mathrm{S}},u_{\mathrm{T}}-\ell_{\mathrm{T}})=\min(\Gc+2\psi\varrho,\,2\psi\varrho+R^2/4+G)$.
\end{pf}

\begin{pf}[Proof of Lemma~\ref{lem:inside}]
By convexity, $\mu_a\le t_a'(\xh_a)\le\bar\sigma_a$. The Cauchy\textendash Schwarz inequality therefore gives $|r\T\xh|\le R\psi\le2\psi\varrho$. Hence
$u_{\mathrm{T}}-\SPTTm(\xh)=\Gc-\tfrac12r\T\xh+\psi\varrho\ge0$ and
$\SPTTm(\xh)-\ell_{\mathrm{S}}=\tfrac12r\T\xh+\psi\varrho\ge0$.
Using the bound on $\Gc$ from the proof of Lemma~\ref{lem:width} also gives
$\SPTTm(\xh)-\ell_{\mathrm{T}}\ge\psi\varrho-\tfrac12r\T\xh+S(r)+R^2/4\ge S(r)\ge0$.
Thus, $\SPTTm(\xh)\in I$.
\end{pf}

\begin{pf}[Proof of Proposition~\ref{prop:sharp}]
Let $q$ be as defined in Step~5 and let $\varrho_\sigma^2:=G+\sum_{\mathcal{P}^+}r_a^2/(4\sigma_a)\le\varrho^2$. Step~1 gives $\cs\T\delta\le G+r\T\delta-E$, while completing the square in Step~5 gives
$E-r\T\delta\ge q^2-\sum_{\mathcal{P}^+}r_a^2/(4\sigma_a)$.
Consequently, $\cs\T\delta\le\varrho_\sigma^2-q^2$, and since $\cs\T\delta\ge0$, we have $q\le\varrho_\sigma$.

The Cauchy\textendash Schwarz inequality in Step~5, before substituting $q\le\varrho$, also gives $\xh\T\Delta\le\tfrac12r\T\xh+\psi q$. Hence, for $q\in[0,\varrho_\sigma]$,
$\TSTT(\xh)-\Js=\xh\T\Delta+\cs\T\delta\le\tfrac12r\T\xh+\psi q+\varrho_\sigma^2-q^2$.
The maximum of $\psi q+\varrho_\sigma^2-q^2$ over this interval is $\psi\varrho_\sigma$ if $\psi\ge2\varrho_\sigma$ and $\varrho_\sigma^2+\psi^2/4$ otherwise. This maximum is continuous and nondecreasing in $\varrho_\sigma$. Since $\varrho_\sigma\le\varrho$, we obtain
$\TSTT(\xh)-\Js\le\tfrac12r\T\xh+p^*_{\mathrm{T}}$.
Combining this bound with $\Js\ge\ell_{\mathrm{S}}$ from Theorem~\ref{thm:enclosure} gives $\Js\in I^*$.

Since $\varrho^2+\psi^2/4-\psi\varrho=(\varrho-\psi/2)^2\ge0$, we have $p^*_{\mathrm{T}}\ge\psi\varrho$, with equality when $\psi\ge2\varrho$.
Moreover, $\ell^*-\ell_{\mathrm{T}}=\varrho^2+\psi\varrho-p^*_{\mathrm{T}}$, which equals $\varrho^2$ when $\psi\ge2\varrho$ and $\psi(\varrho-\psi/4)\ge0$ otherwise. Thus, $I^*\subseteq I$.

The width of $I^*$ is $\min(u_{\mathrm{T}}-\ell^*,u_{\mathrm{T}}-\ell_{\mathrm{S}})$, where $u_{\mathrm{T}}-\ell^*=\psi\varrho+p^*_{\mathrm{T}}\ge2\psi\varrho$ and $u_{\mathrm{T}}-\ell_{\mathrm{S}}=\Gc+2\psi\varrho$ by the proof of Lemma~\ref{lem:width}. When $\psi\ge2\varrho$, we have $u_{\mathrm{T}}-\ell^*=2\psi\varrho\le u_{\mathrm{T}}-\ell_{\mathrm{S}}$, so $|I^*|=2\psi\varrho$.
\end{pf}

\emph{The case $\cb=\ch$.}
In this case, $r=0$, $R=0$, $E_{\max}=G=\Gc(\xh)$ and $\varrho=G^{1/2}$, so the steps $\bar z_a$, the slopes $\bar\sigma_a$ and the externality weight $\psi$ coincide with those of Section~\ref{sec:plain}. Since $S(0)=0$ and $\SPTTm(\xh)+G=\TSTT(\xh)$, both $I_{\mathrm{S}}$ and $I_{\mathrm{T}}$ reduce to
$\bigl[\SPTTm(\xh)-\psi G^{1/2},\,\TSTT(\xh)+\psi G^{1/2}\bigr]=I_0$.
The refined lower bound is $\ell^*=\SPTTm(\xh)+G-p^*_{\mathrm{T}}$, which equals $\SPTTm(\xh)-\psi G^{1/2}+G$ if $\psi\ge2G^{1/2}$ and $\SPTTm(\xh)-\psi^2/4$ otherwise. Thus, $\ell^*=\SPTTm(\xh)-p^*(\psi,G)$, recovering the tighter lower bound of Theorem~\ref{thm:plain}.

Corollary~\ref{cor:benefit} follows directly from $B_s=\Js_0-\Js_s$ by interval subtraction. The error bound $\mathcal{E}_s$ holds for any interval containing $B_s$, regardless of whether it contains $\Bh_s$. The midpoint minimizes $\max_{B\in[L_s,U_s]}|\tilde b-B|$ over $\tilde b\in\R$, with minimum value $\tfrac12(U_s-L_s)$. By interval subtraction, $U_s-L_s=(b_0-a_0)+(b_s-a_s)=2(w_0+w_s)$. Corollary~\ref{cor:compare} follows similarly from $B_i-B_j=\Js_j-\Js_i$.
\subsection{Proofs of Corollary~\ref{cor:optimal} and Proposition~\ref{prop:law}}

\begin{pf}[Proof of Corollary~\ref{cor:optimal}]
\emph{Upper bound.} Consider Example~\ref{ex:tworoute} with $\sigma=1$, $z=G^{1/2}$ and $d=\psi+G^{1/2}$. The resulting feasible point has gap $G$ and externality weight $\psi$, with $\Js-\SPTTm(\xh)=\sigma z d=\psi G^{1/2}+G$. Any valid rule must therefore satisfy $p_+(\psi,G)\ge\psi G^{1/2}+G$.

For BPR costs of any power $\beta\ge1$, take $z>0$ and $d:=z(1+\psi/G^{1/2})$. Let route~1 have cost $t_1(x)=1+Kx^\beta$, where $K:=G/\bigl(z(d^\beta-(d-z)^\beta)\bigr)$, and route~2 have constant cost $t_1(d)$, corresponding to a BPR function with $\alpha=0$. At the feasible point assigning flow $z$ to route~2, the gap is $z\bigl(t_1(d)-t_1(d-z)\bigr)=G$. Hence $\bar z_1=z$, $\bar\sigma_1=G/z^2$ and $\psi=\bar\sigma_1^{1/2}(d-z)$, giving $\Js-\SPTTm(\xh)=d\,G/z=\psi G^{1/2}+G$.

\emph{Lower bound.} The rule with $p_-=p^*$ and $p_+=\psi G^{1/2}+G$ is valid by Theorem~\ref{thm:plain}, since $\TSTT(\xh)=\SPTTm(\xh)+G$. To establish attainment, consider two OD pairs sharing an affine link $l$ with cost $t_l(x)=A+\sigma x$. OD pair~2 has demand $d_2$ and uses only $l$. OD pair~1 has demand $d_1$ and a constant-cost bypass with cost $A+\sigma d_2$, so its route through $l$ is tied with the bypass at zero flow of OD pair~1. At a feasible point where OD pair~1 assigns flow $z\le\min(d_1,d_2)$ to $l$, we have $\psi^2=\sigma(d_2+z)^2$, $G=\sigma z^2$ and $\SPTTm(\xh)-\Js=\sigma zd_2=\psi G^{1/2}-G$. For $\psi<2G^{1/2}$, take $z=d_2\le d_1$ and introduce an additional OD pair with two constant-cost links to account for the remaining gap, giving $\SPTTm(\xh)-\Js=\psi^2/4$. These constructions attain the lower offset for every $(\psi,G)$.

\emph{Strictly convex costs.} Suppose that every nonconstant link cost function is strictly convex and that $\SPTTm(\xh)>\Js$. The proof of Theorem~\ref{thm:plain} gives $\SPTTm(\xh)-\Js\le\xh\T\Delta-E$, implying $\xh\T\Delta>E\ge0$. Consequently, at least one link satisfies $\xh_a>\xs_a$ and $\sigma_a>0$. Strict convexity on this link gives $\sigma_a<t_a'(\xh_a)\le\bar\sigma_a$. Since $\sigma_a\le\bar\sigma_a$ on every other link (Lemma~\ref{lem:secant}), we have $\psi_\sigma^2:=\sum_a\sigma_a\xh_a^2<\psi^2$. The Cauchy\textendash Schwarz inequality then gives
$\SPTTm(\xh)-\Js\le\psi_\sigma E^{1/2}-E\le p^*(\psi_\sigma,G)<p^*(\psi,G)$,
because $G\ge E>0$ and $p^*$ is strictly increasing in $\psi$ when $G>0$. Thus, the lower bound is not attained under strict convexity of the nonconstant costs.

To show that the lower bound can nevertheless be approached, add $\nu x^2$, with $\nu>0$, to $t_l$ in the preceding constructions while keeping $(\psi,G)$ fixed. For $\psi\ge2G^{1/2}$, fix $z$ and solve the equations defining $\psi$ and $G$ for $(\sigma,d_2)$, whose Jacobian at $\nu=0$ is nonsingular. For $\psi<2G^{1/2}$, fix $z=d_2$, solve for $\sigma$, and assign the remaining gap to the additional OD pair. In both cases, the solution varies continuously with $\nu$, and $\SPTTm(\xh)-\Js$ tends to $p^*(\psi,G)$ as $\nu\to0$. Hence $p_-\ge p^*$ even for rules restricted to networks with strictly convex nonconstant link costs.
\end{pf}

\begin{pf}[Proof of Proposition~\ref{prop:law}]
By convexity, $\bar\sigma_a\ge t_a'(\xh_a)$, and the BPR function satisfies $x\,t_a'(x)=\beta_a\bigl(t_a(x)-t_a^0\bigr)$. Links with $\xh_a=0$ or $t_a^0\alpha_a=0$ contribute neither to $D$ nor to $\sum_at_a'(\xh_a)\xh_a^2$. Therefore,
$\psi^2\ge\sum_a\beta_a\xh_a\bigl(t_a(\xh_a)-t_a^0\bigr)\ge\beta_{\min}D(\xh)$.
If no link satisfies both $\xh_a>0$ and $t_a^0\alpha_a>0$, then $D(\xh)=0$ and the inequality holds with $\beta_{\min}=1$.

Lemma~\ref{lem:width} and Proposition~\ref{prop:sharp} give $|I|\ge|I^*|\ge2\psi\varrho\ge2\psi G^{1/2}$. By Corollary~\ref{cor:optimal}, every valid $(\psi,G)$-rule has width at least $\psi G^{1/2}+G+p^*$, which equals $2\psi G^{1/2}$ when $\psi\ge2G^{1/2}$ and $(G^{1/2}+\psi/2)^2\ge2\psi G^{1/2}$ otherwise. A certified half-width of at most $\varepsilon\,\TSTT(\xh)$ therefore requires $\beta_{\min}D(\xh)\,G\le\varepsilon^2\TSTT(\xh)^2$, yielding \eqref{eq:law}.

Suppose the relative gap decreases geometrically from $g_0$, by a constant factor at each iteration. The number of iterations required to satisfy \eqref{eq:law} is then proportional to $\log\bigl(g_0\beta_{\min}f/\varepsilon^2\bigr)$, while that required for an estimate whose error is of the order of the gap is proportional to $\log(g_0/\varepsilon)$, with the same proportionality constant. The ratio of these iteration counts tends to two as $\varepsilon\to0$.
\end{pf}
\subsection{Proofs of Lemma~\ref{lem:parts}, Theorem~\ref{thm:best} and Corollary~\ref{cor:margin}}

\begin{pf}[Proof of Lemma~\ref{lem:parts}]
By definition, $\Gc=\ch\T\xh-S(\ch)=\ch\T\delta+\bigl(\ch\T\xs-S(\ch)\bigr)$.
Since $\cs\T\xs=\Js$, we have $\ch\T\delta=\Delta\T\delta+\cs\T\delta=E+M$, giving $\Gc=E+M+N$.
Monotonicity gives $E\ge0$, the equilibrium condition \eqref{eq:vi} gives $M\ge0$, and feasibility of $\xs$ gives $N\ge0$. The margin $M=\cs\T\xh-\Js$ depends on $\xs$ only through $\cs$ and $\Js$, both of which are unique by Lemma~\ref{lem:unique}.
\end{pf}

\begin{pf}[Proof of Theorem~\ref{thm:best}]
For $v\in\Feas$, define $\cb_v:=\ch-2\mu\circ(\xh-v)$ and $G_v:=\cb_v\T v-S(\cb_v)\ge0$. We establish three facts:
(1) For every $\cb\in\mathbb{F}(\xh)$, $\varrho^2(\cb)\ge\gamma(\xh)$.
(2) For every $v\in\Feas$, $\cb_v\in\mathbb{F}(\xh)$ and $\varrho^2(\cb_v)=\Phi(v)+G_v$.
(3) A flow $v$ maximizes $\Phi$ if and only if $G_v=0$.

Let $u:=\xh-v$. Facts~(1) and~(2) follow from the identity
$-r_au_a+r_a^2/(4\mu_a)+\mu_au_a^2=\bigl(r_a/(2\mu_a^{1/2})-\mu_a^{1/2}u_a\bigr)^2\ge0$ for $\mu_a>0$, with equality when $r_a=2\mu_au_a$. Fact~(3) follows from the first-order optimality condition for maximizing a concave function over a convex set.

For fact~(1), consider
$\cb\T u+R^2/4=\ch\T u-\sum_ar_au_a+\sum_{\mathcal{P},r_a\ne0}r_a^2/(4\mu_a)$
and compare it with $\Phi(v)=\ch\T u-\sum_a\mu_au_a^2$ link by link. On links of $\mathcal{P}$ with $\mu_a>0$, the difference is the nonnegative square above. On links of $\mathcal{P}$ with $\mu_a=0$, admissibility requires $r_a=0$, so the difference is zero. On $\mathcal{Z}$, $u_a=-v_a\le0$ and $r_a\ge0$, giving $-r_au_a\ge0=-\mu_au_a^2$. Maximizing over $v\in\Feas$ therefore yields $\varrho^2(\cb)=G(\cb)+R^2/4\ge\gamma(\xh)$.

For fact~(2), the residual $r=2\mu\circ u$ vanishes wherever $\mu_a=0$, so $\cb_v$ is admissible. The equality condition of the completed square gives $R^2/4=\sum_a\mu_au_a^2$. Hence
$\varrho^2(\cb_v)=\sum_a\mu_au_a^2+\cb_v\T u+G_v=\Phi(v)+G_v$.

For fact~(3), $\Phi$ is concave and differentiable with $\nabla\Phi(v)=-\cb_v$. Thus, $v$ maximizes $\Phi$ over $\Feas$ if and only if $\cb_v\T(y-v)\ge0$ for every $y\in\Feas$, which is equivalent to $G_v=0$.

At any maximizer $v^*$, fact~(2) gives $\varrho^2(\cb_{v^*})=\gamma$, and fact~(1) establishes that this is the minimum over all admissible backward costs. For any two maximizers $v_1,v_2$, the quadratic form of $\Phi$ gives
$\Phi\bigl((v_1+v_2)/2\bigr)=\tfrac12\Phi(v_1)+\tfrac12\Phi(v_2)+\tfrac14\sum_a\mu_a(v_{1,a}-v_{2,a})^2$.
Since the midpoint is feasible and cannot exceed the maximum, $\mu\circ v^*$ and hence $\cb^*$ are identical for all maximizers. Finally, $\gamma\ge\Phi(v)$ by definition, while facts~(1) and~(2) give $\gamma\le\varrho^2(\cb_v)=\Phi(v)+G_v$.
\end{pf}

The residual-minimizing backward cost $\cb^*$ is obtained by solving the affine assignment with link costs $\ch_a+2\mu_a(x_a-\xh_a)$. Dijkstra's algorithm can be used for the shortest-path computations when these costs are nonnegative for all nonnegative flows, which holds if and only if $t_a(\xh_a)\le2t_a(0)$ on every link.

The following lemma establishes the bounds on $\gamma$ used in Corollary~\ref{cor:margin}.

\begin{applemma}\label{lem:gammabounds}
For every equilibrium, $\gamma(\xh)\ge M$ and $\gamma(\xh)\ge\tfrac14E+\tfrac12M$. If $\Delta_a=0$ on every link with
$\mu_a=0$, including every link of $\mathcal{Z}$, where $\mu_a:=0$, then
$\gamma(\xh)\le M+\tfrac14\sum_{\mathcal{P},\mu_a>0}\Delta_a^2/\mu_a$. If, in addition, the costs are affine and $M=0$,
then $\gamma(\xh)=E/4$.
\end{applemma}

\begin{pf}
Consider the feasible segment $v_\lambda:=(1-\lambda)\xh+\lambda\xs$ for $\lambda\in[0,1]$. Along this segment,
$\Phi(v_\lambda)=\lambda\,\ch\T\delta-\lambda^2\sum_a\mu_a\delta_a^2\ge\lambda(E+M)-\lambda^2E$,
since $\ch\T\delta=E+M$ and $\mu_a\le\sigma_a$ on $\mathcal{P}$ by (F2). Setting $\lambda=1$ and $\lambda=1/2$ gives the two lower bounds.

This follows the segment argument of \citet[Theorem 5.1]{ChenEtAl1997}, who use the modulus of a strongly pseudomonotone map to bound the squared distance to the solution. Here, the inequality $\mu_a\le\sigma_a$ bounds the quadratic term by the energy without requiring such a modulus.

Under the stated condition on $\Delta$, $\cs$ is an admissible backward cost with residual $\Delta$. Fact~(1) in the proof of Theorem~\ref{thm:best} therefore gives $\gamma\le\varrho^2(\cs)=G(\cs)+R^2(\cs)/4$, where $G(\cs)=\cs\T\xh-S(\cs)=M$. For affine costs, $\mu_a=\sigma_a$ on $\mathcal{P}$, so the upper bound becomes $M+E/4$. When $M=0$, the lower and upper bounds coincide at $E/4$.
\end{pf}

\begin{pf}[Proof of Corollary~\ref{cor:margin}]
By Proposition~\ref{prop:sharp}, $I^*(\cb)\subseteq I(\cb)$ and $|I^*(\cb)|\ge2\psi\varrho$. Since $\psi\ge\psi_0$, Theorem~\ref{thm:best} implies that both intervals have width at least $2\psi_0\gamma^{1/2}$. The bounds in terms of $M$ and $E$ then follow from Lemma~\ref{lem:gammabounds}.

At $\cb^*$, we have $\varrho^2=\gamma$ and $R^2/4\le\gamma$, so
$E_{\max}(\cb^*)=(R/2+\varrho)^2\le4\gamma$.
By item~1 of Theorem~\ref{thm:enclosure}, $E\le4\gamma$.

To establish attainment, consider two parallel links with $t_1(x)=t_2(x)=x$, demand 2 and feasible flow $\xh=(4/3,2/3)$. The equilibrium is $\xs=(1,1)$, giving $M=0$ and $E=2/9$. The final case of Lemma~\ref{lem:gammabounds} gives $\gamma=E/4$, establishing that the constant~4 is attained.
\end{pf}
\subsection{Proofs of Propositions~\ref{prop:stop} and~\ref{prop:allocation}}

\begin{pf}[Proof of Proposition~\ref{prop:stop}]
Item~1 follows from Theorem~\ref{thm:enclosure} and Corollary~\ref{cor:benefit}. Both results hold at every exactly feasible point and every admissible backward cost, independently of the checkpoints and computational limits. For computed intervals, validity is preserved by the outward rounding described in Section~\ref{sec:verified}, which can only widen the intervals.

For item~2, the plain certificate has $r=0$, so $R=0$, $\varrho=G^{1/2}$ and $E_{\max}=G$. By Lemma~\ref{lem:width}, its half-width is $\psi G^{1/2}+G/2$, which is denoted by $w_s$ in this proof. Let $\bar G$ be the largest gap along the sequence of certificates, which is finite because the gaps tend to zero. For any link with nonconstant cost, convexity gives $t_a(x+z)-t_a(x)\ge t_a(z)-t_a(0)$ for $x\ge0$. Thus, $\bar z_a$ is bounded above by the smallest $z$ satisfying $\bigl(t_a(z)-t_a(0)\bigr)z\ge\bar G$, denoted by $z^+_a$, which is independent of the computed point. Every feasible link flow is bounded above by the total demand $\bar d$, so $\bar\sigma_a$ is bounded above by the right derivative of $t_a$ at $\bar d+z^+_a$. On links with constant cost, $\bar\sigma_a=0$. Hence $\psi$ remains bounded along the sequence, and $w_s\to0$ as $G\to0$.

Until its first stop, the idealized procedure with the stopping test computes the same certificates as the procedure without it. Along this sequence, the width $2(w_0+w_s)$ of the current benefit interval tends to $2w_0<\epsabs\TC_0\le\hat\tau_s$. By Lemma~\ref{lem:inside}, the estimate $\Bh_s$ lies within the current benefit interval. Since the retained intersection is contained in this interval, $\mathcal{E}_s$ is bounded above by its width. The stopping criterion is therefore satisfied after finitely many certificates. Intersecting successive intervals can only increase $\hat\tau_s$.

For item~3, suppose $B_s\ne0$ and $U_s-L_s<|B_s|$. The interval then excludes zero and satisfies $\min(|L_s|,|U_s|)\ge|B_s|-(U_s-L_s)$. As established above, $\mathcal{E}_s$ is bounded above by the width of the current benefit interval. Denote this width by $W_s=2(w_0+w_s)$. Since $U_s-L_s\le W_s$, the stopping criterion $\mathcal{E}_s\le\epsrel\min(|L_s|,|U_s|)$ is satisfied whenever $W_s(1+\epsrel)\le\epsrel|B_s|$. The stated condition guarantees that this inequality holds after finitely many certificates because $W_s\to2w_0$. If $B_s=0$, every interval contains zero, so $\hat\tau_s=0$, and the stopping criterion can be satisfied only if $[L_s,U_s]=\{0\}$.

For item~4, whenever Algorithm~\ref{alg:cia} returns a certificate, its interval is valid by item~1 and is accompanied by the outcome of the stopping test. A finite sequence of checkpoints or a computational limit may terminate the procedure before the stopping criterion is satisfied, in which case the returned certificate is flagged accordingly.
\end{pf}

\begin{pf}[Proof of Proposition~\ref{prop:allocation}]
The objective function is strictly convex and decreasing in each width, so all constraints are active at the optimum. Substituting $w_s=\tau_s-w_0$ gives a strictly convex function of $w_0$ on $(0,\min_s\tau_s)$ that tends to $+\infty$ at both endpoints. Its derivative,
$-c_0/w_0+\sum_sc_s/(\tau_s-w_0)$, increases strictly from $-\infty$ to $+\infty$ and therefore has a unique root, given by \eqref{eq:allocation}. When all cost coefficients and tolerances are equal, the solution is $w_0=\tau/(m+1)$. The computational work on the base, $c\log\bigl(\bar w_0(m+1)/\tau\bigr)$, therefore grows logarithmically with $m$.
\end{pf}

The constraint $w_0+w_s\le\tau_s$ in Proposition~\ref{prop:allocation} is necessary but not sufficient for $\mathcal{E}_s\le\tau_s$. The certified error bound $\mathcal{E}_s$ is at least the half-width of the retained benefit interval and at most the width of the current benefit interval when the latter contains both $\Bh_s$ and $[L_s,U_s]$ (Section~\ref{sec:benefitint}). For the midpoint estimate $\Bm_s$, whose certified error is $w_0+w_s$ when $w_s$ denotes the half-width of the retained scheme interval, the constraint is both necessary and sufficient. Comparisons between schemes sharing the same base are independent of the base interval (Corollary~\ref{cor:compare}).

\subsection{The Gap Near an Equilibrium with Tied Routes}

The following lemma explains the difference between the observed total-cost errors and their certified bounds (Section~\ref{sec:conservative}). It identifies three nonnegative components of the gap near equilibrium: the changes in used-route costs relative to the smallest cost change among routes that are shortest at equilibrium, the flow assigned to routes with positive excess costs at equilibrium, weighted by those excess costs, and a quadratic term in the flow perturbation.

\begin{applemma}\label{lem:firstorder}
Let the link costs be affine, $t(x)=A+Bx$ with $B=\mathrm{diag}(b_a)\ge0$, and let $h^*$ be an equilibrium route flow
with link flow $\xs=\Lambda h^*$, equilibrium OD costs $\pi^*_w$, shortest routes $\Routes^*_w$ and used routes
$\mathcal{U}_w\subseteq\Routes^*_w$. For a move $\delta h$ with $\sum_{k\in\Routes_w}\delta h_k=0$ for every $w$ and
$h^*+\delta h\ge0$, let $\Delta x:=\Lambda\delta h$, $(B\Delta x)_k:=\sum_{a\in k}b_a\Delta x_a$ and
$m_w:=\min_{k\in\Routes^*_w}(B\Delta x)_k$. Then
\begin{align}
\Gc(\xs+\Delta x)&\ge\sum_w\sum_{k\in \mathcal{U}_w}h^*_k\bigl((B\Delta x)_k-m_w\bigr) \nonumber\\
&\quad+\sum_w\sum_{k\in\Routes_w}\delta h_k\bigl(c^*_k-\pi^*_w\bigr)+\Delta x\T B\Delta x, \label{eq:firstorder}
\end{align}
where $c^*_k$ is the cost of route $k$ at $\xs$. Each term is nonnegative, and equality holds when the shortest routes at
$\xs+\Delta x$ lie in $\Routes^*_w$ for every $w$.
\end{applemma}

\begin{pf}
At $x=\xs+\Delta x$, the cost of route $k$ is $c^*_k+(B\Delta x)_k$. Hence
$t(x)\T x=\sum_wd_w\pi^*_w+\sum_kh^*_k(B\Delta x)_k+\sum_k\delta h_kc^*_k+\sum_k\delta h_k(B\Delta x)_k$.
The last sum equals $\Delta x\T B\Delta x$, and the third equals $\sum_w\sum_k\delta h_k(c^*_k-\pi^*_w)$, since the route flow perturbations sum to zero within each OD pair.

The shortest-route cost of each OD pair at $x$ is at most $\pi^*_w+m_w$, with equality under the stated condition. Subtracting $\sum_wd_w(\pi^*_w+m_w)$ and using $\sum_{k\in\mathcal{U}_w}h^*_k=d_w$ gives \eqref{eq:firstorder}.

The first term is nonnegative because $\mathcal{U}_w\subseteq\Routes^*_w$. The second is nonnegative because flow can be removed only from routes used at equilibrium, for which $c^*_k=\pi^*_w$. The third is nonnegative because $B\ge0$.
\end{pf}

A flow perturbation between routes that are shortest at equilibrium contributes nothing to the second term. However, the first term grows linearly with the magnitude of the perturbation when it increases the cost of a used route relative to the least costly route among those tied at equilibrium. When either of the first two terms is positive along a fixed direction of perturbation, the gap has a first-order component, and the total-cost error is at most of the order of the gap. In Example~\ref{ex:tworoute}, both terms vanish and only the quadratic term remains, giving rise to the square-root relationship between the gap and the total-cost error.
\section{Verified Evaluation of the Enclosure}\label{app:verified}

All quantities entering the interval of Theorem~\ref{thm:enclosure} are evaluated using verified arithmetic, ensuring that rounding can only widen the interval. The certificate applies to the problem defined by the stored binary representations of the network parameters and demand. The evaluation proceeds as follows. If any step cannot be completed, no certificate is returned.

\begin{enumerate}
\item Exact feasibility. The solver stores flows by origin on acyclic subnetworks. Negative flows are clipped to zero, and flow conservation at each node and demand satisfaction for each OD pair are restored exactly using 128-bit integers at a fixed scale of $2^{80}$. The repair verifies that each origin's subnetwork is acyclic, that flow conservation holds exactly at every node, that no artificial link carries flow, and that no flow leaves a zone through which the source data prohibit routes, except at the origin. The repaired link flow $\xh$ is therefore a sum of flows on admissible routes and belongs exactly to $\Feas$. A repair requiring negative flow is rejected.

\item Backward cost. The point $y$ used in \eqref{eq:jn} is evaluated as stored, with tiny negative entries clipped to zero. For the plain certificate of Section~\ref{sec:schemes}, $y$ is obtained by rounding the repaired point $\xh$ to double precision, so $\cb$ corresponds to the link costs at $\xh$ up to rounding. Formula~\eqref{eq:remainderformula} gives a residual of second order in the rounding error. For every plain certificate in Section~\ref{sec:experiments}, $R^2$ was less than $10^{-50}G$.

\item Link quantities. The quantities $\ch$, $\cb$, $\mu$ and $r$ are evaluated using interval arithmetic with outward rounding, with integer powers computed by repeated multiplication. For a BPR link with integer power $\beta_a$, the residual of the Josephy\textendash Newton cost is evaluated using the cancellation-free formula
\begin{equation}\label{eq:remainderformula}
r_a = t_a^0\alpha_a C_a^{-\beta_a}(\xh_a-y_a)^2\sum_{j=0}^{\beta_a-2}(\beta_a-1-j)\,\xh_a^{\,j}\,y_a^{\,\beta_a-2-j},
\end{equation}
avoiding the loss of numerical accuracy from directly evaluating $t_a(\xh_a)-\cb_a$ when $\xh_a-y_a$ is small. If the Josephy\textendash Newton cost is negative and clipped to zero (item~5), the residual is instead $\ch_a$. Taking the minimum of \eqref{eq:remainderformula} and $\ch_a$ therefore gives the residual $\ch_a-\cb_a$ for the clipped backward cost on every link.

\item Weighted residual. An upper bound on $R^2$ is computed. If any link has a secant with lower endpoint zero and a residual with positive upper endpoint, no certificate is returned.

\item Backward gap. An upper bound on $G$ is obtained by subtracting a rigorous lower bound on $S(\cb)$ from the upper bound on $\cb\T\xh$. The lower bound is computed using Dijkstra's algorithm on the lower endpoints of the link cost intervals, with every addition rounded down. The search respects the restrictions on routes through zones, ensuring that $S$ is evaluated over the route sets defining $\Feas$. Clipping the backward costs at zero ensures their nonnegativity, allowing Dijkstra's algorithm to provide a valid dual bound.

\item Certificate quantities. Upper bounds are computed for $E_{\max}$, the steps $\bar z_a$, the slopes $\bar\sigma_a$, the externality weight $\psi$ and the backward residual $\varrho$. Each step $\bar z_a$ is obtained using verified root evaluation.

\item Total-cost estimates. A lower bound on $\SPTTm(\xh)$ is obtained from the dual bound computed at the lower endpoints of $\ch$. An upper bound is obtained by evaluating the cost of the shortest-route flow found, with upward rounding. The value of $\TSTT(\xh)$ is evaluated with outward rounding.

\item Certified intervals. The intervals $I_{\mathrm{S}}$ and $I_{\mathrm{T}}$ are computed and intersected. The term $-S(r)\le0$ is omitted from the upper endpoint of $I_{\mathrm{S}}$. This does not change the intersection $I$, since its upper endpoint is determined by $I_{\mathrm{T}}$ (Lemma~\ref{lem:width}). Disjoint intervals indicate an inconsistency in the inputs or implementation, and no certificate is returned. The benefit interval is then obtained by interval subtraction.
\end{enumerate}

A certificate is also refused if any demand is not a multiple of $2^{-80}$, any flow or bound is nonfinite, any OD pair has no admissible route, any link cost falls outside Assumption~\ref{ass:costs}, any step $\bar z_a$ cannot be verified, or the computed upper bound on $G$ is negative. Flow repair and shortest-path bounds are implemented in compiled code, while interval arithmetic uses vectorized operations with exact summation. The compiled certificate evaluation produced bit-for-bit identical intervals to those of an independent reference implementation of the same procedure on all 459 test cases.
\section{Experimental Settings}\label{app:settings}

Table~\ref{tab:settings} summarizes the experimental settings for Section~\ref{sec:experiments}. The settings specific to the Newton route are provided in Table~S3 of the Supplementary Material.

\begin{table}[!ht]
\centering
\caption{Experimental settings, hardware and software.}\label{tab:settings}
\footnotesize
\begin{adjustbox}{max width=\linewidth}
\begin{tabular}{@{}>{\raggedright\arraybackslash}p{0.23\textwidth}>{\raggedright\arraybackslash}p{0.73\textwidth}@{}}
\toprule
Item & Setting \\
\midrule
\multicolumn{2}{@{}l}{\textit{Certified incremental appraisal}} \\
Base & Algorithm B with a cold start to relative gap $10^{-7}$, followed by plain certification at checkpoints $10^{-8}, 3\times10^{-9}, 10^{-9}, \dots, 3\times10^{-13}, 10^{-13}$ (11 values) \\
Base checkpoint & First checkpoint satisfying the half-width target $\tau/(m+1)$ (Proposition~\ref{prop:allocation}), or the final checkpoint if the target is not met. Selected gaps: $10^{-11}$ for one scheme, $3\times10^{-13}$ for ten schemes and $10^{-13}$ for full programs \\
Scheme initialization & Warm start from the stored base solver state. The method uses the final state of the plain base. Both blind tests and the comparison of certification routes use the Newton base's solver state (Section~\ref{sec:impl}) \\
Scheme checkpoints $g_j$ & $10^{-6}, 3\times10^{-7}, 10^{-7}, \dots, 3\times10^{-12}, 10^{-12}$ (13 values) \\
Scheme limits & 1,200 iterations, 64,000 s of solver time, or gap stagnation. From iteration 30 onward, stagnation occurs when the best gap exceeds half the best gap recorded 30 iterations earlier \\
Certificate and repair & Theorem~\ref{thm:plain} at the computed link costs, with backward cost~\eqref{eq:jn} evaluated at the double-precision rounding of $\xh$ after exact flow repair \\
Tolerance & $\epsabs=10^{-5}$, $\epsrel=0.10$, and $\TC_0$ equal to the reference base TSTT \\
Continuation after the stop & For experimental evaluation only, until two additional checkpoints are certified or a tolerance ten times smaller is met \\
\midrule
\multicolumn{2}{@{}l}{\textit{References and comparison protocols}} \\
Comparison protocols & A$(10^{-6})$, W$(10^{-6})$ and W$(10^{-7})$ (Section~\ref{sec:refs}), executed within the same computational task as the corresponding certified scheme assignment \\
References & Blind sets: cold-start Algorithm B to relative gap $10^{-14}$ or its time limit, with a second solver setting to estimate $u_{\mathrm{ref}}$, which is below $\tau_s/10$ for every scheme. Held-out set: Algorithm B at relative gap $10^{-10}$, with uncertainty below $\tau_s/10$. Development sets: Algorithm B at relative gaps near $10^{-12}$ for regional single-link schemes and $10^{-10}$ for area schemes, and high-accuracy references from the method's development for the benchmark networks, with largest uncertainty $0.0012\,\tau_s$ \\
\midrule
\multicolumn{2}{@{}l}{\textit{Reproducibility}} \\
Assignment algorithm & Algorithm B in tap-b (commit 3dc12b3), extended to support affine link costs and save solver states at checkpoints \\
Exact repair & Integer arithmetic at scale $2^{80}$ using 128-bit integers \\
Cost accounting & Solver time up to the stopping checkpoint plus the time required to evaluate all certificates up to that checkpoint. The solver is paused during certificate evaluation, so that the two computations never share a core. Paused time is excluded from solver time \\
Hardware & Intel Core i7-13700F, 16 GB memory, Ubuntu 24.04 under WSL2, one thread per run, with three concurrent runs for the regional networks and the comparison of certification routes \\
Software & Python 3.9.25 with NumPy 1.26.4, certificate program compiled with GCC 15.2.0 \\
Compiler flags & \texttt{-O2} and \texttt{-ffp-contract=off} \\
\bottomrule
\end{tabular}
\end{adjustbox}
\floatnote{The scheme settings apply to the plain route and both blind tests. The scheme assignments in both blind tests use the Newton base's solver state (Section~\ref{sec:impl}).}
\end{table}

\clearpage
\suppstart
\begin{center}
{\large\bfseries Supplementary Material for ``Certified benefit intervals for network changes under user
equilibrium''}
\end{center}

\noindent Contents
\suppcontents

\medskip
\noindent Numbers without the prefix S refer to the paper, namely its sections, equations, tables, theorems and
appendices.
\section{Positioning Against Prior Work}\label{supp:positioning}
\addtocontents{stoc}{\protect\tocnote{Table~S1 compares prior work and the proposed approach across six properties relevant to certified appraisal.}}

Table~\ref{tab:positioning} compares the literature reviewed in Section~\Mref{sec:literature} of the main paper across six properties relevant to certified appraisal. The classification is based on the results established in the full texts of the cited works. A blank cell indicates that the cited work does not establish the corresponding property. The methods reviewed do not establish computable bounds on equilibrium total travel cost or scheme benefits from approximate deterministic user equilibrium assignments. The proposed approach combines all six properties.

\begin{table}[H]
\centering
\caption{Prior work and the proposed approach across six properties relevant to certified appraisal.}\label{tab:positioning}
\footnotesize
\setlength{\tabcolsep}{4pt}
\begin{adjustbox}{max width=\linewidth}
\begin{tabular}{@{}>{\raggedright\arraybackslash}p{0.32\textwidth}*{6}{>{\centering\arraybackslash}p{0.095\textwidth}}@{}}
\toprule
Work & User equilibrium & Bounds an output & Computable & Weak assumptions & Bounds a benefit & Verified arithmetic \\
\midrule
Convergence of assignments \citep{RoseEtAl1988,BoyceEtAl2004,InoueMaruyama2012,PatilEtAl2021} & \checkmark &  &  & n/a &  &  \\
Error bounds for stochastic user equilibrium \citep{BagchiBoyles2025} & (\checkmark)$^a$ & \checkmark & (\checkmark)$^b$ &  &  &  \\
Sensitivity analysis and its precision \citep{TobinFriesz1988,Patriksson2004,JosefssonPatriksson2007,BarGeraEtAl2013} & \checkmark &  &  &  &  &  \\
Stability of equilibria \citep{DafermosNagurney1984} & \checkmark &  &  &  & (\checkmark)$^d$ &  \\
Error bounds for strongly monotone variational inequalities \citep{Pang1987,LarssonPatriksson1994,FacchineiPang2003} &  &  & \checkmark &  &  &  \\
Error bounds for monotone affine problems \citep{MangasarianShiau1986,FacchineiPang2003} &  &  &  & \checkmark &  &  \\
Weak sharp solutions \citep{MarcotteZhu1998} &  &  &  &  &  &  \\
Gap and merit functions \citep{Fukushima1992,AltangerelEtAl2007} &  &  & n/a & n/a &  &  \\
Output bounds and estimates for partial differential equations \citep{ParaschivoiuEtAl1997,BeckerRannacher2001} &  & \checkmark & (\checkmark)$^c$ & n/a &  &  \\
Verified linear programming \citep{NeumaierShcherbina2004,KeilJansson2006,Rump2010} &  & \checkmark & \checkmark & n/a &  & \checkmark \\
Appraisal guidance \citep{DfT2026TAGM31} & \checkmark &  &  & n/a &  &  \\
\midrule
This paper & \checkmark & \checkmark & \checkmark & \checkmark & \checkmark & \checkmark \\
\bottomrule
\end{tabular}
\end{adjustbox}
\floatnote{User equilibrium: addresses user equilibrium in traffic assignment. Bounds an output: bounds or estimates the error of a scalar output, rather than the distance to a solution or the error in flows. Computable: can be evaluated at the computed solution using known quantities and constants. Weak assumptions: requires neither strong monotonicity, nondegeneracy nor uniqueness of the solution. Bounds a benefit: bounds the difference in equilibrium total costs between network configurations. Verified arithmetic: accounts rigorously for floating-point errors. A check mark indicates that the property is established, a parenthesized check mark indicates partial fulfillment, and a blank cell indicates that the property is not established. The partial entries are $^a$for logit stochastic user equilibrium only, $^b$under proximity to the solution without a computable check of this condition, $^c$against a fine reference mesh rather than the exact solution, or as an error estimate rather than a rigorous bound, and $^d$for changes in equilibrium flows only. n/a: not applicable to the problem or result considered.}
\end{table}
\section{The Supergradient Certificate}\label{supp:j2}

\addtocontents{stoc}{\protect\tocnote{Proposition~S1 establishes a certificate using a second feasible flow. Table~S2 reports its checks in exact rational arithmetic.}}

Corollary~\Mref{cor:margin} of the main paper establishes that every single-interval certificate of Theorem~\Mref{thm:enclosure} has a width bounded below by a function of the equilibrium margin. We consider a certificate that incorporates an additional feasible flow $y$ through the supergradient inequality $S(\cs)\le\cs\T y$. This inequality provides additional information about the equilibrium margin and leads to the following result.

\begin{proposition}\label{prop:j2}
For every $\cb\in\mathbb{F}(\xh)$ and every $y\in\Feas$, there is an interval $I_2(\cb,y)$, computable from $\xh$, $y$
and $\cb$, that contains $\Js$ and lies inside $[\ell^*,u_{\mathrm{T}}]$. At the residual-minimizing backward cost $\cb^*$ and $y=v^*$, under conditions
stated below, one of which involves the equilibrium, its half-width is at most $\psi E^{1/2}$ times a factor set by
ratios of link slopes,
and it contains no margin term. For the family $(\cb_y,y)$, however, $|I_2|\ge2\psi\min\bigl(\varrho,\,G_y^{1/2}\bigr)$.
\end{proposition}

Proposition~\ref{prop:j2} extends the analysis of the square-root dependence of certificate width on the gap (Section~\Mref{sec:price} of the main paper) to the family of pairs $(\cb_y,y)$. The supergradient certificate removes the equilibrium margin from the width bound under the stated conditions, and its half-width is at least $\psi\min(\varrho,G_y^{1/2})$. Achieving a half-width of at most $\varepsilon$ therefore requires either $\varrho\le\varepsilon/\psi$, as for the single-interval certificate, or an auxiliary gap $G_y\le(\varepsilon/\psi)^2$. In the latter case, the square-root dependence applies to the gap of the auxiliary assignment problem. The rate result at $(\cb^*,v^*)$ characterizes the dependence of the certificate width on the equilibrium energy.

An alternative use of an auxiliary assignment is the Newton route, in which the affine assignment constitutes a Josephy\textendash Newton step for the scheme itself. The certificate is evaluated at the resulting point, where the remainder is of second order under the conditions of Proposition~\ref{prop:remainder}.

\paragraph{Construction of the interval.}
Take $\cb\in\mathbb{F}(\xh)$ and $y\in\Feas$, and define
$G_y:=\cb\T y-S(\cb)\ge0$, $\zeta_a:=\xh_a-y_a$, $A:=\tfrac12\sum_{\mathcal{P}}r_a\zeta_a-\sum_{\mathcal{Z}}r_ay_a$,
$T:=R^2/4-A+G_y$ and $\psi_\zeta:=\bigl(\sum_a\bar\sigma_a\zeta_a^2\bigr)^{1/2}$, where $\bar\sigma_a$ is defined as in Section~\Mref{sec:quantities} of the main paper.

Let $\bar q:=\min\bigl(\varrho,\,(\psi_\zeta+(\psi_\zeta^2+4T)^{1/2})/2\bigr)$, and define $p(q):=\psi q-q^2$ for $q\le\psi/2$ and $p(q):=\psi^2/4$ otherwise. The endpoints of the certificate are
\begin{align}
\ell_2 &:= \TSTT(\xh)-\tfrac12r\T\xh-\varrho^2-p(\bar q), \label{eq:j2}\\
u_2 &:= \TSTT(\xh)-\tfrac12r\T\xh+\psi\bar q-\max\bigl(0,\,G+A-G_y-\psi_\zeta\bar q\bigr). \nonumber
\end{align}
Then $\Js\in I_2:=[\ell_2,u_2]$ for every equilibrium, and $I_2\subseteq[\ell^*,u_{\mathrm{T}}]$. Step~(iv) of the following proof establishes that a real $q$ satisfying $q^2-\psi_\zeta q-T\le0$ exists at every equilibrium. Since an equilibrium exists, $\psi_\zeta^2+4T\ge0$, and $\bar q$ is well defined.

\begin{pf}
Let $q$ be as defined in Step~5 of the proof of Theorem~\Mref{thm:enclosure} (Appendix~\Mref{app:proofs} of the main paper).

(i) Step~1 gives $\ch\T\delta\le r\T\delta+G$, where $\ch\T\delta=E+M$. Step~5 also gives $E-r\T\delta\ge q^2-R^2/4$. Combining these inequalities yields $M\le\varrho^2-q^2$.

(ii) The supergradient inequality $S(\cs)\le\cs\T y$ gives $M\ge\cs\T\zeta=\ch\T\zeta-\Delta\T\zeta$, with $\ch\T\zeta=\cb\T\zeta+r\T\zeta=G-G_y+r\T\zeta$. Writing $\Delta=\omega+r/2$ on $\mathcal{P}$ and $\Delta=\omega$ on $\mathcal{Z}$, and using $\xh_a=0$ on $\mathcal{Z}$, gives $r\T\zeta-\Delta\T\zeta=A-\omega\T\zeta$. Hence $M\ge G+A-G_y-\omega\T\zeta$.

(iii) By the Cauchy\textendash Schwarz inequality and $\sigma_a\le\bar\sigma_a$, we have $|\omega\T\zeta|\le\psi_\zeta q$. Consequently, $M\ge G+A-G_y-\psi_\zeta q$.

(iv) Since $M\ge0$, the inequalities in (i) and (iii) give $q\le\varrho$ and $q^2-\psi_\zeta q-T\le0$. Thus, $q\le\bar q$.

(v) The total-cost error satisfies $\TSTT(\xh)-\Js=\xh\T\Delta+M$, where $\xh\T\Delta\in[\tfrac12r\T\xh-\psi q,\tfrac12r\T\xh+\psi q]$. The upper bound $\psi q+\varrho^2-q^2\le\varrho^2+p(\bar q)$ gives the lower endpoint of $I_2$. The expression $-\psi q+\max(0,G+A-G_y-\psi_\zeta q)$ is nonincreasing in $q$, so evaluating it at $\bar q$ gives the upper endpoint.

(vi) Since $p(\bar q)\le p(\varrho)$ and $\varrho^2+p(\varrho)=p^*_{\mathrm{T}}$, the lower endpoint is at least $\ell^*$. Similarly, $\bar q\le\varrho$ ensures that the upper endpoint is at most $u_{\mathrm{T}}$. Hence $I_2\subseteq[\ell^*,u_{\mathrm{T}}]$.
\end{pf}

\begin{table}[!t]
\centering
\caption{Checks of Proposition~\ref{prop:j2} using exact rational arithmetic.}\label{tab:suppchecks}
\footnotesize
\setlength{\tabcolsep}{4pt}
\begin{adjustbox}{max width=\linewidth}
\begin{tabular}{@{}>{\raggedright\arraybackslash}p{0.31\textwidth}>{\raggedright\arraybackslash}p{0.16\textwidth}rrrrl@{}}
\toprule
Requirement & Evaluated on & Evaluations & \multicolumn{1}{p{0.08\textwidth}}{\raggedleft Verified by check} & Violated & Inconclusive & Check \\
\midrule
\multicolumn{7}{@{}l}{\textit{Validity}} \\
$\Js\in I_2$ & every triple & 36,441 & 36,441 & 0 & 0 & a3 \\
$I_2\subseteq[\ell^*,u_{\mathrm{T}}]$, to within $10^{-30}$ & every pair & 28,746 & 28,746 & 0 & 0 & a3 \\
(i) $M\le\varrho^2-q^2$ & every triple & 36,441 & 36,441 & 0 & 0 & a3 \\
(iii) $\omega_a=0$ where $\sigma_a=0$ & every triple & 36,441 & 36,441 & 0 & 0 & a3 \\
(iii) $\sigma_a\le \bar\sigma_a$ on every link & every triple & 36,441 & 36,431 & 0 & 10 & a3 \\
(iii) $M\ge G+A-G_y-\psi_\zeta q$ & every triple & 36,441 & 35,021 & 0 & 1,420 & a3 \\
(iv) $q\le\bar q$ & every triple & 36,441 & 33,193 & 0 & 3,248 & a3 \\
\midrule
\multicolumn{7}{@{}l}{\textit{Width}} \\
$|I_2|\ge2\psi\bar q$ & every pair & 17,919 & 6,873 & 0 & 11,046 & w \\
$\bar q\ge\min\bigl(\varrho,\max(T_+^{1/2},\psi_\zeta/2)\bigr)$ & every pair & 17,919 & 11,772 & 0 & 6,147 & w \\
$T=G_y$ & pairs $(\cb_y,y)$ & 9,955 & 9,955 & 0 & 0 & w \\
$T=G_y$ and $G+A-G_y=\Phi(y)$ & pairs $(\cb_v,v)$ & 6,388 & 6,388 & 0 & 0 & a3 \\
$\bar q\ge\min(\varrho,G_y^{1/2})$ & pairs $(\cb_y,y)$ & 9,955 & 8,310 & 0 & 1,645 & w \\
$|I_2|\ge2\psi\min(\varrho,G_y^{1/2})$ & pairs $(\cb_y,y)$ & 9,955 & 9,038 & 0 & 917 & w \\
\midrule
\multicolumn{7}{@{}l}{\textit{Rate at $(\cb^*,v^*)$}} \\
$T=0$ and $\varrho^2=\gamma$ & every point & 3,194 & 3,194 & 0 & 0 & a3 \\
$V=R^2/4$ & every point & 3,260 & 3,260 & 0 & 0 & a4 \\
$|I_2|=2\psi\bar q$ & points with $\bar q\le\psi/2$ & 2,897 & 2,897 & 0 & 0 & a4 \\
$2V\le\Delta\T\zeta-E+2\sum_a\mu_a\zeta_a\delta_a$ & every point and equilibrium & 4,157 & 4,157 & 0 & 0 & a4 \\
$V^{1/2}\le(1+\kappa_\sigma^{1/2}/2)E^{1/2}$ & first condition holds & 3,919 & 3,919 & 0 & 0 & a4 \\
$\psi_\zeta^2\le\bar\kappa V$ & both conditions hold & 3,416 & 3,416 & 0 & 0 & a4 \\
half-width of $I_2$ at most $\psi\bar\kappa^{1/2}(1+\kappa_\sigma^{1/2}/2)E^{1/2}$ & both hold, $\bar q\le\psi/2$ & 3,215 & 3,215 & 0 & 0 & a4 \\
\bottomrule
\end{tabular}
\end{adjustbox}
\floatnote{Each row corresponds to an inequality or identity in Proposition~\ref{prop:j2}, steps (i) to (iv) of its proof, or the subsequent analysis of width and rate. Each check is applied to all instances satisfying the corresponding hypotheses. Check a3 uses 3,194 points with nine pairs $(\cb,y)$ each, giving 28,746 pairs and, considering every equilibrium of each instance, 36,441 triples of a point, a pair and an equilibrium. Check w uses 1,991 points with nine pairs each, giving 17,919 pairs, including 9,955 pairs $(\cb_y,y)$. Check a4 uses 3,260 points and every corresponding equilibrium, giving 4,157 pairs of a point and an equilibrium. Here $T_+:=\max(T,0)$. An inequality is verified when the rational brackets of its two sides establish the stated ordering, and is inconclusive when the brackets overlap, as may occur when the bound holds with equality. The analytical proof establishes Proposition~\ref{prop:j2}, and the checks evaluate its statements on these instances.}
\end{table}

\paragraph{Interval width.}
Let $T\ge0$. Since $G+A-G_y=\varrho^2-T$, the expression inside the maximum in $u_2$ equals $\varrho^2-\bar q^2\ge0$ when $\bar q$ is the root of the quadratic. When $\bar q=\varrho$ does not exceed the root, the expression is nonpositive. In both cases,
$|I_2|=u_2-\ell_2=\psi\bar q+\bar q^2+p(\bar q)$,
which equals $2\psi\bar q$ when $\bar q\le\psi/2$ and $(\bar q+\psi/2)^2\ge2\psi\bar q$ otherwise.

The root is at least $T^{1/2}$. For the family $(\cb_y,y)$, we have $r=2\mu\circ\zeta$ on $\mathcal{P}$ and $r=0$ on $\mathcal{Z}$, so $A=\sum_a\mu_a\zeta_a^2=R^2/4$ and $T=G_y\ge0$. Consequently, $\bar q\ge\min(\varrho,G_y^{1/2})$ and $|I_2|\ge2\psi\bar q\ge2\psi\min(\varrho,G_y^{1/2})$ for every $\psi$.

\paragraph{Rate at the residual-minimizing backward cost.}
At $\cb^*$ and $y=v^*$, we have $\zeta=\xh-v^*$, $G_y=0$, $T=0$, $G+A=\gamma$ and $\bar q=\min(\gamma^{1/2},\psi_\zeta)$. Let $V:=\sum_a\mu_a\zeta_a^2$. Since $v^*$ maximizes $\Phi$, the first-order optimality condition gives $\cb^{*\top}(\xs-v^*)\ge0$. Combining this inequality with $\ch\T\delta=E+M$ and $\ch\T\zeta=\cs\T\zeta+\Delta\T\zeta\le M+\Delta\T\zeta$ yields
$2V\le\Delta\T\zeta-E+2\sum_a\mu_a\zeta_a\delta_a$,
in which the equilibrium margin cancels.

Suppose first that $\Delta_a=0$ on every link with $\mu_a=0$. Then
$\Delta\T\zeta\le(\kappa_\sigma E)^{1/2}V^{1/2}$ and $\sum_a\mu_a\zeta_a\delta_a\le(EV)^{1/2}$,
where $\kappa_\sigma:=\max\{\sigma_a/\mu_a:\mu_a>0,\delta_a\ne0\}$, with $\kappa_\sigma:=0$ if the set is empty, in which case $\Delta=0$. It follows that $V^{1/2}\le(1+\kappa_\sigma^{1/2}/2)E^{1/2}$.

Suppose also that $\bar\sigma_a\zeta_a=0$ wherever $\mu_a=0$, a condition that can be checked once $v^*$ is known. Then $\psi_\zeta^2\le\bar\kappa V$, where $\bar\kappa:=\max\{\bar\sigma_a/\mu_a:\mu_a>0,\zeta_a\ne0\}$, with $\bar\kappa:=0$ if the set is empty, in which case $\psi_\zeta=0$. When $\bar q\le\psi/2$, the half-width of $I_2$ is therefore at most
$\psi\bar\kappa^{1/2}(1+\kappa_\sigma^{1/2}/2)E^{1/2}$,
which contains no explicit margin term.

Since $\mu_a:=0$ on $\mathcal{Z}$, the first condition applies to every link unused at $\xh$ and requires $t_a(\xs_a)=t_a(0)$ on these links. Both this condition and $\kappa_\sigma$ depend on the unknown equilibrium, so the rate characterizes the behavior of $I_2$ without providing an additional computable error bound.

\paragraph{Numerical checks.}
The statements of this section were evaluated using exact rational arithmetic on instances with known equilibria (Table~\ref{tab:suppchecks}), with square roots enclosed between rational bounds. No violations were found, and evaluations with overlapping rational bounds are reported as inconclusive.

The interval was also evaluated in double precision at 255 stopping points from the base and 50 schemes on Anaheim. At every point, $I_2$ for the pair $(\cb_v,v)$ contained the reference equilibrium total cost. Its width equaled that of the best single interval at relative gaps from $10^{-4}$ to $10^{-7}$ and was a median factor of 2.8 smaller at $10^{-8}$. These double-precision intervals were not computed using verified arithmetic and therefore do not constitute certificates.
\section{The Newton Route, a Tested Alternative}\label{supp:newton}
\addtocontents{stoc}{\protect\tocnote{Certification through Newton steps, as evaluated in the first blind test, with Proposition~S2 on the linearization remainder and experimental settings in Table~S3.}}

This section describes an alternative certification procedure based on Josephy\textendash Newton steps from a certified base point. The procedure was evaluated in the first blind test of Section~\Mref{sec:blind}. It incorporates affine approximations into the certificate and shows how the square-root dependence extends to the gap of the linearized assignment problem. Section~\ref{supp:routes} compares it with the plain warm-start method of Section~\Mref{sec:schemes}.

\subsection{Newton Steps for the Base}

In the first blind test, the base equilibrium was approximated to a relative gap of $10^{-5}$ and refined using two Josephy\textendash Newton steps. Step $k$ linearizes the link cost functions $t$ at a point $y^{k-1}$, giving the affine costs
\begin{equation}\label{eq:affine}
x_a \mapsto t_a(y^{k-1}_a) + t_a'(y^{k-1}_a)\,\bigl(x_a-y^{k-1}_a\bigr), \qquad a\in\Links.
\end{equation}
The assignment problem with these costs is an affine variational inequality, so each step constitutes a Josephy\textendash Newton step for the equilibrium problem
\citep{Josephy1979,Robinson1980,IzmailovSolodov2010}. Each affine assignment is solved to a tight affine relative gap using the bush-based algorithm, warm-started from the solver state of the preceding assignment.

The resulting point $\xh_0$ is certified by Theorem~\Mref{thm:enclosure}, using the Josephy\textendash Newton cost~\Meqref{eq:jn} at the preceding linearization point $y$ as the backward cost. With this choice, $G$ is the gap of $\xh_0$ in the linearized problem, and $r$ is the linearization remainder, which is nonnegative on every link by convexity. The implementation clips the backward costs at zero to allow the use of Dijkstra's algorithm for evaluating $S(\cb)$ while preserving the nonnegativity of the residual (Appendix~\Mref{app:verified} of the main paper).

\subsection{Newton Steps from the Base's Certified Point}

For each scheme $s$, the Newton steps start from the certified base point. Step $k$ linearizes the scheme's cost functions $t^{(s)}$ at $y^{k-1}$, with $y^0:=\xh_0$, giving the corresponding affine costs~\eqref{eq:affine}. This incorporates the linearization underlying equilibrium sensitivity analysis \citep{TobinFriesz1988} into the certification procedure.

Each affine assignment is warm-started from the solver state of the preceding step, which for $k=1$ is the base's state. The solver is paused at a decreasing sequence of affine relative gaps. At each checkpoint, the iterate is repaired to an exactly feasible point $\xh$ and certified by Theorem~\Mref{thm:enclosure}, using the affine costs as the backward cost, with clipping where required. The benefit interval, benefit estimate, certified error bound and stopping criterion are defined as in Section~\Mref{sec:schemes}.

\subsection{The Linearization Remainder}

The residual-minimizing backward cost of Theorem~\Mref{thm:best} and the Josephy\textendash Newton cost are both associated with affine assignment problems. The former uses the secant weights $2\mu$ at the computed point $\xh$ to minimize the backward residual $\varrho$. The latter uses the tangent of the cost functions at a linearization point $y$ to obtain a new assignment.

The following proposition establishes conditions under which the weighted linearization remainder $R$ in \Meqref{eq:R} is of second order in the Newton step.

\begin{proposition}\label{prop:remainder}
Let each $t_a$ be twice continuously differentiable on $(0,\infty)$, let $\xh$ be a point of a Newton step linearized
at $y$, and let $H_a\in[0,\infty]$ be the supremum of $t_a''$ strictly between $y_a$ and $\xh_a$, with $H_a:=0$ when
$\xh_a=y_a$. Then $0\le r_a\le\tfrac12 H_a(\xh_a-y_a)^2$ on every link, and
\begin{equation}\label{eq:remainder}
R^2 \le \sum_{a\in\mathcal{P}} \frac{H_a^2\,(\xh_a-y_a)^4}{4\mu_a}.
\end{equation}
If the affine assignment is solved exactly and the backward cost is the unclipped cost~\Meqref{eq:jn}, so that
$G=0$, then $|I|\le\psi R+R^2/4$. For fixed $y$ and $\xh\to y$,
the remainder $R$ is $O(\|\xh-y\|^2)$ whenever $H_a^2/\mu_a$ stays bounded on the links with $r_a>0$. This holds on
every such link with $y_a>0$ and $t_a(y_a)>t_a(0)$, on every such link with $y_a=0$ and $t_a'(0)>0$ whose $t_a''$ is
bounded near zero, and on every BPR link with $\beta_a\ge3$.
\end{proposition}

The condition on BPR powers is needed. For a BPR link with $y_a=0$ and $2\le\beta_a<3$, the contribution to $R^2$ is exactly $t_a^0\alpha_a\xh_a^{\beta_a+1}/C_a^{\beta_a}$. The corresponding contribution to $R$ is therefore $O(\|\xh-y\|^{(\beta_a+1)/2})$, which is slower than second order. The networks in Section~\Mref{sec:experiments} use $\beta_a=4$. The second-order result holds for fixed $y$ and is not uniform in $y$, since the bound on $H_a^2/\mu_a$ may grow without limit as the linearization flow of a link approaches zero.

The Newton step reduces the linearization remainder to second order under the conditions of Proposition~\ref{prop:remainder}, and the interval width is $2\psi\varrho$ plus a term of at most the raw gap (Lemma~\Mref{lem:width}), where $\varrho=(R^2/4+G)^{1/2}$ now contains the affine gap $G$. The Newton route therefore retains the square-root dependence of the certificate width on the convergence gap, with the gap evaluated in the linearized assignment problem.

\subsection{The Step Rule}

The reduction in interval width achievable by solving an affine assignment more accurately diminishes as the linearization remainder begins to dominate the affine gap. Following the forcing principle of inexact Newton methods \citep{DemboEtAl1982}, we use the certificate's remainder to determine when to introduce a new linearization.

At a point satisfying $G\le\eta R^2$ for some $\eta>0$,
$\varrho^2=R^2/4+G\le\tfrac14R^2(1+4\eta)\le\tfrac14R^2(1+2\eta)^2$.
Consequently, $\psi\varrho\le\tfrac12\psi R\,(1+2\eta)$, so reducing the affine gap while holding the remainder and externality weight fixed can reduce the energy component of the interval width by at most a factor of $1+2\eta$.

A Newton step therefore ends at the first checkpoint satisfying $G\le\eta R^2$. The next step is linearized at the repaired point of that checkpoint and warm-started from its solver state. Since further assignment iterations also change the computed point and its remainder, this rule is a heuristic motivated by the bound at the current point.

If a step reaches its final checkpoint or computational limit, a new step is also initiated when $R^2/4>G$ at the last certified checkpoint, indicating that the remainder dominates the affine gap. Otherwise, the procedure terminates. The procedure also terminates after $K_{\max}$ steps.

Item~1 of Proposition~\Mref{prop:stop} applies to this procedure independently of the choice of linearization points, since certificate validity requires only that the backward cost satisfy the conditions of Theorem~\Mref{thm:enclosure}.

\subsection{Proof of Proposition~\ref{prop:remainder}}

\begin{pf}
For the Josephy\textendash Newton cost at $y$, the residual is
$r_a=t_a(\xh_a)-t_a(y_a)-t_a'(y_a)(\xh_a-y_a)$.
Convexity ensures that $r_a\ge0$. By Taylor's theorem, $r_a=\tfrac12t_a''(\xi_a)(\xh_a-y_a)^2$ for some $\xi_a$ strictly between $y_a$ and $\xh_a$ when $\xh_a\ne y_a$. Thus, $r_a\le\tfrac12 H_a(\xh_a-y_a)^2$. Clipping the backward cost at zero can only reduce the residual, so the bound also holds for the clipped cost. Squaring, dividing by $\mu_a$ on the links with positive residual in $\mathcal{P}$, and summing gives \eqref{eq:remainder}.

When $G=0$, we have $\varrho=R/2$, and Lemma~\Mref{lem:width} gives
$|I|\le2\psi\varrho+R^2/4+G=\psi R+R^2/4$.

Fix $y$ and let $\xh_a\to y_a$. If $y_a>0$ and $t_a(y_a)>t_a(0)$, then $\mu_a$ converges to the positive secant $\bigl(t_a(y_a)-t_a(0)\bigr)/y_a$, and $H_a$ converges to $t_a''(y_a)$ along nonzero steps. If $y_a=0$ and $t_a'(0)>0$, convexity gives $\mu_a\ge t_a'(0)$ for positive $\xh_a$, and $H_a$ is bounded when $t_a''$ is bounded near zero.

The condition $t_a(y_a)>t_a(0)$ cannot generally be omitted: for $t_a(x)=1+\max(0,x-1)^3$ at $y_a=1$, we have $H_a^2/\mu_a\to\infty$ and $R=\Theta(\|\xh-y\|^{3/2})$ along positive steps. Similarly, the bound on $t_a''$ near zero is needed, as $t_a(x)=1+x+x^{3/2}$ at $y_a=0$ gives the same order.

For a BPR link with $y_a=0$, we have
$H_a=t_a^0\alpha_a\beta_a(\beta_a-1)\xh_a^{\beta_a-2}/C_a^{\beta_a}$ and
$\mu_a=t_a^0\alpha_a\xh_a^{\beta_a-1}/C_a^{\beta_a}$.
Hence $H_a^2/\mu_a$ is proportional to $\xh_a^{\beta_a-3}$ and remains bounded for $\beta_a\ge3$. For $\beta_a>1$ and $y_a=0$, the linearized cost is $t_a^0$, so $r_a=t_a^0\alpha_a(\xh_a/C_a)^{\beta_a}$ and
$r_a^2/\mu_a=t_a^0\alpha_a\xh_a^{\beta_a+1}/C_a^{\beta_a}$,
giving the order stated after the proposition.
\end{pf}
\subsection{Settings}

Table~\ref{tab:newtonsettings} summarizes the settings specific to the Newton route used in the first blind test and the comparison in Section~\ref{supp:routes}. All other settings are given in Table~\Mref{tab:settings} of the main paper.

\begin{table}[H]
\centering
\caption{Experimental settings of the Newton route.}\label{tab:newtonsettings}
\footnotesize
\begin{adjustbox}{max width=\linewidth}
\begin{tabular}{@{}>{\raggedright\arraybackslash}p{0.23\textwidth}>{\raggedright\arraybackslash}p{0.73\textwidth}@{}}
\toprule
Item & Setting \\
\midrule
Base & Algorithm B to relative gap $10^{-5}$, followed by two Newton steps to affine relative gaps $10^{-12}$ and $10^{-14}$, with certification at the final point \\
Steps & Affine assignment with costs~\eqref{eq:affine} solved by Algorithm B, warm-started from the preceding step's solver state, at checkpoints $10^{-6}, 3\times10^{-7}, 10^{-7}, \dots, 3\times10^{-11}, 10^{-11}$ (11 values) \\
Limits and step rule & 300 iterations or 16,000 s per step. Maximum number of steps $K_{\max}=4$ for regional networks and 3 for benchmark networks, where no scheme reached the limit. Branching parameter $\eta=0.01$ \\
Continuation after the stop & Until a tolerance ten times smaller is met, the branching rule terminates the procedure, or the step limit is reached \\
Cost accounting & Solver time up to the stopping checkpoint, including the additional time required to construct the affine costs and evaluate all certificates up to that checkpoint \\
\bottomrule
\end{tabular}
\end{adjustbox}
\end{table}

\section{Additional Results}\label{supp:results}
\addtocontents{stoc}{\protect\tocnote{Tables~S4 to~S9 provide additional experimental results in the order cited in the main paper.}}

This section provides additional results and analyses supporting the experiments in Section~\Mref{sec:experiments} of the main paper.

\subsection{The Held-Out Test}\label{supp:heldout}

Table~\ref{tab:heldout} reports the results for all protocols evaluated in the held-out test of Section~\Mref{sec:practice}. Each warm-start protocol initializes the scheme assignment from the base's solver state at its stopping point. Let $n_0$ denote the number of iterations required for the base to reach the target relative gap $g$. The iteration-matching protocol runs the scheme for exactly $n_0$ iterations from the same initialization as the base. The benefit-stability protocol stops at the first iteration at which the relative gap is at most $g$ and the benefit estimate has changed by at most 1\% over the preceding three iterations.

\begin{table}[H]
\centering
\caption{Accuracy and computational cost of all protocols in the held-out test of 55 schemes.}\label{tab:heldout}
\footnotesize
\begin{adjustbox}{max width=\linewidth}
\begin{tabular}{@{}lrrr@{}}
\toprule
Protocol & Inaccurate benefits & Unfinished runs & Cost relative to A$(g)$ \\
\midrule
\multicolumn{4}{@{}l}{\textit{Target relative gap $g=10^{-4}$}} \\
A, gap stopping & 25 & 0 & \textemdash \\
B, iteration matching & 27 & 0 & 1.000 \\
W, shared-base warm start & 50 & 1 & 0.094 \\
W with $\lceil n_0/4\rceil$ additional iterations & 50 & 0 & 0.240 \\
W with $n_0$ additional iterations & 49 & 0 & 1.132 \\
W with benefit-stability stopping & 7 & 47 & 0.997 \\
\midrule
\multicolumn{4}{@{}l}{\textit{Target relative gap $g=10^{-5}$}} \\
A, gap stopping & 1 & 0 & \textemdash \\
B, iteration matching & 1 & 0 & 1.080 \\
W, shared-base warm start & 0 & 1 & 0.117 \\
W with $\lceil n_0/4\rceil$ additional iterations & 1 & 0 & 0.293 \\
W with $n_0$ additional iterations & 1 & 0 & 1.686 \\
W with benefit-stability stopping & 0 & 32 & 0.775 \\
\midrule
\multicolumn{4}{@{}l}{\textit{Target relative gap $g=10^{-6}$}} \\
A, gap stopping & 0 & 0 & \textemdash \\
B, iteration matching & 0 & 0 & 1.000 \\
W, shared-base warm start & 0 & 1 & 0.117 \\
W with $\lceil n_0/4\rceil$ additional iterations & 1 & 0 & 0.332 \\
W with $n_0$ additional iterations & 1 & 0 & 1.575 \\
W with benefit-stability stopping & 0 & 12 & 0.697 \\
\bottomrule
\end{tabular}
\end{adjustbox}
\floatnote{A benefit estimate is classified as inaccurate when its absolute difference from the reference benefit exceeds $\tau_s$. A run is unfinished when it fails to reach the required stopping criterion within its computational limit. Cost is the median scheme-level cost relative to A$(g)$. W denotes the shared-base warm-start protocol W$(g)$ of Section~\Mref{sec:refs}.}
\end{table}

\subsection{Computational Checks of the Theory}\label{supp:checks}

Table~\ref{tab:checks} provides the detailed results of the checks summarized in Section~\Mref{sec:designed} of the main paper.

\begin{table}[H]
\centering
\caption{Checks of theoretical results on instances with known equilibria.}\label{tab:checks}
\footnotesize
\begin{adjustbox}{max width=\linewidth}
\begin{tabular}{@{}>{\raggedright\arraybackslash}p{0.31\textwidth}>{\raggedright\arraybackslash}p{0.27\textwidth}rrl@{}}
\toprule
Statement & Test family & Evaluations & Violations & Verification script \\
\midrule
Theorem~\Mref{thm:enclosure}, random certificates & 300 networks & 9,000 & 0 & audit\_theorem1 \\
Theorem~\Mref{thm:enclosure}, adversarial search & 60 searches of 150 steps, 4 targets & 36,000 & 0 & audit\_theorem1 \\
Lemma~\Mref{lem:width}, width identity & General, Josephy\textendash Newton and clipped Josephy\textendash Newton costs & 7,297 & 0 & t1 \\
Lower bounds, exact equilibria & Multiple-OD instances & 22,500 & 0 & t2 \\
Proposition~\Mref{prop:sharp}, refined lower bound & $r=0$, $r\ge0$, mixed signs & 36,000 & 0 & t8 \\
Theorem~\Mref{thm:enclosure} and Proposition~\Mref{prop:sharp}, BPR & Instances with designed exact equilibria & 13,500 & 0 & t9 \\
Proposition~\Mref{prop:law} & Integer and non-integer powers from 1 to 6, in 60-digit floating-point arithmetic & 3,000 & 0 & t4 \\
Lemma~\Mref{lem:firstorder} & Flow perturbations between routes & 6,000 & 0 & t6 \\
\bottomrule
\end{tabular}
\end{adjustbox}
\floatnote{Each evaluation tests a stated bound or identity on an individual instance. The checks use exact rational arithmetic, except the check of Proposition~\Mref{prop:law}, whose non-integer BPR powers require floating-point arithmetic. No violations were found. These computational checks complement the analytical proofs in Appendix~\Mref{app:proofs} of the main paper. The final column identifies the corresponding verification script in the code archive.}
\end{table}
\subsection{The First Blind Test}\label{supp:blindK}

Table~\ref{tab:blindK} reports the results of the first blind test of Section~\Mref{sec:blind} by network. The protocol established three prespecified acceptance criteria. H1 requires every certified benefit interval at every checkpoint of every step to contain $B_{\mathrm{ref}}$ within $u_{\mathrm{ref}}$. H2 requires at least 95\% of the schemes in scope to meet $\hat\tau_s$ within the step limit. H3 requires at least 95\% of these schemes to achieve the tolerance at a computational cost no greater than ten times that of the paired A$(10^{-6})$ run. All three criteria were satisfied.

\begin{table}[H]
\centering
\caption{Results of the first blind test using the Newton route on two regional networks.}\label{tab:blindK}
\footnotesize
\begin{adjustbox}{max width=\linewidth}
\begin{tabular}{@{}lrrrrrrcrr@{}}
\toprule
 &  &  & Valid & In & $\hat\tau_s$ met & \multicolumn{3}{c}{Cost relative to A$(10^{-6})$} & Cost relative to \\
\cmidrule(lr){7-9}
Network & Schemes & Certificates & (H1) & scope & (H2) & Median & Quartiles & Maximum & W$(10^{-7})$ \\
\midrule
Chicago regional & 48 & 1,101 & 48 & 43 & 43 & 2.93 & 2.15 to 3.72 & 4.70 & 10.3 \\
Philadelphia & 48 & 1,032 & 48 & 48 & 48 & 2.54 & 1.68 to 3.10 & 3.88 & 10.1 \\
All & 96 & 2,133 & 96 & 91 & 91 & 2.64 & 1.75 to 3.33 & 4.70 & 10.1 \\
\bottomrule
\end{tabular}
\end{adjustbox}
\floatnote{The prespecified test uses the Newton route of Section~\ref{supp:newton}. The Valid column counts schemes for which every certified interval contains the reference benefit within $u_{\mathrm{ref}}$ (H1). Schemes with multipliers above 0.01 are in scope. The $\hat\tau_s$ column counts schemes in scope that meet the computable tolerance within four Newton steps (H2). Computational costs are measured at the stopping checkpoint relative to the paired cold-start assignment A$(10^{-6})$ and shared-base warm-start assignment W$(10^{-7})$. All 91 schemes in scope satisfy the cost criterion of H3.}
\end{table}

The following descriptive analyses were conducted after the blind results were opened, without changing the prespecified stopping criterion or benefit estimate. Stopping on the midpoint estimate $\Bm_s$, using $U_s-L_s\le2\hat\tau_s$, gives the same stopping checkpoint as stopping on $\Bh_s$ for all 91 schemes in scope on both routes and all 205 development schemes of the plain route. Retaining the intersection of successive intervals, as in Algorithm~\Mref{alg:cia}, also gives the same stopping checkpoints and decisions across these runs. The reported results of the plain route therefore correspond to those of Algorithm~\Mref{alg:cia}. At the stopping checkpoints of the Newton route, the distance between $\Bh_s$ and the midpoint is less than or equal to 1\% of the interval half-width.

The recorded assignment trajectories also allow the effect of alternative tolerances to be examined. Increasing the absolute tolerance by a factor of ten allows every scheme to meet the stopping criterion at a median cost of 0.63 times that of A$(10^{-6})$. At $\epsabs=10^{-6}$, most trajectories terminate under the step rule of Section~\ref{supp:newton} before satisfying the tighter tolerance. At the original stopping checkpoints, the affine gap accounts for the larger component of $\varrho^2$ in 84 of the 91 schemes. The linearization remainder dominates in the remaining seven schemes, all of which stop during the first Newton step.

The median computational times per certificate are 1.25~s for exact flow repair, 6.54~s for shortest-path computations, 0.17~s for interval arithmetic and 2.68~s for evaluating the point estimates. Certificate evaluation accounts for a median of 13.9\% of the total computational cost at the stopping checkpoint.

\subsection{The Base and Whole Programs}\label{supp:base}

Table~\ref{tab:r2} reports the base certificates obtained using the plain assignment procedure, and Table~\ref{tab:programme} compares reconstructed program-level costs using the plain and Newton base certificates.

\begin{table}[H]
\centering
\caption{Base certification without Newton steps.}\label{tab:r2}
\footnotesize
\begin{adjustbox}{max width=\linewidth}
\begin{tabular}{@{}lrrrrr@{}}
\toprule
 & \multicolumn{3}{c}{Plain base} & Target & Newton base \\
\cmidrule(lr){2-4}
Network & Half-width & Relative gap & Cost (s) & half-width $\tau/(m+1)$ & half-width \\
\midrule
Sioux Falls & 2.63 & $10^{-13}$ & 0.1 & 1.53 & 1.04 \\
Anaheim & 0.252 & $10^{-13}$ & 0.3 & 0.29 & 0.0674 \\
Chicago-Sketch & 3.82 & $3\times10^{-13}$ & 3.1 & 3.75 & 1.16 \\
Chicago regional & 11.5 & $10^{-13}$ & 4,105.3 & 6.87 & 5.2 \\
Philadelphia & 82.6 & $10^{-13}$ & 1,884.2 & 68.5 & 27.5 \\
\bottomrule
\end{tabular}
\end{adjustbox}
\floatnote{Algorithm~B is continued on the base's own costs from a solver state at relative gap $10^{-7}$, with plain certificates evaluated at checkpoints down to $10^{-13}$. The plain-base columns report the smallest half-width attained and the corresponding relative gap and computational cost. The target $\tau/(m+1)$ follows the allocation of Proposition~\Mref{prop:allocation} for $m=48$ schemes under its cost model. Exceeding this target does not invalidate the benefit intervals, but requires narrower scheme intervals to achieve the prescribed accuracy. The final column reports the half-width of the Newton base certificate used in the first blind test.}
\end{table}

For programs of this size, the proposed method initializes each scheme from the final solver state of the plain base assignment, at relative gap $10^{-13}$. Both blind tests instead use the Newton base's solver state (Section~\Mref{sec:impl}). To assess the effect of this difference, the 40 schemes of the regional development set were rerun from the plain base's solver state, retaining the same plain base interval and all other experimental settings. Their results were compared with those obtained from the recorded assignments evaluated using the same plain base interval, according to a comparison protocol fixed before the rerun.

All 40 schemes met $\hat\tau_s$, all 327 certificates contained $B_{\mathrm{ref}}$ within $u_{\mathrm{ref}}$, and the same 18 schemes met T10. The median cost relative to the recorded assignments was 0.96, with quartiles of 0.92 and 0.98 and a range of 0.73 to 1.30. The median ratio of iterations to the stopping checkpoint was 1.00. Of the 40 schemes, 36 stopped at the same checkpoint, three one checkpoint earlier and one a checkpoint later.

\begin{table}[H]
\centering
\caption{Reconstructed program-level computational costs for the two blind sets using the plain and Newton base certificates.}\label{tab:programme}
\footnotesize
\begin{adjustbox}{max width=\linewidth}
\begin{tabular}{@{}lrrrr@{}}
\toprule
 & \multicolumn{2}{c}{Chicago regional} & \multicolumn{2}{c@{}}{Philadelphia} \\
\cmidrule(lr){2-3}\cmidrule(l){4-5}
 & Plain base & Newton base & Plain base & Newton base \\
\midrule
Certified base for one scheme (s) & 3,272 & 6,178 & 1,594 & 5,211 \\
Certified base for all schemes (s) & 4,781 & 6,178 & 2,307 & 5,211 \\
Cold base to $10^{-6}$, for A$(10^{-6})$ (s) & \multicolumn{2}{c}{498} & \multicolumn{2}{c}{276} \\
\midrule
\multicolumn{5}{@{}l}{\textit{First blind set, reconstructed program costs relative to A$(10^{-6})$}} \\
\quad One scheme & 4.06 & 6.56 & 3.54 & 9.37 \\
\quad Ten schemes & 2.39 & 2.58 & 1.94 & 2.79 \\
\quad All schemes in scope & 1.95 & 1.97 & 1.56 & 1.74 \\
\quad All schemes in scope, relative to W$(10^{-7})$ & 6.72 & 6.77 & 6.51 & 7.26 \\
\quad Share of base cost, all schemes & 0.092 & 0.118 & 0.092 & 0.185 \\
\midrule
\multicolumn{5}{@{}l}{\textit{Second blind set, reconstructed program costs, 36 schemes}} \\
\quad Relative to A$(10^{-6})$ & 1.68 & 1.69 & 1.55 & 1.79 \\
\quad Relative to W$(10^{-7})$ & 2.87 & 2.88 & 2.35 & 2.71 \\
\quad Share of base cost & 0.129 & 0.166 & 0.122 & 0.239 \\
\bottomrule
\end{tabular}
\end{adjustbox}
\floatnote{Program-level cost comprises a single base assignment and the computational costs of all schemes at their stopping checkpoints. The plain base uses cold-start Algorithm~B to relative gap $10^{-7}$ and continues to the checkpoint selected according to Proposition~\Mref{prop:allocation} for the program size, including the cost of every certificate evaluated. The Newton base uses a cold-start assignment to $10^{-5}$ followed by the two Newton steps of the first blind test and their certificates. Scheme costs are obtained from the plain route initialized from the Newton base's solver state. Program costs are therefore reconstructed from separate computational runs. Ratios for programs of one and ten schemes use the mean scheme cost, while those for all schemes use the total scheme costs. All computations use the workstation specified in Table~\Mref{tab:settings}.}
\end{table}

\subsection{Comparison of the Two Routes}\label{supp:routes}

Table~\ref{tab:routes} compares the plain and Newton routes on the development schemes (Section~\Mref{sec:cost}). The prespecified decision set comprises the 22 regional single-link development schemes with multipliers above 0.01 and six area schemes on Chicago regional. The other ten area schemes of the regional development set were run after this comparison, with the plain route only, and are not part of it. For each scheme, both routes are evaluated within the same computational job, using the same base interval, certificate implementation, cost accounting and hardware. Both routes use $\hat\tau_s$ as the stopping criterion.

The selection rule uses the median ratio of the plain route's computational cost to that of the Newton route across the decision set. The Newton route is selected if this ratio is at least 1.5 and it fails to meet the tolerance on no more schemes than the plain route, or if the plain route fails on more schemes than the Newton route. Otherwise, the plain route is selected.

The plain route is less expensive on 24 of the 28 schemes in the decision set. On the remaining four schemes, comprising two near-null schemes and two with multipliers of 1.1, its cost is at most 1.044 times that of the Newton route. The median computational costs relative to A$(10^{-6})$ are 1.78 for the plain route and 2.73 for the Newton route. On the six area schemes, the cost of the plain route ranges from 0.24 to 0.81 times that of the Newton route. The prespecified rule therefore selects the plain route.

For the two near-closure schemes on Chicago regional, which are outside the decision set, the Newton route reaches its step limit without meeting the tolerance, while the plain route meets it after 1,216 and 1,533~s. On the benchmark networks, both routes meet the tolerance for all 175 schemes, with median cost ratios of the plain route to the Newton route ranging from 0.86 to 0.98. All 3,722 Newton certificates and 1,373 plain certificates contain their respective reference benefits within $u_{\mathrm{ref}}$. In the cold-start comparison on the ten schemes of Section~\Mref{sec:cost}, the plain route costs a median of 1.34 times as much as the Newton route.

Both routes use Theorem~\Mref{thm:enclosure}. The Newton route incurs additional costs from constructing the linearized problems and performing several affine assignments, with 20 of the 28 schemes in the decision set requiring more than one Newton step.

\begin{table}[H]
\centering
\caption{Computational performance of the plain and Newton routes on development schemes.}\label{tab:routes}
\footnotesize
\setlength{\tabcolsep}{5pt}
\begin{adjustbox}{max width=\linewidth}
\begin{tabular}{@{}lrrrrcrrr@{}}
\toprule
 &  & \multicolumn{2}{c}{$\hat\tau_s$ met} & \multicolumn{2}{c}{Plain relative to Newton} & Plain relative to & \multicolumn{2}{c@{}}{Plain certificates} \\
\cmidrule(lr){3-4}\cmidrule(lr){5-6}\cmidrule(l){8-9}
 & Schemes & Newton & Plain & Median & Quartiles & A$(10^{-6})$ & Computed & Within $u_{\mathrm{ref}}$ \\
\midrule
\textbf{Prespecified decision set} & 28 & 28 & 28 & 0.71 & 0.50 to 0.95 & 1.78 & 256 & 256 \\
\midrule
\multicolumn{9}{@{}l}{\textit{Development schemes in the comparison, by network}} \\
Chicago regional & 18 & 16 & 18 & 0.71 & 0.45 to 0.85 & 1.82 & 160 & 160 \\
Philadelphia & 12 & 12 & 12 & 0.66 & 0.51 to 0.98 & 1.80 & 116 & 116 \\
Sioux Falls & 40 & 40 & 40 & 0.86 & 0.50 to 1.01 & 4.43 & 310 & 310 \\
Anaheim & 50 & 50 & 50 & 0.96 & 0.92 to 0.98 & 4.07 & 269 & 269 \\
Chicago-Sketch & 85 & 85 & 85 & 0.98 & 0.85 to 1.00 & 1.83 & 518 & 518 \\
\bottomrule
\end{tabular}
\end{adjustbox}
\floatnote{The prespecified decision set comprises 28 regional development schemes with multipliers above 0.01 and is used to select the certification route according to the prespecified rule. The remaining rows report the development schemes of the comparison by network, including the two near-closure schemes on Chicago regional. Costs are evaluated at each route's stopping checkpoint, and the ratios include the schemes on which a route ended without meeting the tolerance. A$(10^{-6})$ denotes the cold-start scheme assignment to relative gap $10^{-6}$, executed within the same computational job. The final column counts plain certificates whose intervals contain the reference benefit within $u_{\mathrm{ref}}$, providing a check of numerical consistency.}
\end{table}

\end{document}